\documentclass[fleqn,usenatbib]{mnras}

\usepackage{newtxtext,newtxmath}

\usepackage[T1]{fontenc}

\DeclareRobustCommand{\VAN}[3]{#2}
\let\VANthebibliography\thebibliography
\def\thebibliography{\DeclareRobustCommand{\VAN}[3]{##3}\VANthebibliography}

\usepackage{graphicx}	
\usepackage{amsmath}	

\title[Halo heating from fluctuating gas in a model dwarf]{Halo heating from fluctuating gas in a model dwarf}

\author[M. Hashim et. al.]{
Mahmoud Hashim,$^{1}$
Amr A. El-Zant,$^{1}$\thanks{E-mail: amr.elzant@bue.edu.eg}
Jonathan Freundlich,$^{2}$
Justin I. Read$^{3}$
and Fran\c{c}oise Combes$^{4, 5}$
\\
$^{1}$Centre for Theoretical Physics, The British University in Egypt, Sherouk City, 11837 Cairo, Egypt\\
$^{2}$Observatoire Astronomique, Universite de Strasbourg, CNRS, 11 rue de l’Universite, 67000 Strasbourg, France\\
$^{3}$Department of Physics, University of Surrey, Guildford, GU2 7XH, United Kingdom\\
$^{4}$LERMA, Observatoire de Paris, CNRS, PSL Univ., Sorbonne Univ., F-75014 Paris, France\\
$^{5}$Collège de France, 11 place Marcelin Berthelot, F-75005 Paris, France
}

\date{Accepted XXX. Received YYY; in original form ZZZ}

\pubyear{2022}

\begin{document}
\label{firstpage}
\pagerange{\pageref{firstpage}--\pageref{lastpage}}
\maketitle

\begin{abstract}
The cold dark matter (CDM) structure formation scenario faces challenges on (sub)galactic scales, central among them being the `cusp-core' problem. A known remedy, driving CDM out of galactic centres, invokes interactions with baryons, through fluctuations in the gravitational potential  arising from feedback or orbiting clumps of  gas or stars.
Here we interpret core formation in a hydrodynamic simulation in terms of a theoretical formulation, which may be considered a generalisation of Chandrasekhar's theory of two body relaxation to the case when the density fluctuations do not arise from white noise; it presents a simple characterisation of the effects of complex hydrodynamics and `subgrid physics'.
The power spectrum of gaseous fluctuations is found to follow a power law over a range of scales, appropriate for a fully turbulent compressible medium. The  potential fluctuations leading to core formation are nearly normally distributed, which allows for the energy transfer leading to core formation to be described as a standard diffusion process, initially increasing the velocity dispersion of test particles as in Chandrasekhar's theory.
We calculate the energy transfer from the fluctuating gas to the halo and find it consistent with theoretical expectations. We also examine how the initial kinetic energy input to halo particles is redistributed to form a core. The temporal mass decrease inside the forming core may be fit by an exponential form; a simple prescription based on our model associates the characteristic timescale with an energy relaxation time. We compare the  resulting theoretical density distribution with that in the simulation.

\end{abstract}




\section{Introduction}

In the standard cosmological scenario, CDM particles decouple while non-relativistic 
from the primordial thermal bath. Their consequently small speeds allow for a clustering 
that agrees well with an impressive array of observations on large scales \citep{Frenk2012}. 
But this same `coldness' leads to excessive central densities in the 
nonlinear regime. As soon as they could resolve the central regions, simulations showed that 
CDM haloes were endowed with central density cusps, with their density 
diverging towards the centre \citep{Dubinski1991, Warren1992}. It subsequently 
transpired that this reflected nearly  
`universal', self-similar, behaviour; with the density profiles of 
haloes at different masses, identified at various epochs, well fit with by 
simple empirical formulae \citep{nfw}.  
But the  central density cusps seemed too dense to
match those inferred for dark matter dominated galaxies, like low-surface-brightness and 
dwarf galaxies \citep{Moore1994, Flores1994}.

This `cusp-core' problem is 
central to the low-redshift small-scale issues associated with CDM structure formation, which have attracted much attention during the past decades, as its resolution may simultaneously alleviate other puzzles 
associated with small scale structure in CDM simulations, such as the 
so-called `too-big-to-fail' problem (\citealp{ReadTBF2006, BoylanTBF2011, OgiyaBurkTBF2015}; for general reviews see, e.g., \citealp{delPopolo2017,Bullock_B2017, Salucci:2018hqu}). 
CDM-based simulations also predict a general 
over-abundance of small haloes compared to 
small galaxies. {However, down to the resolution 
level of current cosmological simulations and expected minimal galaxy mass,} this issue has {been claimed to be}  
less severe as small Milky Way satellite galaxies are being discovered and counted in (e.g., \citealp{nadler2020}).

Assuming that the data has not been misinterpreted (e.g.~\citealp{Oman2019}), 
solutions to the aforementioned problems can be principally divided into two sets;
those modifying the underlying dark matter particle physics model and those invoking  
gravitationally mediated CDM interaction with baryons. The most popular of the first category have been warm dark matter and self interacting dark matter, and the more recently topical fuzzy dark matter. The simplest models of the latter type have problems explaining core scaling relations in galaxies \citep{DHHertz18, Burkert20, BarBlumRotII21}, and warm dark matter does not generally succeed in producing 
central density cores at all, as cold collapse occurs unless the dark matter is `warm' enough to prevent the 
halo at hand from forming in the first place \citep{Maccio2012b}. 
Warm dark matter may make baryonic core formation easier, due to lower halo concentrations, but 
this is debated since warm dark matter also acts to suppress early star formation \citep{Governato2015}.
In general particle physics based modifications to the standard structure 
formation scenario invoking warm dark matter or fuzzy dark matter, 
have been more successful in alleviating the apparent 
over-abundance of small scale structure in CDM-based structure formation. 
But this, instead of being a strong point, 
has now joined Lyman-$\alpha$ constraints (e.g., \citealp{ViDvielWDM2017, VidVielFDM}) 
and strong lensing (e.g., \citealp{GilmanLensMisssat2020})
in setting limits on the efficacy of such 
models~\citep{SimonMisssat2007, KopGilMisssat2008, KimMisssat2018, ReadMisssat2019, NewtonMisssat2021,nadler2020, Nadler2021}.
{ In addition, models where small haloes are invariably expected to come with significant cores, such as in the fuzzy dark matter scenario, 
may be constrained by the survival of Milky Way satellites
with eccentric trajectories 
that take them to the host halo's central region (\citealp{Errani_Penn2023}).} 
Interacting dark matter-based models on the other hand  
generically suffer from eventual gravothermal contraction of cores, which are not easily avoided for all relevant spatial and time scales \citep{Burkert2000, Kochanek2000, CoreColSIDM21}, ({ although, 
fine tuning aside, this may actually be considered
a phenomenological strong point of a subset of such models; \citealp{YangNad_SIDMDivers2022}).}

A non-negligible scattering crossection between standard model 
particles and dark matter has also been proposed (\citealp{FamaueyBarint2020, SallucciBarint2020}),  
although this option has not been as fully investigated so far,

On the other hand, gravitational coupling between dark matter and baryons has long been known to be effective in transforming CDM cusps into cores. 
Its main shortcoming however comes from the complex physics and the associated uncertainties in the 
parameters involved. Indeed, cusp-core transformation through such coupling come in three different forms: 
one time mass blowout due to a single burst of energy feedback (\citealt{Navarro1996a}, and more recently \citealt{Freundlich2020} and \citealt{Li2022}); 
the pumping of  energy from (gaseous or stellar) baryonic 
clumps to CDM via dynamical friction \citep{Zant2001, Zant2004}; and density and 
potential fluctuations in feedback-driven gas during galaxy formation \citep{Read2005, Pontzen2014}. 
Despite the apparent similarity of the 
first and  third mechanisms (as they both invoke feedback), 
it is the last two that are in fact more closely related at a deeper level --- as they both involve a long-lived 
fluctuations progressively heating the CDM on timescales much longer than that of a  single feedback starburst, 
as discussed in \citet{EZFC}. 

Dynamical friction heating is nevertheless expected to be observationally distinct from supernova heating since the latter correlates with star formation \citep[e.g.][]{Read2019} while the former does not, and both may act in tandem during galaxy formation \citep{Orkney2021, Dekel2021, OgiyaDF2022}.
In the particular case of dwarf galaxies, there is mounting observational evidence for dark matter heating from impulsive, `bursty', star formation, ubiquitous in dwarfs below a stellar mass of $M_* \sim 10^8$\,M$_\odot$ \citep[e.g.][]{Collins2022}.  Indeed, there is an observed anticorrelation between the inner dark matter density of dwarfs and their stellar-to-halo mass ratio $M_*/M_{200}$ \citep{Read2019, Bouche2022}, which is a proxy for the amount of star formation -- and therefore dark matter heating -- that has taken place \citep[e.g.][]{Penarrubia2012, DiCintio2014, Freundlich2020b}. This may favour a scenario of feedback-driven core formation in such galaxies. 

As conjectured in \cite{EZFC}, it may be possible to understand the process of feedback-driven core formation with repeated bursts
from first principles, bypassing the uncertain complexities of `gastrophysics' and its various 
subgrid implementations. 
Indeed the stochastic dynamical model presented there suggested that 
the process of core formation depended primarily on just two parameters, namely the gas mass fraction and the strength of the fluctuations characterised by 
the normalisation of a power-law power spectrum. It is our purpose here to study 
the properties of feedback-driven fluctuations in a full hydrodynamic simulation where a cups-core transformation 
occurs, with the aforementioned model forming an interpretive framework and guide. Thus 
testing, in the process, its assumptions and predictions. 


\section{Modelling halo heating from gas fluctuations}
\label{sec:level2}

\subsection{Physical setting}
\label{sec:physicalset}

The general picture envisioned, in \citet{EZFC}, and here, is that of a gas settling 
into a CDM halo. As it contracts, a critical density is reached, 
leading to star formation and consequent starburst.  
The star formation process assumed is akin 
to that described in \citet{Teyssier2013} and \citet{Readsim2016}. Namely, the 
threshold is considered 
high enough so that most star formation does not occur in a few bursts, 
but instead repetitively over a long timespan. 
This leads to sustained
density and mass fluctuations in the gas that appear amenable to 
a description in terms of a stationary stochastic process 
over the timescale of the simulation.

If, to a first 
approximation, the gas density is assumed to be isotropic and homogeneous when averaged 
over large spatial or time scales,  
with average density $\rho_0$, then the mass 
fluctuations within a spatial scale $R$ can (as in characterising cosmic structure) 
be characterized by a dispersion 
\begin{equation}
\sigma _R^2 = \frac{1}{2 \pi ^2} \int _0^\infty W^2(k,R) \mathcal {P}(k) k^2 {\rm d}k,
\label{eq:RMS}
\end{equation}
where  $\mathcal {P}(k)$  is the equal time power spectrum of the density fluctuations 
$\delta \rho ({\bf r}, t)/\rho_0 - 1$, and $W$ is a Fourier filter function.  
If the fluctuations furthermore constitute a stationary Gaussian random process, this is all 
what one needs to know to completely characterise them.
The stochastic dynamics can likewise be described completely in terms 
of the first and second moment statistics of the force field, 
which are easily obtainable from the Poisson equations: in particular, 
the $k$-modes of the potential fluctuations are related to those 
of the density by $\phi _{\boldsymbol k} = -4 \pi G \rho _0 \delta _{\boldsymbol k} k^{-2}$, as discussed in~\cite{EZFC} and, in another context, in~\cite{EZFCH}. 
As shown in the latter work, this formulation can also describe 
standard two-body relaxation. As such, it can be used to calculate the 
effect of fluctuations 
arising from the presence of a collection of massive 
particles on a system of lighter ones. 
Relaxation in this context `heats' the light particles, by 
increasing their velocity dispersion, while 
the heavier particles lose energy via dynamical friction. 
The general theoretical framework used here thus 
also applies to dynamical friction heating
of the halo in the presence of massive clumps. The 
relevant power spectrum of density fluctuations, in that case, is flat (white noise)
over scales larger than the maximum size of the (monolithic) clumps.
Thus the description in terms of dynamical friction\ heating is more relevant 
when the gas fluctuations can be described in terms of a system of long lived distinct clumps with sizes significantly 
smaller than the region where the core forms.

When feedback is strong, and the gas fully turbulent on scales relevant to core formation, the spectrum of density 
fluctuations may be assumed to be approximated by a power law over such scales. 
As in the white noise case, the fluctuations  
lead to energy transfer from the gas to the halo particles, which drives the 
transformation of the latter's problematic central cusp into a core. 
The energy transfer may still be  described in particularly simple terms, reminiscent of Chandrasekhar's theory of two body relaxation; 
for a given halo and power spectrum $\mathcal {P} \propto k^{-n}$,
the associated relaxation time $t_{\rm relax}$ is predicted to principally 
depend just on the normalization of that 
spectrum at the minimal wavenumber $\mathcal{P} (k_m)$ (at which the power law breaks), 
and the average gas density $\rho_0$. 
More concretely, for a halo particle moving with an unperturbed speed 
$v_p$ in a field of gas fluctuations carried by bulk flows moving 
with characteristic speed $v_r$ relative to it, one finds
\begin{equation}
t_{\rm relax} = \frac{n v_r v_p^2}{8 \pi (G \rho _0)^2\mathcal {P} (k_m)}.
\label{eq:relax}
\end{equation} 
This is the timescale for the random velocity gained by a CDM particle, 
as a result of its motion 
in the stochastic force field born of gas fluctuations, 
to reach the unperturbed characteristic speed $v_p$. 
{It is analogous to the standard relaxation time. In fact equation (\ref{eq:relax})
reduces to Chandrasekhar's formula  
for a white noise spectrum ($n = 0$),
with  Coulomb logarithm determined by the ratio of maximal to minimal wavenumumber cutoffs (\citealp{EZFCH}, Section~2). 
Also, as equation~(\ref{eq:relax}) is derived using Fourier transformation, 
which is transitionally invariant and therefore necessarily 
leads to loss of spatial information, it 
is (like Chandrasekhar's formula) strictly  
valid  only for an infinite homogeneous system 
with a constant level of fluctuations. 
In applying it to a real system, 
we will thus assume  an approach that is  akin to that widely invoked with the use of Chandrasekhar’s theory in an inhomogeneous system, where it is customary to assume that the variation of the relaxation time with position 
is accounted for by the density variation. In our formulation of Chandrasekhar's theory, referred to above, 
this would mean keeping $n = 0$ and the ratio of maximal to minimal wavenumber fixed, while adjusting the density to account 
for radial variations in the relaxation time. 
Analogously,  we will keep the power spectrum parameters 
in equation (\ref{eq:relax}) fixed and assume that the spatial variation of the energy input to the dark matter from the gas fluctuations can be accounted for by variation in the density.}

To account for the radial change in the average
gas density in a realistic system, we define the average density, at a given time, 
within a sphere with radial coordinate $r$ by
\begin{equation}
 \rho_0 (r, t) = \langle \rho_g ({\bf r},~t) \rangle_{|{\bf r}|<r},    
\label{eq: rhoavdef}
\end{equation}
where $\rho_g ({\bf r}, t)$ is the local gas density and the average is evaluated within  the 
volume enclosed by $r = |{\bf r}|$. A further average over time results in $\rho_0 (r)$.
If the averaging is not evaluated over a sphere centered at the origin, but over spheres 
with centres at radial coordinate $r$ and radius $r_{\rm av}$,  an analogous 
average  may also be defined (as in equation~\ref{eq:semiloc}).  
{In Section~\ref{sec:enin}, we examine three different approximations
for an effective $\rho_0$, in order to account for  
changes of relaxation time with radius 
while using equation~(\ref{eq:relax}):
one associated with what we will call global energy transfer, i.e. with $\rho_0$ a constant corresponding to an average inside an energy transfer saturation radius 
$r_{\rm sat} \gtrsim r_s$, which turns out to be plausible, given the slow variation of average gas density inside $\sim r_s$ and subsequent sharp drop (Fig.~\ref{fig:densityprofiletime_all}); a local approximation, which does not work; and 
one involving averages over radius $r_{\rm av} = r_s \simeq {\rm 1 kpc} \approx 1/k_m$, which we will find works best
(Fig.~\ref{fig:Lovsglb}). This is perhaps not surprising given the aforementioned gas density variation, and given that equation (\ref{eq:relax}) implies that modes near $k_m$ contribute 
most to the relaxation process. 
(It in fact arises from a low-$k$ ‘effective theory’, as discussed further in the penultimate paragraph of Appendix~\ref{Sec:Additional_Gauss})}.   

The characteristic velocity $v_r$ appearing in equation~(\ref{eq:relax}) is estimated  through the `sweeping' approximation, long 
used in turbulence theory (\citealp{TaylSwepp1938,KraichSweep1964,TennekSweep1975}) 
and invoked by~\cite{EZFC}.  
It assumes that the equal-time spatial statistics of the fluctuation field are swept ('frozen in') into the time domain through large scale fluid flows.  
In this picture $v_r$ consists 
of random and regular velocity components $U$ and $V$, such that $v_r = \sqrt{U^2+V^2}$ that are characteristic of those flow.  
In our case, since 
the halo particles are also moving with respect to the gaseous flows, the characteristic 
$v_r$ may be considered to include such motions.~\footnote{In principle, each fluctuating mode may have its own sweeping speed and the velocity distribution of the particles can be taken into account, as in \cite{EZFCH}. We do not consider this more complex case here.} We test this 
assumption in Section~\ref{sec:disp}.

In the context of the formulation outlined in this section, 
the timescale of significant central halo transformation, from cusp to core, should  scale as in equation~\ref{eq:relax}, 
with parameters as defined above. 
Controlled simulations (using the Hernquist-Ostriker code (\citealt{Hernquist1992}),
whereby Gaussian random noise was applied to particles of live CDM haloes,
produced cores on the expected timescale, when the haloes were kept strictly spherical \citep{EZFC}.~\footnote{When this condition was relaxed, the timescales for core formation 
were found to be about an order of magnitude shorter.
But the scaling with the normalisation of the power spectrum and average density, as reflected in equation~(\ref{eq:relax}), remained. As noted in Section~\ref{sec:massmod},  we find no evidence of such accelerated (relative to the relaxation time) core formation rate here.}

\subsection{Numerical implementation}

\begin{figure*}
\centering
	\includegraphics[width=\textwidth]{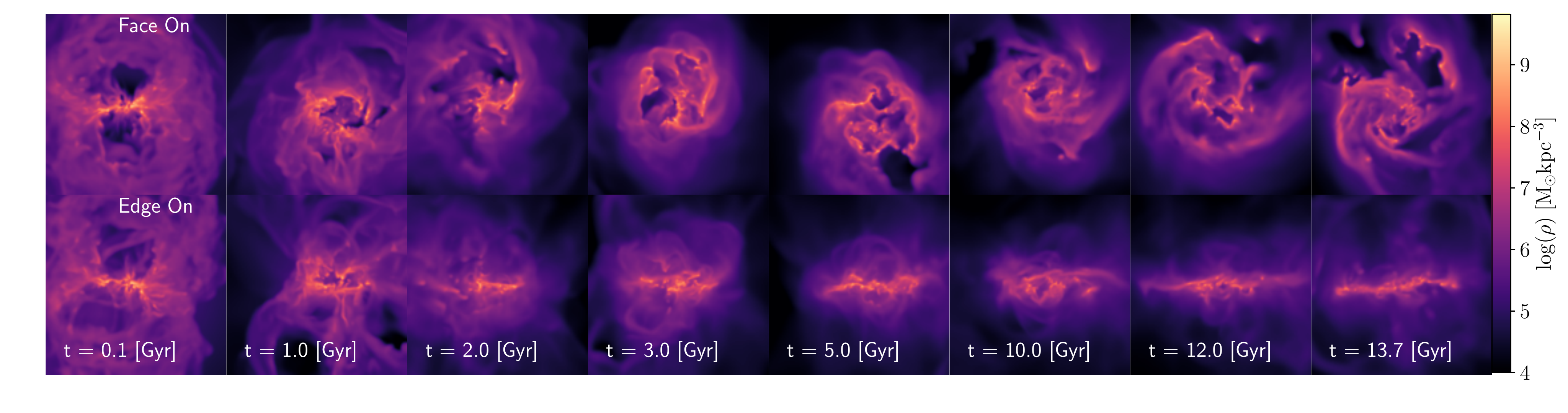}
    \caption{Gas projected density map inside a box of length 5 kpc (2.5 kpc on a side from the centre of the halo). The gas contracts from the initial NFW profile, then much of it is expelled from the inner region during the first few hundred Myr. It gradually settles into a disk-like configuration, but its distribution fluctuates over time.}
    \label{fig:Gas_Map}
\end{figure*}

We wish to examine feedback-driven gas fluctuations in a full hydrodynamic simulation, 
to find out whether their characteristics and their effects 
may fit the picture outlined above. 
To facilitate the isolation of the principal features, 
we focus on the case of an isolated galaxy. 
Its parameters (Table \ref{table:halo})
are chosen to correspond to those of dwarf galaxies, where
dark matter dominates even in the central regions and 
the discrepancy between the inferred density and a centrally concentrated halo 
profile seems particularly clear. 

The simulation of the isolated  dwarf is set up and run with the RAMSES adaptive mesh refinement code \citep{Teyssier2002}, and already presented in \citet{Readsim2016}. We use `Run M9c224e6' from that paper, which has 281 time outputs regularly spaced over a timespan of 13.7 Gyr. The initial condition for this simulation assumes a NFW dark matter halo \citep{nfw} with total mass inside the virial radius $M_{200} = 10^9$\,M$_\odot$ and concentration parameter 
$c_{200} = 22.23$. 
A fraction $f_b = 0.15$ of 
that mass ($M_g (r_{\rm vir}) = 0.15 \times 10^9\,M_\odot$), 
is in gas that is initially in hydrostatic equilibrium, with a metallicity of $10^{-3}$\,Z$_\odot$ and some angular momentum set to match median expectations in a $\Lambda$CDM cosmology.
While the simulation assumes that the dwarf galaxy contains the Universal baryon fraction out to its virial radius, the ratio between the gas mass initially within 
the region of interest for core formation ($\sim 1~{\rm kpc}$) 
and the mass within the virial radius 
is much lower, and thus consistent with observational constraints (e.g.~\citealp{ReadBarFrac2005}). 
At the start of the simulation, the gas rapidly cools and collapses,
then re-expands under the influence of feedback. In the process, 
the central gas mass 
fraction further decreases (cf. Section~\ref{sec:gasdens}). 

The sub-grid model for star formation and feedback is described in detail in \citet{Readsim2016}. 
Briefly, star formation follows a Schmidt relation with a star formation efficiency per free fall time of $\epsilon_{\rm ff} = 0.1$ and a density threshold for star formation of $\rho_* = 300$\,atoms\,cm$^{-3}$. The stellar feedback model is as in \citet{Agertz2013} and includes a model for Type II and Tyle Ia supernovae, stellar winds and radiation pressure. The simulation resolution was chosen to capture momentum driving from individual supernovae events, with a  gas spatial resolution (at highest mesh refinement) of $\Delta x \sim 4$\,pc and a mass resolution in gas, stars and dark matter of $m_{\rm g} = 60\,{\rm M}_\odot$ and $m_* = M_{\rm DM} = 250\,{\rm M}_\odot$, respectively. This allows individual star formation events to be resolved, injecting stellar feedback to the interstellar medium in the correct locations at the correct times.

\begin{table}
	\centering
	\caption{Initial profile parameters.}
	\label{tab:nfw}
	\begin{tabular}{lll} 
		\hline
		
	    NFW scale length & $r_s$ & $0.88\; {\rm kpc}$ \\
	    NFW characteristic density & $\rho_s$ & $5.34 \times 10^7~ {\rm M_{\odot} ~kpc^{-3}}$ \\
	    Concentration parameter & c & $22.23$ \\
		Mass within the virial radius & $M_{\rm vir}$ & $10^9\; {\rm M_{\odot}}$ \\
		Baryonic mass fraction & $f_b$ & $0.15$ \\
		\hline
	\end{tabular}
	\label{table:halo}
\end{table}

\section{Characterizing the gas fluctuations}

\begin{figure}
\centering
	\includegraphics[width=\linewidth]{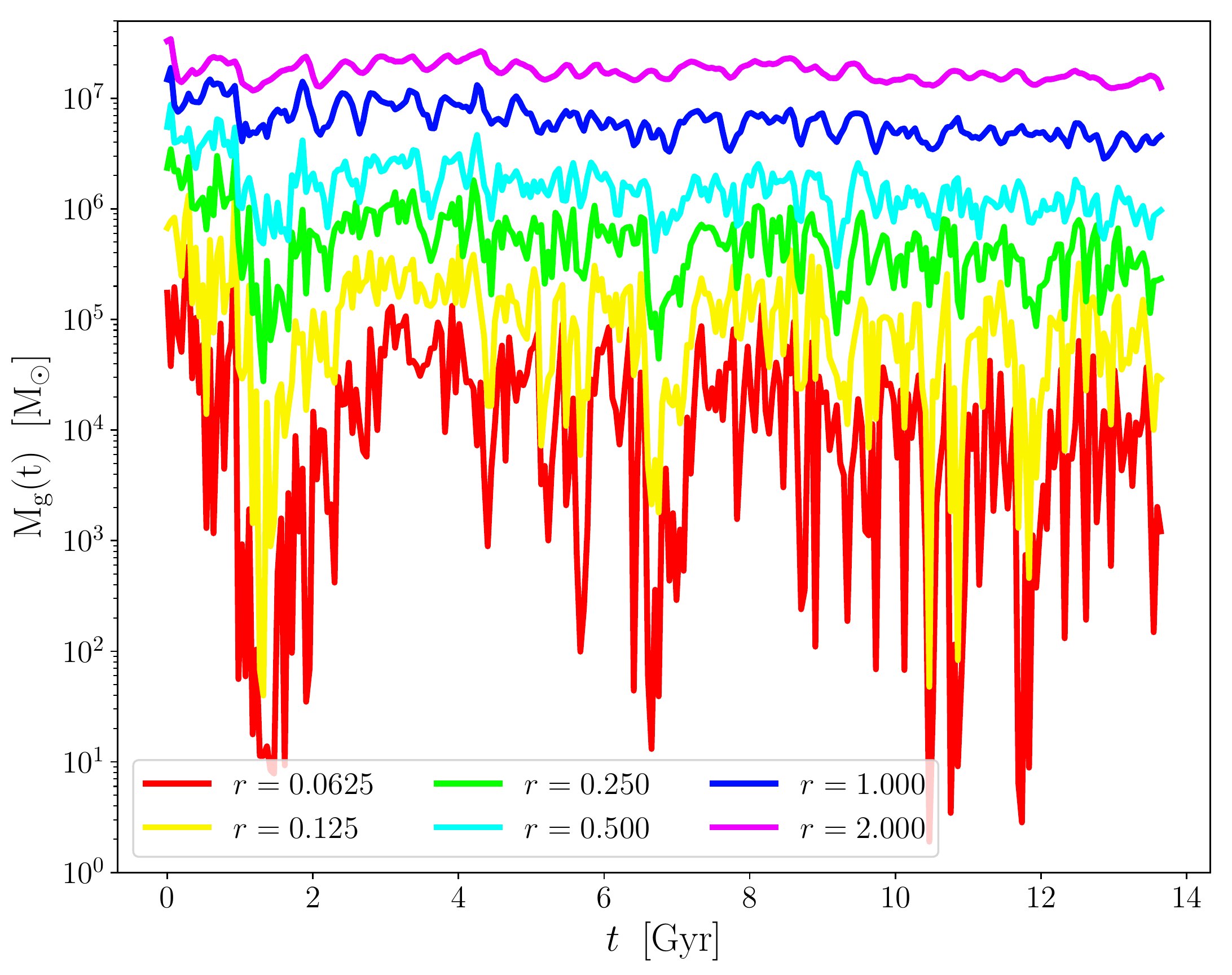}
    \caption{Time variation of the gas mass enclosed within different radii $r$ (in kpc), suggesting a quasi-stationary stochastic process.}
    \label{fig:MTsiers}
\end{figure}

\begin{figure*}
    \centering
	\includegraphics[width= 0.48\textwidth, ]{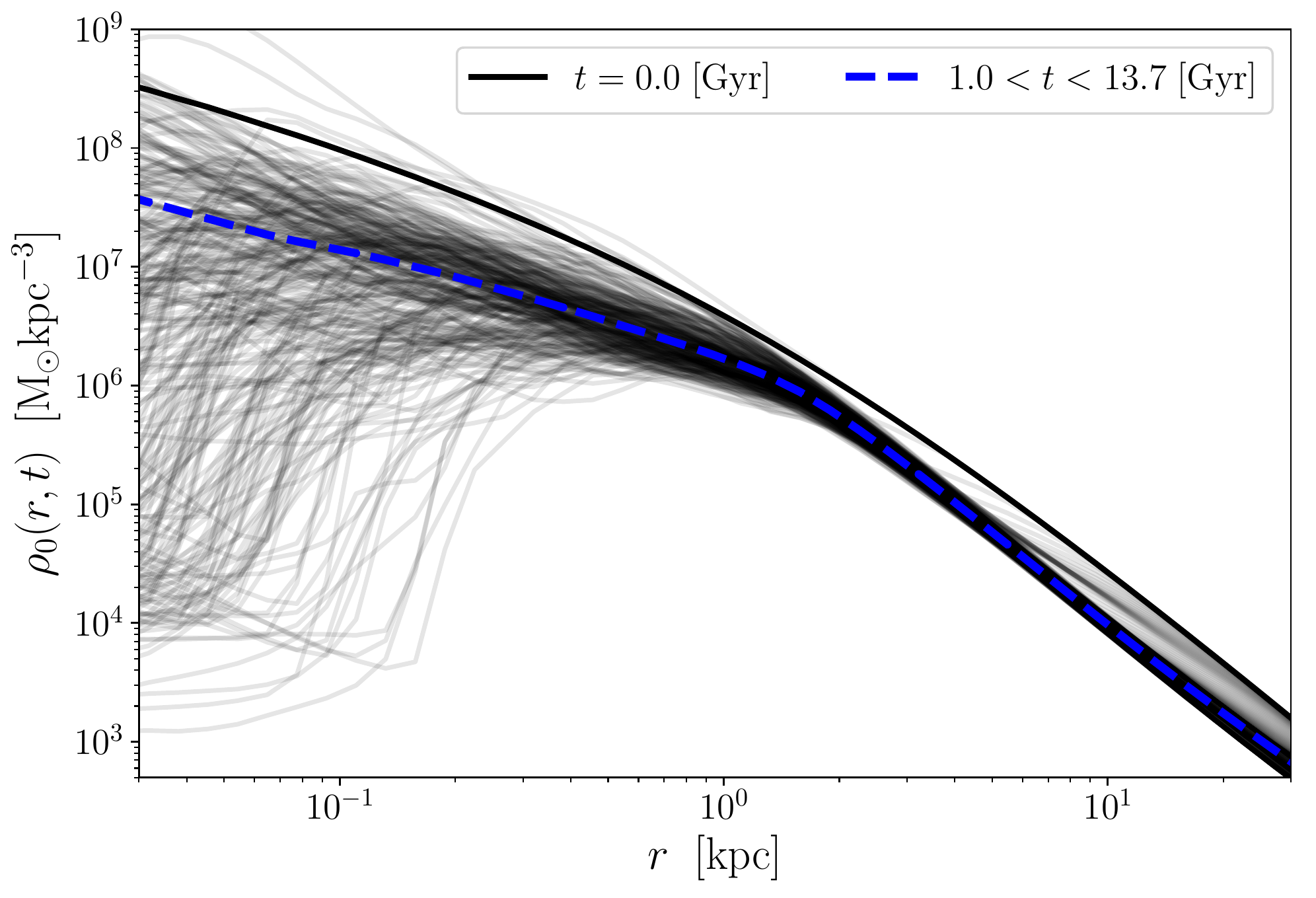}
     \includegraphics[width= 0.49\textwidth, ]{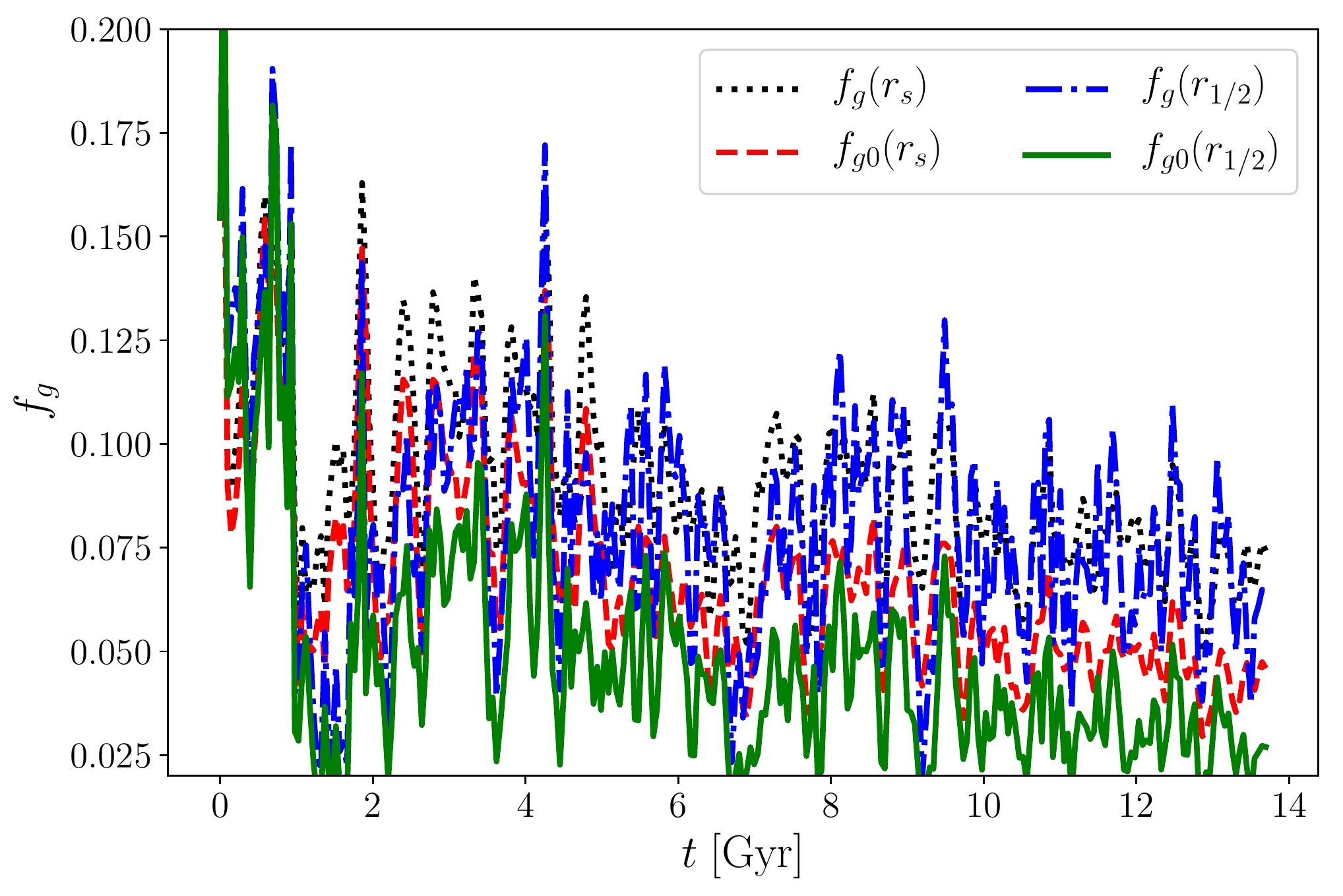}
   
    \caption{{\it Left:} Average gas density inside radius $r$, $\rho_0  (r, t)$, for all snapshots, obtained by using equation~(\ref{eq: rhoavdef}). The dashed line is a time average over the indicated interval, which corresponds to the simulation time beyond an initial non-equilibrium phase through which much of the gas is blown out 
    of the region inside $r_s \simeq 1~{\rm kpc}$.  
    The time averaged profile is only mildly varying inside $r_s$. {\it Right:} Ratio of the gas to combined gas and dark matter masses $f_g =M_g /(M_g + M_d)$, with $f_g (r_s)$ denoting the fraction inside $r_s$; thus here $M_g = M_g (<r_s, t)$ and  $M_d = M (<r_s, t)$, while $f_{g0}$ denotes the ratio of the gas mass to the initial total mass enclosed: $f_{g0} (r_s) = M_g (<r_s, t) /[M_g (r_s, 0) + M_d (< r_s, 0)]$.
    Similarly $f_g (r_{1/2})$ and $f_{g0} (r_{1/2})$ refer to those ratios     
    inside the half mass radius of the stellar component that forms from the repeated starbursts, $r_{1/2} \simeq 0.5 {\rm kpc}$, and which physically correlates with the scale of the relatively steady gas density associated with strong fluctuations.} 
    \label{fig:densityprofiletime_all}
\end{figure*}

In order to apply the physical model described above, leading to equation~(\ref{eq:relax}), we need 
to know the average gas density $\rho_0$. 
We also need to 
evaluate the power spectrum of the gas fluctuations and 
find out whether it may indeed be adequately
described by a power-law, in which case we need 
to estimate values for its normalization $\mathcal{P} (k_m)$ and 
index $n$, and  verify that the spatial spectrum is relevant in describing the temporal fluctuations. This is the objective of this section. 

\subsection{Gas density and mass over time} 
\label{sec:gasdens}

Fig.~\ref{fig:Gas_Map} shows gas density contrast maps, within a 
5 kpc box, 2.5 kpc on a side from the halo centre\footnote{Unless otherwise stated, 
all centering here refers to the halo centre located using the shrinking sphere method.}. 
The gas 
initially contracts from an NFW profile with scale length $r_s$ 
until the critical threshold for star formation is reached. The fluid is then  driven by the nascent feedback, much of it driven completely out 
of the region delineated by $r_s$; the gas mass inside $\sim$$r_s$ decreases to less than half its initial value
(but the dark matter mass is largely unaffected, decreasing only by about $20 \%$ within $r_s$, 
and thus the ratio decreases significantly, as shown in the right hand panel of Fig.~\ref{fig:densityprofiletime_all}). 
After that, the gas 
mass inside $\sim$$r_s$ is relatively well conserved.  
Variations in the gas mass enclosed within a given radius then principally 
arise from a pattern of fluctuations (gaseous 
blobs) materialising on a large range of scales. 
The gas progressively settles into a 
disk-like configuration, but the mass fluctuations 
remain sustained and steady. 

The assumption that the fluctuations form  a quasi-steady stochastic 
process is suggested by Fig.~\ref{fig:MTsiers}, showing the mass variation 
within different radii over time. 
The mass fluctuations persist in time 
but diminish with radius, as higher mass scales are reached. 
Significant variations on larger time-scales persist however. 
They correspond to large scale flows. In the context 
of the model of \cite{EZFC} (recapped in Section~\ref{sec:physicalset}), 
such large scale motions `sweep' the smaller scale fluctuations with the characteristic speed $v_r$ (equation~\ref{eq:relax}). 
For the description of the effect of the gas fluctuations on the halo particles in terms of a standard diffusion process (leading to relaxation timescales in the form of equation~\ref{eq:relax}) to be complete, the statistics of the stationary stochastic process should also be entirely described by averages and dispersions, as in a Gaussian random process. This is examined in Appendix~\ref{Sec:Additional_Gauss}.


The gas density inside spheres of radius $r$ is shown in  
Fig.~\ref{fig:densityprofiletime_all}, as a functions of time.  
After the initial outflow during the first few hundred Myr, 
the averaged (over time)
gas profile is only mildly varying with radius inside the
initial NFW scale length ($r_s 
= 0.88~{\rm kpc}$), 
and then starts decreasing rapidly beyond that. 
Any core that forms is expected to be of the order of the scale length of the cusp, which renders the assumptions 
in~\cite{EZFC} (discussed in Section 2.1.2 therein),  plausible.  
One may for instance  
attempt to estimate the relaxation time using equation~(\ref{eq:relax}) with 
mean density $\rho_0$ evaluated at radius $\gtrsim r_s$,
as the sharp decrease in density beyond that scale suggests 
that fluctuations in the gas at larger radii would contribute 
little to the overall variations in the gas potential
and associated energy input to the central halo. 

In Section~\ref{sec:relax}, we estimate that this 
may indeed be a good approximation.  Here we 
note that there are two theoretically independent reasons 
why, given the parameters of the simulation, 
a core is expected to form on a scale of order $1~{\rm kpc}$. 
For, even if the gas density does not drop 
beyond $r \gtrsim r_s$, and strong fluctuations are present much beyond that radius, the resulting stochastic perturbations are expected to have a major dynamical effect only up to scale of order $r_s$. This was in fact 
found to be the case  
even when the amplitude of the fluctuations (as fixed by the normalization of the power spectrum)
were increased well beyond the minimum level required to significantly affect 
the inner halo profile (cf. Fig.~6 of~\cite{EZFC}).
The phenomenon is likely 
due to the transition to an isothermal profile in an NFW halo at $r \simeq r_s$, 
which 
is relatively stable against fluctuations, { as the 
differential energy distribution tends to an exponential form
(cf. Appendix~\ref{app:exp}).}
This phenomenon relates to 
the {\it effect} of the fluctuations.The other reason why 
a core that forms out of a cusp here should have radial scale 
of order $1 {\rm kpc}$ has to do with the existence of fluctuations in a realistic system: the extent of the spatial scale where the gas density and density contrasts are large is  
characteristic of the half mass radius $r_{1/2}$ of the stellar component that 
forms from the repeated starbursts. In our case about $0.5 {\rm kpc}$. 
Because of this, we will be able to define (in relation to Fig.~\ref{fig:Ein_main}) 
an energy input saturation radius, of order $\simeq 2 r_{1/2}$,
which results from the decreasing gas density and level of fluctuations. 
Beyond it, there is negligible energy transfer from gas to halo. 

The right panel of Fig.~\ref{fig:densityprofiletime_all} shows the fraction of gas inside both $r_s$ 
and $r_{1/2}$. We note that due to the small variation of the gas density 
averaged over radius, the lines closely correspond to one another. 
In addition, after the gas loss associated with  the initial blowout phase, the gas fraction
decreases to about half its initial value. Thus, any effect due to core formation will result
from fluctuations from a rather small gas fraction in the inner regions.  The 
efficacy of this perhaps surprisingly small 
central gas fraction in modifying the 
dark matter dynamics is in fact a generic prediction 
of our theoretical framework, provided the strength of the  
gas fluctuations is at the level of the fiducial model  
in~\cite{EZFC} (cf. Section~2.1.2 and 4.4.2 there).
Below, we will find that the fluctuation levels
in the present simulation,  as measured by the 
normalisation of the power spectrum,
are in indeed consistent with 
those assumed in that fiducial model. 

\subsection{Power spectra}

\begin{figure}
\centering
	\includegraphics[width=\linewidth]{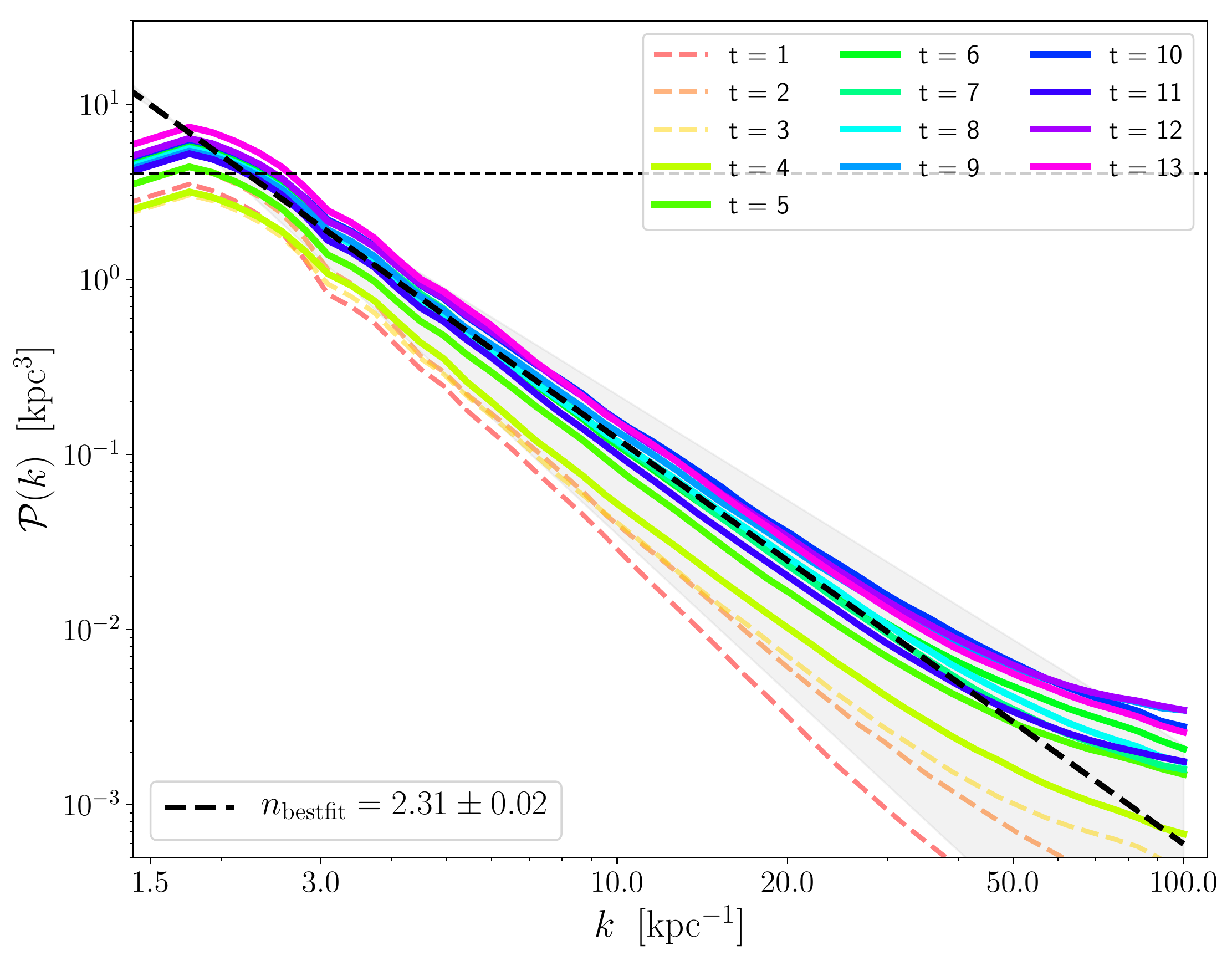}
    \caption{Power spectrum of gas density fluctuations, averaged over 1 
    Gyr intervals. The shaded region corresponds to variation of a power law 
    $\sim k^{-n}$, with  $2 \le  n \le 3$; the horizontal line corresponds 
    to $\mathcal {P} (k_m) = 4~{\rm kpc^3}$. The spectrum follows a power-law form over a range of scales, with a best-fit exponential index $n_{\rm best fit} = 2.31\pm 0.02$ (black dashed line).} 
    \label{fig:powerspecavrg}
\end{figure}

We now proceed to determine the power spectrum of the density 
fluctuations qualitatively 
probed above. We evaluate it in the following way. 
For a given time snapshot, we consider gas cells within a cubical box
of extent 2.5~kpc on each side from the origin.\footnote{Taken here to be the centre of the halo located through the shrinking sphere method. We have verified that the results are robust to displacements of that centre $\lesssim{\rm 1~kpc}$, and similar results are obtained when  the gas centre of mass or shrinking sphere centres are used.} 
We assign each cell, depending on its radial location, to  a spherical 
shell. { The average gas density 
inside the shell $\rho_{\rm sh}$ is then subtracted 
from that at a given cell $\rho_{\rm cl}$. The density contrast in 
each cell $\delta_{\rm cl} = \rho_{\rm cl}/\rho_{\rm sh}-1$ is evaluated, and the Fourier transform is taken over the cells and squared in the usual manner, to obtain the power spectrum (using the Fourier convention employed in~\citealp{EZFC}).}  
This is done for each time snapshot, and the results are averaged over a chosen set of snapshots. 
The outcome is shown in Fig.~\ref{fig:powerspecavrg}.
The spectrum follows a power-law form over a range of scales,  
with an index consistent with that assumed
in the fiducial model of~\cite{EZFC}, where it was assumed to correspond to 2.4.
If we take $k_m \simeq 2 {\rm kpc^{-1}}$, then
$\mathcal {P} (k_m) \simeq 4~{\rm kpc}^3$ (again 
close to what was assumed in the fiducial model of \citealt{EZFC}, 
namely $\mathcal {P} (k_m) = 4.6 ~{\rm kpc}^3$). 

\begin{figure}
\centering
	\includegraphics[width=\linewidth]{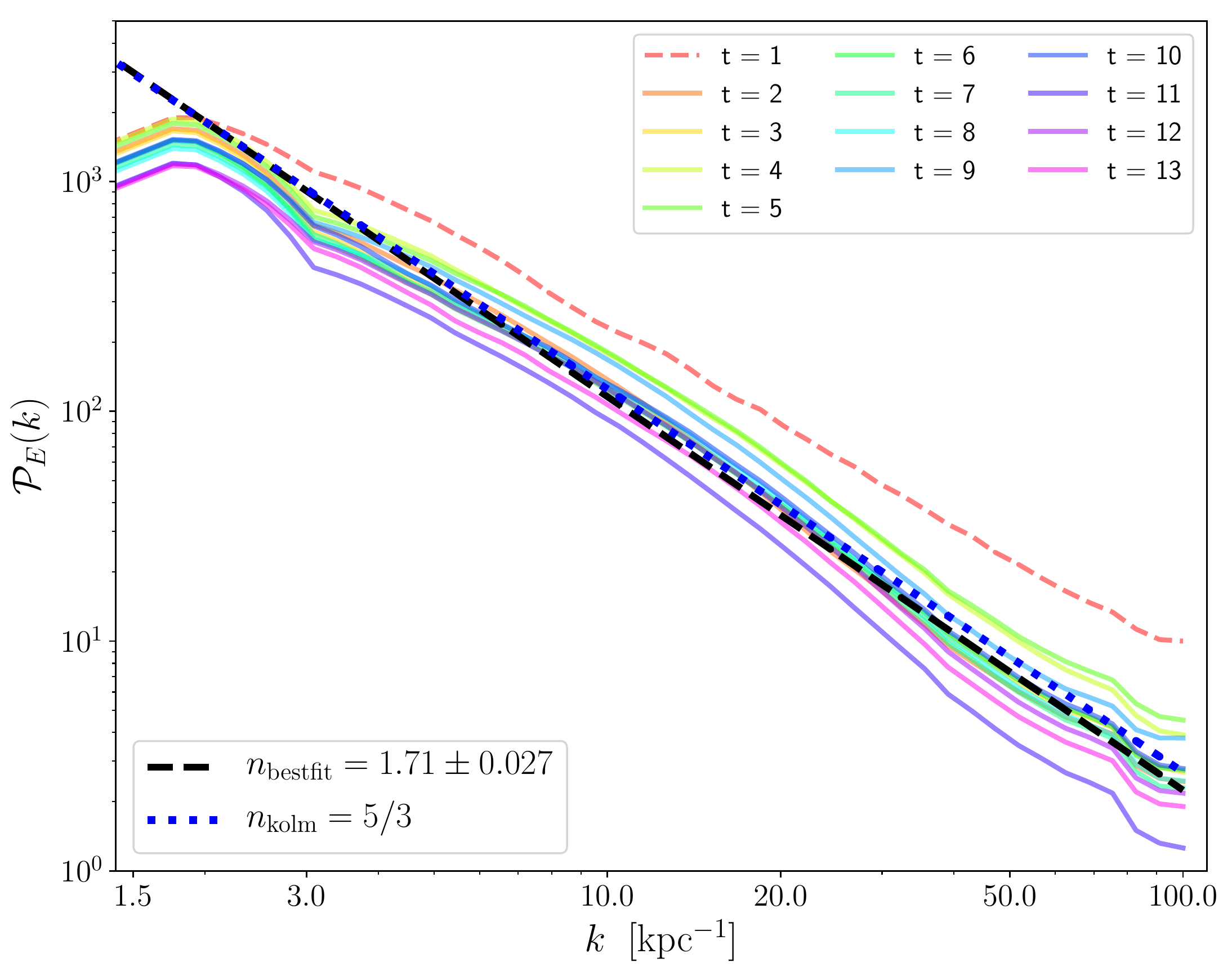}
    \caption{Kinetic energy power spectrum of gas motion 
    with least squares best fit (black dashed line) and Kolmogorov spectrum ($\propto k^{-5/3})$, blue dashed line). The spectrum is calculated through the square of the Fourier transform of $\rho^{1/2}_{g_i} {\bf v}_i$, taken over gas cells $i$ inside a 5~kpc box around the halo (shrinking sphere) centre.
    The agreement with the Kolmogorov form supports feedback-driven fluctuations having characteristics of fully developed turbulence.}
    \label{fig:velocitypowerspecfit}
\end{figure}

The picture of fully turbulent  feedback-driven fluctuations  
is supported by the near Kolmogorov form of the specific kinetic energy spectrum, shown in Fig.~\ref{fig:velocitypowerspecfit}, 
which is followed for a large range in wave numbers.
This spectrum persists 
even as the gaseous system evolves 
from a fully three dimensional configuration to disk-like form.
Indeed~\cite{GrisRomReadTurb2017} found that it is present even in relatively 
quiescent phases, as long as feedback is also present (noting that the spectrum flattened when 
feedback is eliminated). 
In principle,  turbulence may also be driven by self gravitating instabilities rather than feedback
(e.g., \citealp{YUKrumTurb2021, NusSilkTurb2022}). A simple calculation of the gas Toomre parameter in the present case nevertheless suggests an absence of strong gravitational 
instabilities (see however~\citealp{DekelQ2016}). 

In the context of our theoretical framework, 
a further (likely related)
physical distinction may also be reflected in the shape of the density fluctuation power spectrum; 
for, as mentioned in Section~\ref{sec:physicalset}, 
a flat  (white noise) power 
spectrum corresponds to the limit 
when heating through dynamical friction coupling between monolithic clumps and the dark matter 
is the dominant process. In this case the clumps are long lived and distinct, and their maximum 
size is significantly smaller than the scale associated with the 
region where the cusp-core transformation takes place
(and on which the spectrum takes the flat form).
The steep power-law dependence that continues up to large scales, as found here, signals a preeminence of feedback driven fluctuations. 
{Only at the largest scales ($ < k_m$) is a flattening 
of the spectrum recovered. (The flattening at large $k$ denotes the start of the  
transition from the turbulent inertial range to the dissipation range, set here 
by the numerical resolution, given the small kinetic viscosity, 
and may reflect a 'bottleneck phenomenon', resulting from a pile-up of inefficiently dissipated energy as the transition is approached; \citealp{FalkovichBottle1994,Scmidtbottle2004, KristukBottle2007}.)}

Finally, we note that the 
break in the kinetic energy 
power spectrum at $k_m \simeq 2 ~{\rm kpc}^{-1}$ is consistent with that inferred from Fig.~\ref{fig:powerspecavrg}, suggesting a demarcation between the turbulent scaling regime and the large scale flows that may carry the fluctuations into the time domain. We examine
this issue of the transposition of the fluctuation characteristics from the 
spatial to temporal domains further in the next subsection.

\subsection{RMS mass fluctuations}
\label{sec:disp}
\begin{figure}
\centering
	\includegraphics[width= 0.49\textwidth]{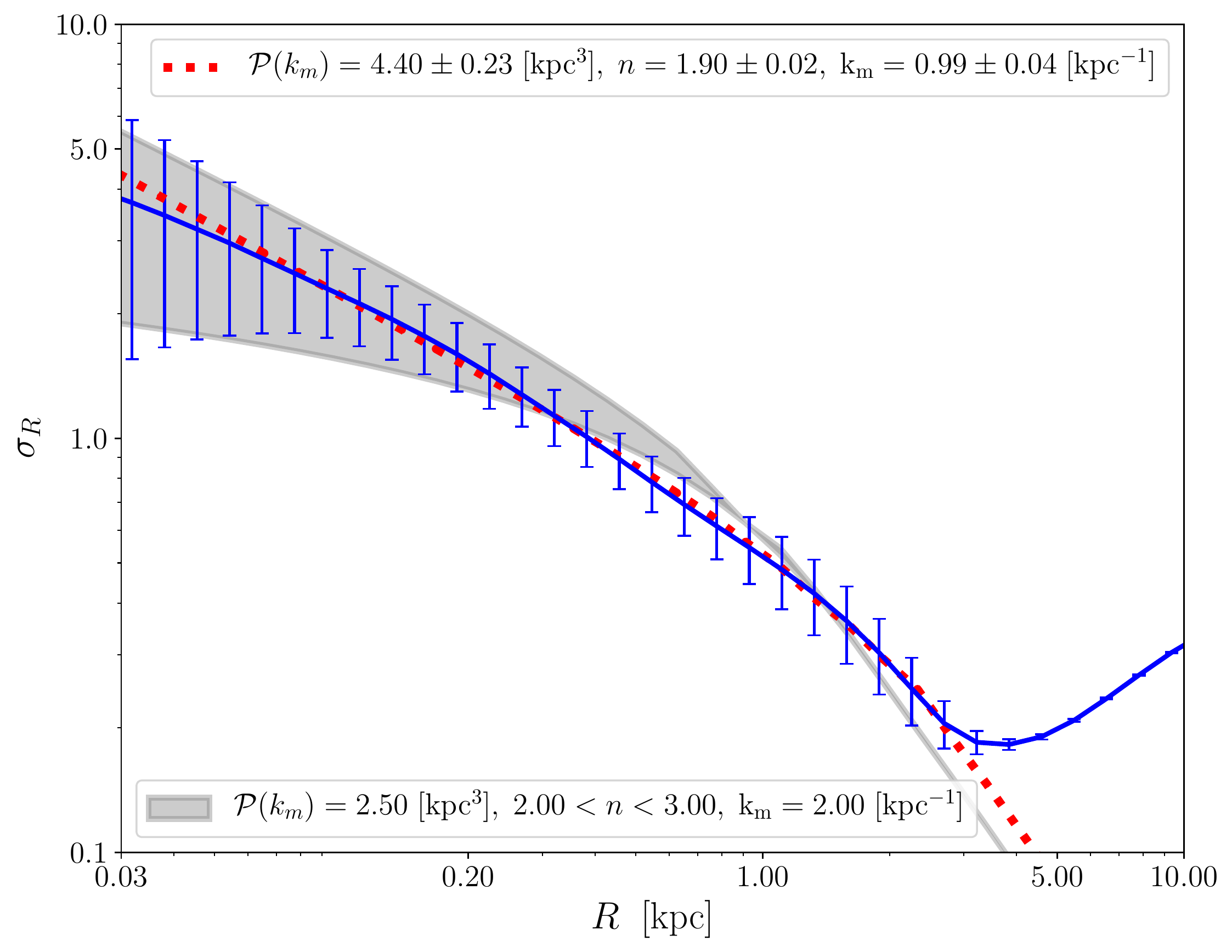}
    \caption{Gas mass dispersion at scale $R$, $\sigma^2_R = \sum_i (M_i (R) - \langle M (R) \rangle_i)^2/\langle M (R) \rangle_i^2$, where $i$ denotes different time snapshots, 
    and $M(R)$ is the mass inside a sphere of radius $R$. 
    The centre of the sphere is randomized; chosen from a homogeneous distribution inside radius 1~kpc ($\simeq r_s$). The process 
    is repeated for 80 realizations and the average $\sigma_R$ over the realizations (solid line) and dispersion (error bars) are evaluated. The dotted line refers to a (least squares) best-fit using equation~(\ref{eq:RMS}) with a power-law power spectrum and a top-hat filter, while the shaded region shows the variation with $n$ in the range of the corresponding shaded region on Fig.~\ref{fig:powerspecavrg}, with moderately smaller normalization at $k_m =2~ {\rm kpc}^{-1}$. All fits assume $\mathcal {P} (< k_m) = \mathcal {P} (k_m)$.
   The  power spectrum parameters, inferred here from simulation data in the time domain,  are generally consistent with those of the equal-time power spectra of  Fig.~\ref{fig:powerspecavrg}.} 
    \label{fig:massdisper}
\end{figure}

Our formulation of the dynamical effect of the gas fluctuations
makes use of the sweeping and random sweeping approximations of turbulence theory, 
mentioned at the end of Section~\ref{sec:physicalset}. At some level, this is 
analogous
to an ergodic assumption, in the sense that statistical properties  of the random field in the spatial 
domain are transferred into the time domain. 
If such an approximation is 
valid in our case, then we should expect the following: if the variance $\sigma_R^2$ on the left hand side of equation~(\ref{eq:RMS})
is calculated in the time domain over the simulation snapshots, it should correspond to the 
result obtained by plugging the equal-time power spectrum $\mathcal{P} (k)$, with parameters 
consistent with what is inferred above (Fig.~\ref{fig:powerspecavrg}), into the right hand side of equation~(\ref{eq:RMS}). 

In a stochastic process that is homogeneous in space and time, with the sweeping assumptions holding, the
$\sigma_R$ measured in the time domain  (over sufficiently long time) 
will be invariant with respect to the centre it is measured from. This cannot be strictly the case here, however, since the average gas density is not strictly homogeneous, {although 
the power spectrum of relative density fluctuations was found to be  rather robust 
against changes in the centre of the box inside which it is calculated (Footnote~4). 
Furthermore, when attempting to match temporal fluctuations to those captured by
spatial power spectra at given 
snapshots one needs to eliminate effects not captured
by the frozen-in spectra.}  
For example, the effect of fluctuations carried by rotational flows will be reduced when calculating 
$\sigma_R$ in spheres anchored close to the centre of rotation, with little mass transiting through the shells (and those carried by large scale radial flows are enhanced).  
To account for such effects, we randomise the centre of the sphere within which the mass is calculated.  The centres are, in practice, sampled from a homogeneous distribution within the radius of interest for core formation (1~kpc from the origin). We then conduct eighty different realizations of such randomisations and evaluate the average of the dispersion over the realisations. 

Fig.~\ref{fig:massdisper}  shows the result.
The best-fit using equation~(\ref{eq:RMS}) involves a power-law power spectrum 
with index $n$  consistent with the lower limit in the shaded region of Fig.~\ref{fig:powerspecavrg}; and also with similar normalization $\mathcal{P} (k_m)$, but at smaller $k_m$. The change in $k_m$, by itself, is not expected to affect the dynamical effect of the fluctuations in the diffusion limit at the basis of our model (with consequence that the relaxation time in equation~\ref{eq:relax} does not directly depend on $k_m$; see also Fig.~8 of~\citealt{EZFC}).  On the other hand, the  shaded area of Fig.~\ref{fig:massdisper} suggests that the mass dispersion is also consistent with larger values of $k_m$, but with mildly smaller normalization (which does moderately affect the relaxation time). 

The general consistency of the parameters inferred here, in comparison with those obtained from the spatial power spectrum (Fig.~\ref{fig:powerspecavrg}), 
supports the contention that the spatial 
fluctuations are transported to the time domain in accordance with the sweeping assumptions (although the separation between the large scale `carrier flows' and the transported fluctuations may not be sharp, and this may contribute to the smaller best fit $k_m$ here).  As we will see below, these parameters are also consistent with the actual energy input, from the fluctuating gas to the halo component, in the simulation studied here; in the sense that the inferred energy input is consistent with theoretical 
estimates using such values.


\section{Stochastic energy transfer and core formation}
\label{sec:relax}

\subsection{The relaxation time and its parameters}
\label{sec:relaxpar}

Now that we have estimates of the parameters 
entering equation~(\ref{eq:relax}), we can apply it in order to 
check whether it predicts significant effect for the model 
galaxy at hand. As a first estimate we simply assume 
$v_p \approx v_r \approx  v_c$, where $v_c = v_c (r_s)$ is the circular
speed at $r_s$ (close to the maximal rotation speed).  If we furthermore 
use a value for $\rho_0$ characteristic of the region 
when the density starts to decrese rapidly ($r \gtrsim r_s$ from Fig.~\ref{fig:densityprofiletime_all}), we find 
\begin{equation}
t_{\rm relax} = 13.2~{\rm Gyr} ~
 \frac{n}{2.5} \left(\frac{v}{20 {\rm ~km/s}} \right)^3 
\left(\frac{\mathcal{P} (k_m) }{3 {\rm ~kpc}^3}\right)^{-1}
\left(\frac{\rho_0}{10^{6} ~{\rm M}_\odot/{\rm kpc}^3}\right)^{-2}\!\!.
\label{eq:relaxnum}
\end{equation}
%
%
This suggests that gas fluctuations may have a significant 
effect on dark matter orbits within $r_s$
on a timescale of the order of a Hubble time. 
If $v_p$ and $v_r$ are associated with the characteristic 
velocity dispersion (rather than $v_c$), 
which is slightly larger (by a factor of about $10 \%$ at $\sim$$r_s$), 
the above  timescale increases by a factor of about $(1.1)^{3/2}$. 

The timescale also increases if
the characteristic speed 
of the halo particles relative to the gas $v_r$ (cf. Section\ref{sec:physicalset})
is associated with 
gas velocities  at $\sim$$r_s$, which are around ${\rm 30 ~km/s}$ (cf. Fig. ~\ref{fig:velav}). 
Because flows may take the gas well beyond $r_s$, 
gas velocities are generally larger than $v_p$. 
Once a rotating disk starts to materialise (Fig.~\ref{fig:Gas_Map}), the regular 
component would also include  gas rotation, in addition to the orbital 
speeds of the halo particles. Below, we will generally 
set $v_r \simeq 30 ~{\rm km/s}$, but note that $v_r$  
should increase with radius inside $r_s$, and that it may generally be larger than the value adopted here 
when all the aforementioned motions and flows are taken 
into account. 

The variation of the relaxation time with radius also depends 
on whether the energy input from gas fluctuations is assumed to 
be primarily global or local:  
the relaxation mechanism may in principle 
be global, in the sense 
of depending only on some effective $\rho_0$, corresponding to 
a space and time average over an appropriately chosen region (e.g., the 
region within which the feedback driven fluctuations are significant), 
or it may be local; depending  on the local gas density $\langle \rho_g \rangle (r)$ (averaged only over time). 
This is an issue we discuss further below, 
in connection to Fig.~\ref{fig:Lovsglb}. 

For a  global energy input, the relaxation time  
depends only on  variation in $v_p$, if $v_r$ is kept fixed.  
with radius.  If $v_p$ is associated  with $v_c (r)$, 
then  within the initial NFW cusp $v_p^2 \sim r/r_s$ increases by 
a factor of 6.25 as $r/r_s$ goes from 0.1 to 1. 
If $v_p$ is taken to correspond to the velocity dispersion 
then $v_p^2 \sim - r/r_s \ln r/r_s$ deep inside the cusp, and increases   
by a factor of about 2.25 between $r/r_s =0.1$ and 1.  

Thus, particles
nearer to the centre are expected to 
be affected first by the fluctuations; if only 
simply because their initial velocities, and therefore the relaxation times required to affect them significantly, tend to be smaller.  
Core formation would thus proceed inside out, with mass distribution 
at larger radii affected at later times. 

In what follows we evaluate
the energy input in the simulation at hand
and compare it with the theoretical picture sketched here; testing the 
`inside out' scheme of core formation and the assumption of dependence 
of the energy transfer principally on the average density inside $\sim$$r_s$ rather than the local density.

\subsection{Energy input}
\label{sec:enin}
\begin{figure}
\centering
	\includegraphics[width=\linewidth]{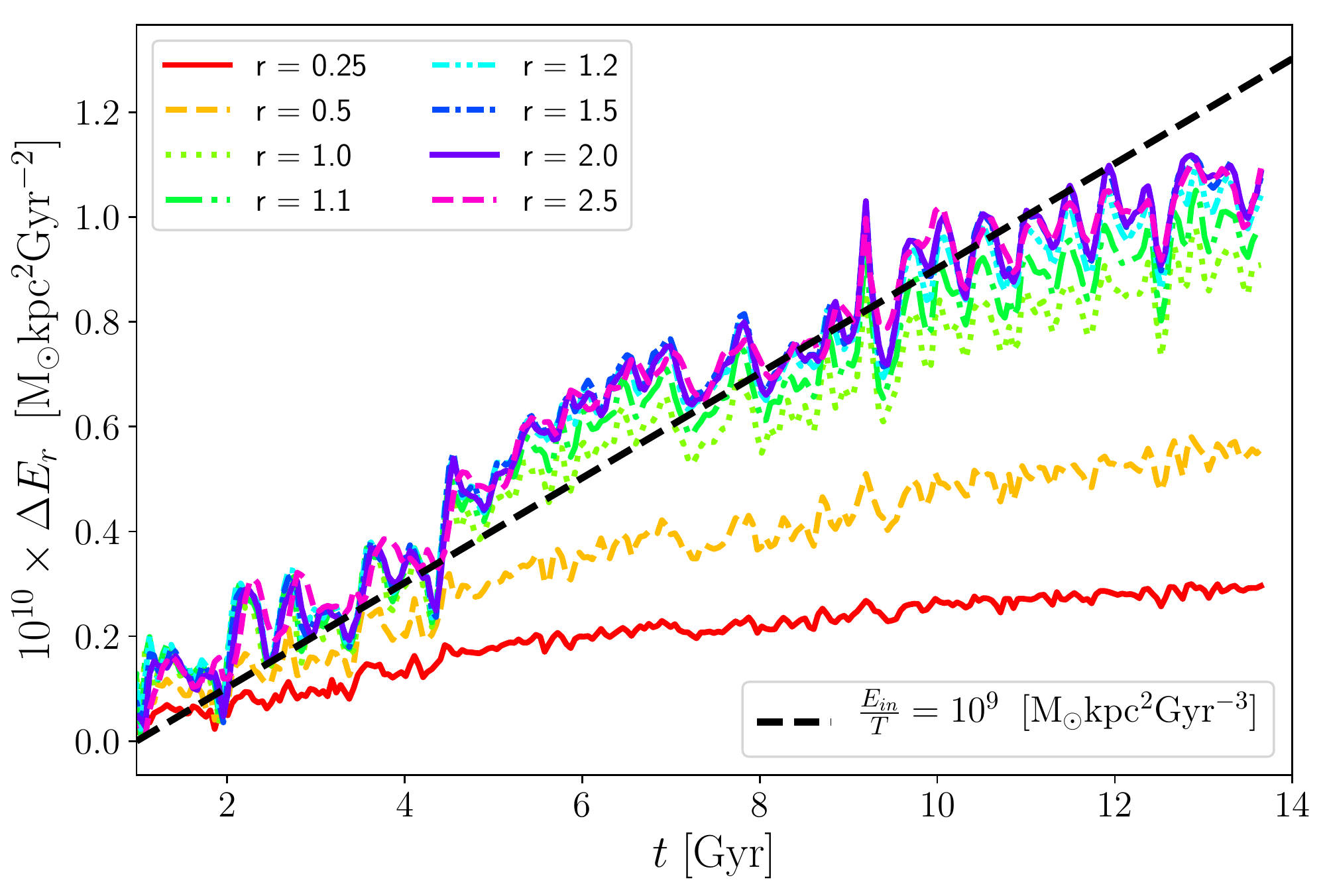}
    \caption{Energy input to halo particles from gas fluctuations inside indicated radii $r$.  
    $E_r$ is evaluated via equation~(\ref{eq:Eind}), 
    and measured from the temporal zero point $t = t_{\rm diff}$, 
    beyond which the diffusion limit may be assumed to hold. In line with the discussion following equation~(\ref{eq:vdisp}), we take $t_{\rm diff} = 1 ~{\rm Gyr}$.
    In order to reduce fluctuations related to the precise choice of zero point, we 
    calculate in practice  $\Delta E_r = E_r (t> t_{\rm diff})  - \langle E_r \rangle_{t_{\rm diff}}$, where the average is taken in the interval $0.5~ {\rm Gyr} \le t \le 1.5 ~{\rm Gyr}$.  The dashed line indicates energy 
transfer through a stationary stochastic  
process, with  parameter values appearing in equation~(\ref{eq:Ein}). 
It assumes that energy input saturates around $r_{\rm sat} = 1.5~ r_s \simeq 1.3~ {\rm kpc}$ 
(corresponding to the converging lines). 
For smaller radii, the lines flatten after a few Gyr due to mass and energy transfer 
resulting from particle migration towards larger radii.}
\label{fig:Ein_main}
\end{figure}
\begin{figure*}
\centering
	\includegraphics[width= 0.49\textwidth]{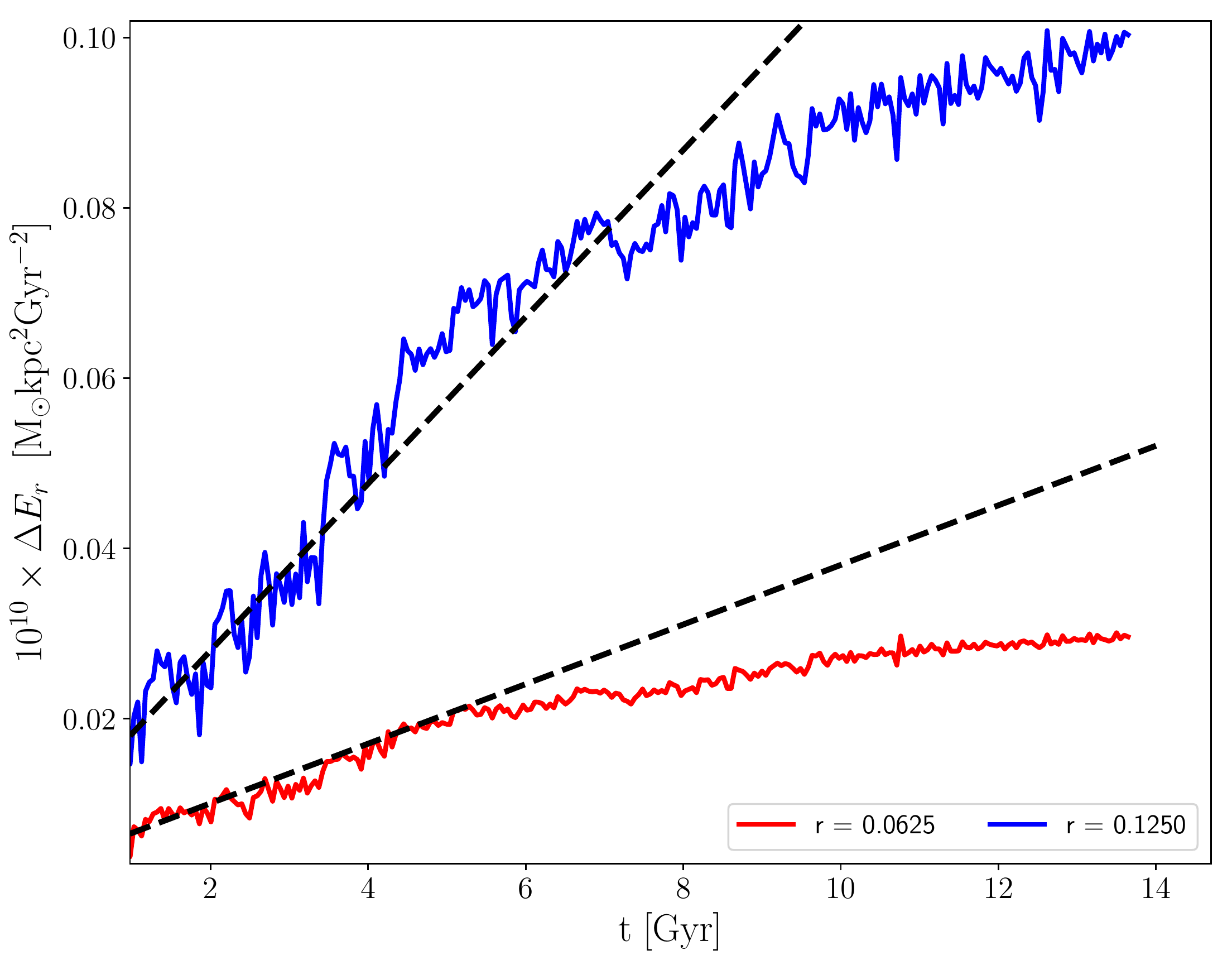}
	\includegraphics[width= 0.49\textwidth]{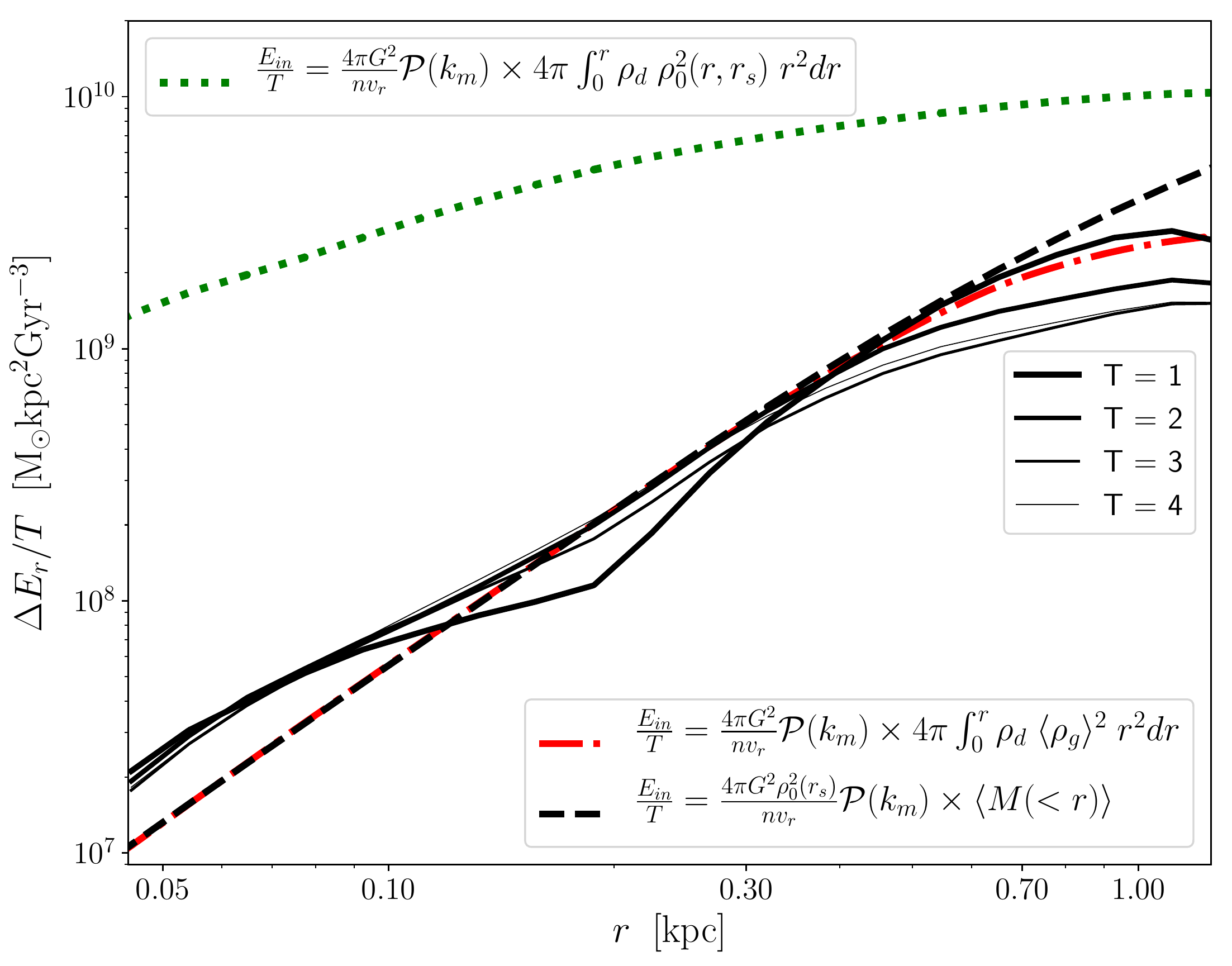}

    \caption{Local vs global energy transfer. {\it Left:} 
    Radial energy transfer as a function of time. {\it Right:} Energy input rate as a function of radius.
    The left hand panel shows that radial energy transfer is slow
    compared to the energy input; the straight lines represent 
    steady diffusive energy input, with no transfer outside the radii enclosed. 
    While this holds, one may estimate the energy input rate at various radii by simply 
    evaluating  the average slope of the generally linear energy increase. The right 
    hand panel shows those slopes over the first 
    1, 2, 3 and 4 Gyr  (solid lines with indicated values of $T$).  
    The dashed line corresponds to estimate assuming global energy transfer; using equation~(\ref{eq:EGlob})
    with $n = 2.5$,   $v_r = 30.0 ~{\rm km/s}$, $\mathcal{P} (k_m) =  3.6~ {\rm kpc^3}$,
    and average gas density within $r_s$, $\rho_0 (r_s) = \langle \rho_0 (r_s, t) \rangle_T$, with $\rho(r, t)$
    measured directly from the simulation (using equation~\ref{eq: rhoavdef}) 
    and averaged over $T = [0: 4] ~{\rm Gyr}$ 
    (giving $\rho_0 = 1.95 \times 10^6 ~M_\odot/{\rm kpc^3}$).   The dotted line  
    corresponds to the assumption of purely local energy input; using the indicated equation, where the local gas density $\rho_g ({\bf r}, t)$ is time-averaged 
    (in the range $T = [0: 4] ~{\rm Gyr}$) over spherical shells at $r$. 
    This vastly overestimates the energy input at all radii. The dashed dotted line represents an intermediate alternative, whereby $\rho_g$ is averaged over spheres of radius $r_s$ and centered at $r$ (as in equation~\ref{eq:semiloc}).} 
\label{fig:Lovsglb}
\end{figure*}

Equation~(\ref{eq:relax}) implies that a particle  
increases its velocity variance with time $T$ as 
\begin{equation}
\langle (\Delta v)^2 \rangle =   
\frac{8 \pi (G \rho_0)^2 \mathcal{P}({k_m})}{n v_r} T.      
\label{eq:vdisp}
\end{equation}
The linear time dependence here is  characteristic of a diffusion 
process. It is valid in the diffusion limit, which means that
$T \gg t_{\rm diff} \equiv  (v_r k_m)^{-1}$~\citep{EZFC}. 
To account for this in practice, we will generally measure 
$T$ and   $\langle (\Delta v)^2 \rangle$ from zero points at 
times $t \gtrsim t_{\rm diff}$. This is necessitated not just by 
the applicability of the diffusion limit, but also by the assumption
of a quasi-steady stochastic process, 
which is clearly invalid during the first few hundred Myr, which are 
characterised by a  highly evolving phase, involving the initial gas contraction 
and the triggering of a 
starburst that expels most of it out of $r_s$.    

Assuming $v_r \simeq 30 ~{\rm kpc/Gyr}$, 
gives $(v_r k_m)^{-1} \lesssim 1/30 {~\rm Gyr}$,
for $1/k_m\lesssim 1~{\rm kpc}$. 
Thus the steady state diffusion limit may be safely assumed to be established beyond 1~Gyr. We will take it to be our zero point for $T = t - t_{\rm diff}$. 

As in standard two body relaxation, one may assume that fluctuations 
initially change just the velocities of the halo particles~\footnote{This is strictly speaking valid when the orbital period of halo 
particles $T_p$ is much longer than the fluctuation timescale $T_f$. Here, they are generally 
of the same order. However the impulsive 
approximation is known to remain reasonable even when $T_p$ is moderately smaller than  $T_f$ (e.g. \citealp{BT,AguWhite1985, PenarrubiaA}). 
For a circular orbit and gas mode 
of wavenumber $k$, $T_p/T_f = k r v_r/v_c (r)$. This is minimized in the initial cuspy potential, and when $k = k_m$, and $v_r \simeq  30 {\rm km/s}$ includes only the gas speed.  
For $k_m = 2 {\rm kpc}^{-1}$ (consistent with Fig.~\ref{fig:powerspecavrg}), this minimal ratio is
$T_p/T_f \simeq 3$ for $r \simeq 1 {\rm kpc}$. It is significantly smaller than unity ($\lesssim 1/3$) only at $r < 30 {\rm pc}$. Adiabatic corrections may be included (\citealp{PenarrubiaA}),
though the corrections  are generally 
modest for our range of $T_p/T_f$ 
(Cf. Fig. 2 of \citealp{PenarrubiaA}). In our context the impulsive regime also in fact places a lower limit on $k_m$, independent of the flattening of the power spectrum in Fig.~\ref{fig:powerspecavrg}, as the effect of fluctuations rapidly decreases beyond it.}

The average energy change per unit mass resulting from this is 
$\langle \Delta E \rangle = \langle \Delta~\frac{1}{2} (\Delta v)^2\rangle = v \langle \Delta v \rangle + \frac{1}{2} \langle (\Delta v)^2 \rangle$. 
In principle, the particles may gain energy through the first term or lose it to the 
fluctuating field through dynamical friction, 
as in any general diffusive dynamical process. But 
in practice the mass of the particles is far too small and 
so the average energy gain per unit mass is simply  
$\langle \Delta E \rangle =\frac{1}{2} \langle (\Delta v)^2 \rangle$. 

By fixing $v_r$, and assuming  
the energy transfer mechanism to be global ---  again, in the sense of 
depending on an effective characteristic density $\rho_0$ rather than on the local $\langle \rho_g \rangle_T (r)$ --- the expected energy input to halo particles 
within radius $r$ may be estimated as  
\begin{equation}
 E_{\rm in} (< r) =~\langle M (<r)  
 \rangle~\langle \Delta E \rangle~=~\langle M (<r)  \rangle~\frac{4 \pi (G \rho_0)^2 \mathcal{P}_{k_m}}{n v_r}~T, 
 \label{eq:EGlob}
\end{equation}
where $\langle M(<r) \rangle$ is the (time) average of the 
halo mass enclosed 
within $r$. 

We discuss the assumption of global energy input below. First, we assume 
it holds at least approximately, and use it to estimate 
the total energy transfer from the gas to the halo.
Given the rapidly decreasing gas density beyond $r_s$ 
(Fig.~\ref{fig:densityprofiletime_all}), one may suppose 
that the energy input 
saturates at $r_{\rm sat} \gtrsim r_s$. 
If the energy transfer depends on global properties, then 
$\rho_0$ may be considered to correspond to the 
space and time average of the gas density
inside $r_{\rm sat}$, i.e.  
$\rho_0 ({\rm sat}) = \langle \rho_0 (r_{\rm sat}, t)\rangle_T$, where
$\rho_0 (r_{\rm sat}, t)$ is given by~(\ref{eq: rhoavdef}), with 
$r = r_{\rm sat}$ (unless otherwise stated, particularly in relation to Fig.~\ref{fig:Lovsglb}, 
time averages are evaluated from $T=0$, i.e. $t = t_{\rm diff}=1~\rm Gyr$,
to the end of the simulation at $t = 13.7 \rm Gyr$).  
For $r_{\rm sat} \gtrsim r_s$, $\langle M (<r) \rangle$
is  close to the initial mass at $T=0$,  
since the dark matter 
mass inside it is essentially conserved.  
We thus write, in terms of the initial mass $M_0 (< r)$,
\begin{multline}
E_{\rm in} (<r_{\rm sat}) = 10^9~M_\odot~{\rm kpc^2~Gyr^{-2}}~\frac{M_0 (< r_{\rm sat})}{10^8 M_\odot}~\frac{T}{\rm Gyr}~\times\\
\left(\frac{n}{2.5}\right)^{-1} \left(\frac{v_r}{30 ~{\rm km/s}} \right)^{-1}
   \left(\frac{\mathcal{P} (k_m)}{3 {\rm~ kpc}^3}\right)
\left(\frac{\rho_0~(r_{\rm sat}) }{10^{6} ~M_\odot ~{\rm kpc}^{-3}}\right)^2,
\label{eq:Ein}
\end{multline}
where we have chosen $v_r$ in accordance with 
the discussion of the previous subsection,
and inserted values characteristic of the initial halo 
mass and average gas density (as defined above) 
around $r_s$. More precisely, the numbers correspond 
to values at $r = r_{\rm sat} = 1.5 r_s \simeq 1.3 ~{\rm kpc}$, 
but we note that the estimate of the total energy input
is not  sensitive to the choice of $r_{\rm sat}$; as 
the product $M_0 (< r) \rho_0^2 (r)$ 
varies  slowly with radius around $r_s$ (indeed, more generally,
$\langle M (<r) \rangle \rho_0^2 (r)$ varies by 
at most a factor of 2 in the range  $0.1~ r_s \le r \le 2 r_s$). 
The power spectrum parameters are 
estimated from the results in figures \ref{fig:powerspecavrg} and 
\ref{fig:massdisper} (a larger $\mathcal{P} (k_m) = 4 {\rm kpc}^3$  
gives the same result if $v_r = 40 {\rm km/s}$, which would take into account enhancement in the speeds of halo halo particles, relative to the gas, due to their own motion).  

We now wish to compare the prediction in equation~(\ref{eq:Ein})
with the actual energy input inferred from the 
simulation.  
For this purpose, we define the quantity 
\begin{equation}
    E_r = \frac{m}{2}  \left( \sum_i v_i^2  +  \Phi_i \right),
    \label{eq:Eind}
\end{equation}
where $m$ is the halo particle mass in the simulation, 
$v_i$ is the 
speed of particle $i$ and $\Phi_i$ is the Newtonian potential 
at its location. The summation 
is evaluated over all particles within radius $r$. 
Increase in this quantity measures the 
energy input inside radius $r$, 
provided there is no dark matter mass or energy outflow from that radius
and the mass distribution beyond it remains constant 
(so that changes in the potential are solely due to modification of the mass distribution 
within $r$)~\footnote{The potential $\Phi_i$ at particle $i$ is due to other halo particles; we ignore changes in the potential arising directly from the gas flows, 
as these constitute a fluctuating contribution with small average, 
particularly after the initial expulsion episode during the first few hundred Myr}. 
We may expect that these conditions hold, at all radii, 
for sufficiently small times, as  
the kinetic energy acquired by the halo particles is still being converted 
into changes in the mass distribution and potential, a process slower than that characterising the initial energy input. 
The conditions should hold for longer times as $r$ increases; and 
we expect that beyond $r_{\rm sat} \gtrsim r_s$,  the energy 
input saturates to a specific value regardless of radius, and little 
energy is transferred beyond that radius, so the change in $E_r$
from its initial value (at $T=0$) corresponds 
to the total energy input from the fluctuating gas. 
In particular, if that input may be estimated 
theoretically using equation~(\ref{eq:Ein}), then one expects 
$\Delta E_{r_{\rm sat}} \simeq E_{\rm in} (< r_{\rm sat})$.

These expectations are confirmed  in Fig.~\ref{fig:Ein_main}.
At all radii, we initially find a general linear increase in $E_r$, 
as expected from  a steady diffusive process, 
with stochastic variations on this general trend. At smaller radii, 
the lines clearly flatten after a few Gyr, as a result of energy 
and mass outflow onto larger $r$. 
A saturation radius $r_{\rm sat} \gtrsim r_s$, beyond which
the lines converge,
can be defined with value consistent with 
$r_{\rm sat} = 1.5 ~r_s \simeq 1.3 ~{\rm kpc}$ as assumed in equation~(\ref{eq:Ein}).
The slope  derived from that relation is also consistent 
with that derived directly from the simulation, as indicated
by the dashed line.   

The above suggests that the total energy input within $r_{\rm sat}$ may 
be adequately described  by assuming a global energy 
transfer mechanism, parameterised  by the average gas density and halo mass 
at $r_{\rm sat}$ inferred from the simulation; 
and with $v_r$ and density contrast  
power spectrum parameters  also consistent with those 
found in the simulated hydrodynamics. 

We now wish to ask to what extent this global approximation 
holds in general; for, as the time averaged 
gas density inside $r_s$ is not strictly constant,  
energy transfer may, in principle,  depend instead 
on the local density. This is a possibility akin to invoking the purely local approximation when evaluating the effect of two body relaxation in stellar dynamics, by
using Chandrasekhar's formula and plugging in the local stellar density. 
This is in principle plausible, but more difficult to interpret 
in the present case; whereas in two body relaxation logarithmic intervals in impact parameters (or spatial scales) equally contribute to the fluctuations leading to relaxation, here fluctuations at the largest scales (characterized by $\mathcal{P} (k_m)$) are more important.

Global energy transfer implies the validity of equation~(\ref{eq:EGlob}).  
Local energy transfer, on the other hand, requires the 
evaluation of the time average $\langle \rho_g (r,t) \rangle^2 (r)$
over spherical shells centered at $r$, 
and integrating it over the dark matter mass in those shells, 
$4 \pi r^2 \rho_d (r)$ (assuming a fixed $v_r$).  
The right panel of Fig.~\ref{fig:Lovsglb} (dotted line) shows that, 
when assuming power spectrum parameters and velocity $v_r$ 
consistent with the simulation, this vastly overestimates the energy input. 
The non-local 
assumption --- with energy input rate within radius $r$ scaling simply as 
$\rho_0^2 M(<r)$ as implied by equation~(\ref{eq:EGlob}) --- 
results in much better agreement, with a moderate discrepancy at small radii that may at least partly 
be accounted for by an increase of $v_r$ with radius (as discussed above and expected from Fig.~\ref{fig:velav}),  
which we do not take into account here for simplicity.

The global approximation, with total energy input within radius $r$ simply 
proportional to $M (<r)$,  clearly
cannot  remain valid as $r \rightarrow r_{\rm sat}$, when the gas density 
and fluctuations rapidly decrease. Indeed, the difference between 
$\rho_0 (r_s)$ that fits the energy transfer rate in 
Fig.~\ref{fig:Lovsglb} 
and $\rho_0(r_{\rm sat})$ used in Fig.~\ref{fig:Ein_main}, principally reflects
the gradual saturation process; the fit in Fig.~\ref{fig:Ein_main}
effectively assumes sudden saturation, 
while in reality the process is gradual. 
To take this into account one may 
invoke an intermediate regime, between the purely local and purely global energy transfer 
limits. We do this by defining 
\begin{equation} 
\rho_0 (r, r_{\rm av}) =  \langle~\rho_g ({\bf r}, ~{\bf r} - {\bf r}_g)~\rangle_{r, |{\bf r} - {\bf r}_g)| < r_{\rm av} , T},
\label{eq:semiloc}
\end{equation}
where the average over the gas density 
is evaluated over time at points ${\bf r}_g$ inside spheres
with centres at radial coordinate $r$ and radii $r_{\rm av}$. 
Thus in this context,  $\rho_0 (r_{\rm sat}) = \rho_0 (0, r_{\rm sat})$ 
and $\rho_0 (r_s) = \rho_0 (0, r_s)$. 
The results for $\rho_0 (r, r_s)$  are shown by the dashed dotted
line in the right hand panel of Fig.~\ref{fig:Lovsglb}.  They suggest 
that the energy transfer process is best considered as non-local, with a 
range $\sim$$r_s$.


Finally, we note that although the energy is assumed to be transferred to halo particles initially   
as kinetic energy, due to modification 
in their velocities (as given by equation~\ref{eq:vdisp}),  
the changes eventually 
affect the self consistent potential.
The resulting average gain in total energy per unit mass
turns out to be amenable to estimation from 
a low energy cutoff that appears 
in the phase space distribution function. 
In Appendix~\ref{sec:cut}, we show how this 
can be related to the energy input calculated here, and connected to the change in the potential that accompanies core formation. 

\begin{figure*}
\centering
	\includegraphics[width= 0.495\linewidth]{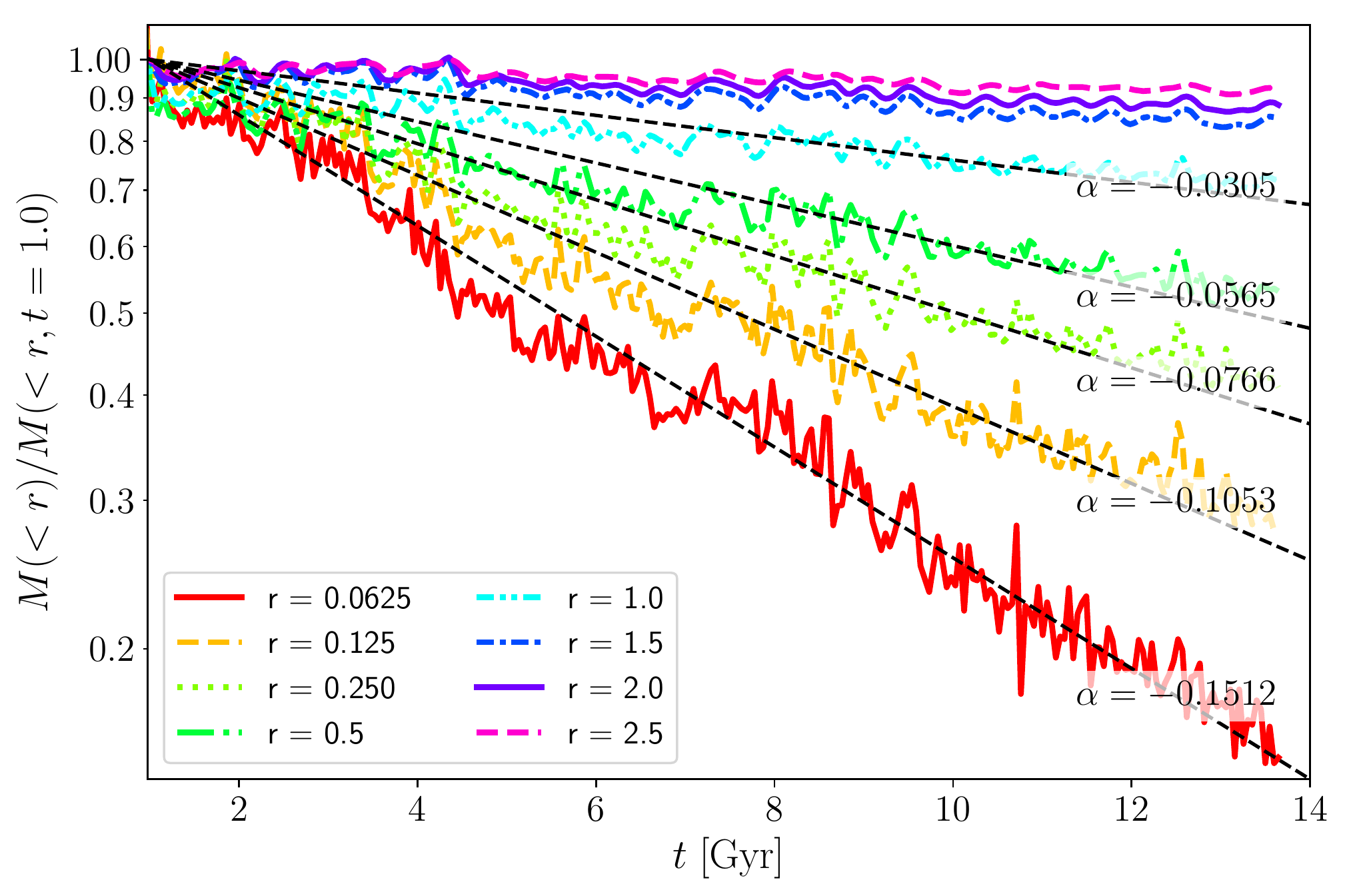}
	\includegraphics[width= 0.495\linewidth]{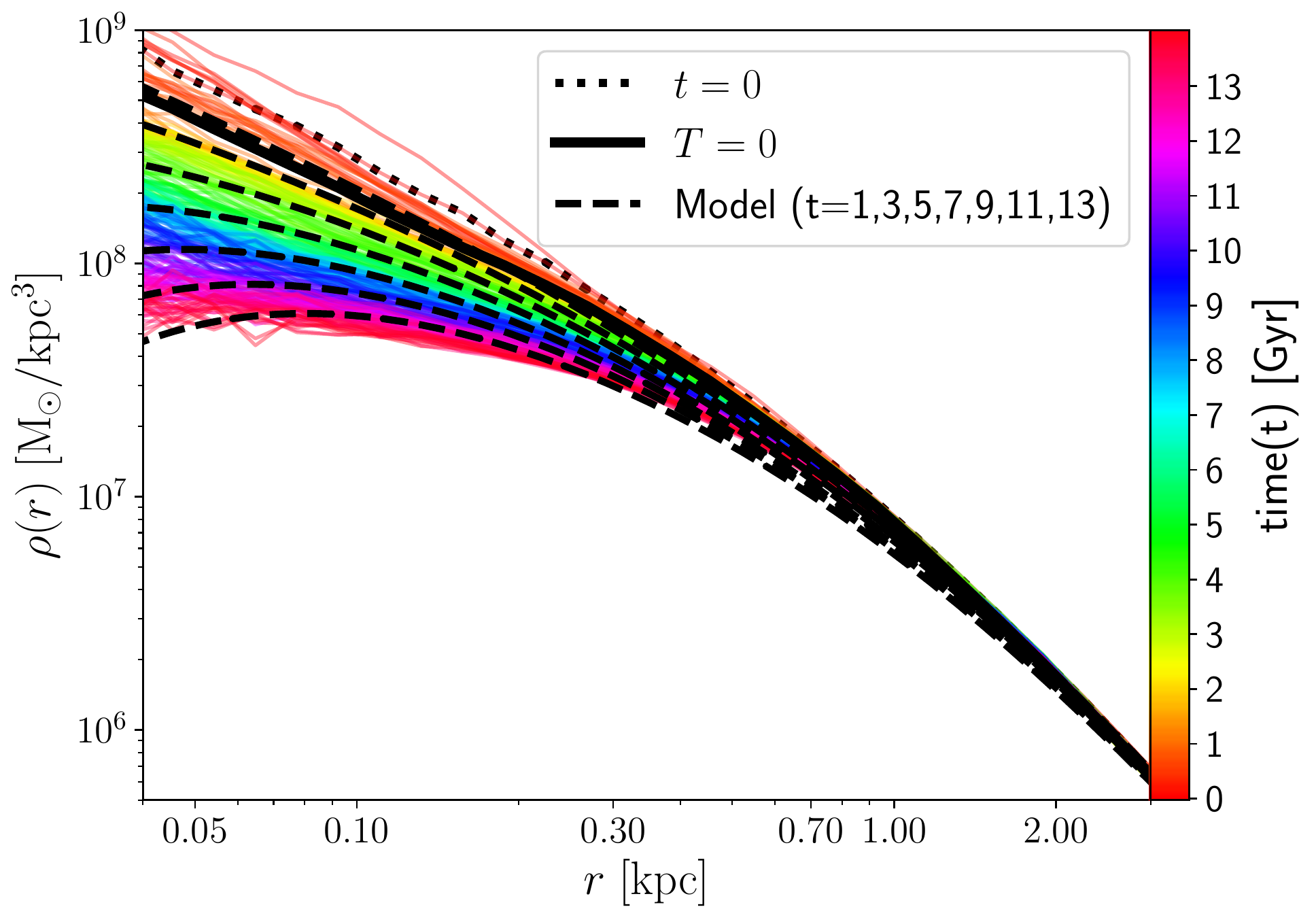}
    \caption{Mass migration and core formation. The left panel shows 
   the ratio between the halo mass within radii $r$ (in kpc) and the corresponding initial mass at $T= 0$
   ($t = t_{\rm diff} = 1 {\rm Gyr}$, as discussed in relation to Fig.~\ref{fig:Ein_main}). 
   The dashed lines are exponential fits, $\propto \exp(\alpha T)$, with numbers corresponding to 
   $\alpha (r)$ in ${\rm Gyr}^{-1}$.  The exponential decay may be derived from a simple theoretical 
   model for the mass transfer; in its context $\alpha$ is predicted to scale with the inverse of an energy
   relaxation time (obtained using Eqs.~\ref{eq:Erelax}. \ref{eq:Dcoef}, and~\ref{eq:Einits}).
    The right panel shows the corresponding evolution in density.
    Theoretical expectations are shown by the dashed lines. 
    They are obtained by differentiating equation~(\ref{eq:ME}),  starting from 
    an NFW fit to the dark matter density in the simulation at $T =0$. To reduce 
  uncertainties  arising from fluctuations in the density profile around $T=0$, we average simulation  
   outputs over the range  $0.5 ~{\rm Gyr} \le t  \le 1.5 ~{\rm Gyr}$ (as in Fig.~\ref{fig:Ein_main}). 
   Model predictions are shown at $T =0$ (fitting the averaged profile), and at  $2, 4, 6, 8, 10$ and 12 Gyr, corresponding to times $1, 3, 5, 7, 9, 11$ and 13 Gyr.}
\label{fig:core}
\end{figure*}

\subsection{Mass migration and core formation}
\label{sec:massmod}

The energy input from the fluctuating gas 
leads to the 
migration of halo particles from the inner radii, which  
decreases the enclosed mass,  
as shown in Fig.~\ref{fig:core}, left panel. 
Straight lines on the log linear scale suggest a general 
exponential decrease in mass with time. 
A full examination of its origin in the context 
of a diffusion model would require a full Fokker Planck formulation, using 
the full (first and second order) diffusion coefficients and  explicitly including changes in the potential due to the evolving mass distribution. 
Here we proceed heuristically, {leaving the examination 
of the reasons why the drastic simplifications invoked 
seem to work to Appendix~\ref{app:powerflux}.} 

We suppose that the mass flux across energy surface $E$ changes the mass within it through the first order energy diffusion coefficient  
\begin{equation}
 D [\Delta E] = \frac{1}{M(< r_{\rm sat})} \frac{E_{\rm in} (r_{\rm sat})}{T},
 \label{eq:Dcoef}
\end{equation}
describing the average rate of change of halo particle energy per unit mass due
the gas fluctuations. {At the energy input rate inferred from Fig.~\ref{fig:Ein_main},
 this is of the order of 
$10 ~{\rm kpc^2~Gyr^{-3}}$ in the simulation at hand (which is consistent with theoretical expectations, given the power spectrum and density inferred from simulation, 
cf. equation~ \ref{eq:Ein},
but is actually empirically independent of the model).} 

The change in mass of particles with energy less than $E$ is 
\begin{equation}
\frac{\partial M (< E)}{\partial t}~\approx -\frac{\partial M (<E)}{\partial E}~D~[\Delta E], 
\label{eq:Mflux}
\end{equation}
where  $M (< E)$ is the mass in halo particles with energy less than $E$ and 
$\partial M (<E)/\partial E$
is the mass-weighed differential 
energy distribution \citep{BT}. 
Furthermore, we approximate this latter quantity  by its average within $E$, such that
\begin{equation}
    \frac{\partial M (<E)}{\partial E} \approx~a_p~\frac{M(<E) - M(0)}{E - E (0)},
\label{eq:a_p}
\end{equation}
where $E (0)$ is the energy of a particle at rest at the centre of the potential (so $E (0)= \Phi (0)$ and $M (0) = 0$), and $a_p$ is a constant numerical factor of order 1. 
This holds exactly inside
pure power-law cusps: e.g. {as derived in Appendix~\ref{app:powerflux}}, 
one finds in such cases that 
$a_p  = (3 + \gamma) / (2 + \gamma)$ for $\rho \propto r^{\gamma}$ (thus 
$a_p = 2$ for a $1/r$ cusp, tends towards $1.5$ for flatter 
profiles, and is larger 
for steeper ones (finally diverging in the case of the  
singular isothermal sphere, where a potential zero point at the 
centre cannot be fixed as above due to logarithmic divergence).

{With this in mind we will assume that 
\begin{equation}
\frac{\partial M (< E)}{\partial t}~= - a~\frac{M(<E) - M(0)}{E - E (0)}
D~[\Delta E],
\label{eq: approxFP}
\end{equation}
where the parameter $a$ absorbs all of the major approximations  
involving the use of the first order diffusion coefficient (equation~\ref{eq:Mflux}), 
the association of the differential energy distribution with its
average value (equation~\ref{eq:a_p}), and 
the neglect
of the effect of the evolution of the self consistent potential.

In Appendix~\ref{app:powerflux}, we discuss how such a formulation may 
be justified if the differential energy distribution may be approximated
by a power law; and how, when one includes 
both diffusion coefficients, a zero Fokker Planck flux may be attained if the 
distribution takes an exponential form.

We now define an energy relaxation time as
\begin{equation}
t_{\rm relax} (E) =  D[\Delta E]^{-1}~[E- E (0)].
\label{eq:Erelax}
\end{equation}
For timescales small compared to the initial relaxation time  
(with $E$ and $E_0$ defined at the initial time), one may take this 
timescale to be constant, evaluated by inserting the initial values 
of $E$ and $E_0$ in the above equation. We will assume that this is 
the case here (as we will see below, the 
energy changes are small enough, during timescales of interest, 
to render this approximation reasonable).}
Then, assuming that equation~(\ref{eq:Mflux}) 
is applicable in the diffusion limit, solving it (starting at time $T=0$) results in 
\begin{equation}
    M (< E) = M_0 (<E)~\exp\left[-a~T/t_{\rm relax}(E)\right], 
\label{eq:ME}
\end{equation} 
where $M_0$ refers to the mass at time $T=0$. 

Within this picture, the numbers on the lines in the left panel 
of Fig.~\ref{fig:core}, denoted by $\alpha$, should correspond to $ a/t_{\rm relax} (E)$, 
if $E$ is associated with particle energies inside the chosen radii. 
To make this correspondence,  we fix $E = \langle E \rangle (r)$ to be the average specific energy at radius $r$ in the initial profile. 
As a first approximation,  we simply associate this 
with the 
initial NFW profile of the configuration.  
The specific potential energy $\Phi (r) - \Phi (0)$ in this case is 
$4 \pi G \rho_s r_s^2 [1- \ln (1+x)/x]$, with $x = r/r_s$, 
and $r_s$ and $\rho_s$ as in Table~1. 
To obtain 
a simple form for the kinetic energy we make use of established empirical relations according to which the pseudo phase space density varies with radius approximately as 
$\rho/\langle v^2 \rangle^{3/2} \sim r^{-1.875}$ \citep{TayNav01}, and normalize
the velocity variance to its value around $r = r_s$, such that   
$\langle v^2 \rangle \simeq 470~ {\rm km^2/s^2}$. One may then write
\begin{equation}
E - E (0)  \simeq  4 \pi G \rho_s r_s^2 \left[1 + \frac{2^{4/3}}{10} \frac{\left(x^{0.875}\right)^{2/3}}{(1+x)^{4/3}} - \frac{\ln (1+x)}{x}\right]. 
\label{eq:Einits}
\end{equation}
Using this formula in conjunction with equation~(\ref{eq:Erelax}), and using $a = 3.4$,
we find  $a/t_{\rm relax} = 0.034, 0.047, 0.071, 0.11, 0.17~{\rm Gyr}^{-1}$ at $r = 1, 0.5, 0.25, 0.125, 0.0625~{\rm kpc}$ respectively. 
These numbers agree to better than $20 \%$ with the values of $\alpha$ 
indicated on the fitting lines in Fig.~\ref{fig:core} (left panel). 
The corresponding energy relaxation times are about $100, 72.3, 47.9, 30.9$ and 
$20 ~{\rm Gyr}$. 

A more careful analysis, 
taking into account that the quasi-steady 
state diffusive process is well established only  after $t \gtrsim t_{\rm diff} =  1 \rm Gyr$, 
suggests $a \simeq 2.5$ (cf. below). The corresponding energy relaxation times are therefore
lower by a factor of about 1.4 than those quoted above.   
Even then, these relaxation times are still 
significantly larger 
than what is obtained from the velocity variance (equation~\ref{eq:relaxnum}). 
This is because 
$E - E(0)$ is generally  larger 
than the average kinetic energy (from equation~\ref{eq:Einits}, by a factor of about four at $r_s$). 
Indeed, changes in kinetic energy lead to relatively little change
in total energy. In fact the relative change in $E_r$ at 1 kpc (as calculated from equation~\ref{eq:Eind}, starting at $T=0$), is $\Delta E_{r}/ E_{r} \simeq 0.2$, over  12.7 Gyr. Initially the rate of relative change in energy 
is much larger at smaller radii 
(as illustrated by the differences in $E_{\rm in}/T$ inferred from 
Figs.~\ref{fig:Ein_main} and~\ref{fig:Lovsglb}). 
But it subsequently saturates, as
the energy is redistributed in the system, with mass and energy flowing towards outer radii (Fig.~\ref{fig:Lovsglb}, left panel). When thus redistributed, 
the modest changes in total energy lead to a modification
of the self consistent potential, including the minimal possible 
energy in it, as discussed in Appendix~\ref{sec:cut}. 

This leads to  
core formation, as observed in the right 
panel of Fig.~\ref{fig:core}, There, we note that  we find  no evidence 
of any accelerated 
core formation, 
relative to the velocity relaxation time (equation~\ref{eq:relaxnum}), 
as found in~\cite{EZFC}, when non-spherical modes were used in conjunction with the Hernquist-Ostriker code. 

Having obtained the evolution of the enclosed mass within a given radius, 
it is also possible to define a theoretical density 
$\rho = r^{-2} d M/d r$ and derive 
a closed form formula for it using Eqs.~(\ref{eq:ME}), (\ref{eq:Erelax}), 
and~(\ref{eq:Einits}). The result may then be 
compared with the dark matter density evolution in the simulation.
More care is however required here in  defining the initial 
energies in equation~(\ref{eq:Einits}), as we must start our 
comparison at $T = t - t_{\rm diff} =0$, i.e., when the diffusion limit
at the basis of our model is valid (cf. the discussion in relation to 
Fig.~\ref{fig:Ein_main}). For this purpose, 
we average simulation outputs 
in the time range $0.5~ {\rm Gyr} \le  t \le 1.5~ {\rm Gyr}$, and then fit the resulting dark 
matter density with an NFW profile. 
The corresponding parameters are $r_s = 1.17 ~{\rm kpc}$ and $\rho_s = 2.7 \times 10^7~ M_\odot/{\rm kpc^3}$. 
When the latter is adjusted by adding a gas fraction of about $0.075$ 
(assumed for simplicity to be also NFW with same $r_s$, but taking into account the initial central gas expulsion), the multiplicative factor 
$\rho_s r_s^2$ in front of the bracket in equation~(\ref{eq:Einits}) remains approximately 
the same as in the case of the $t=0$ profile. After some trials, we found that values 
of $a$ in the range 2.5 to 2.8 provide  reasonable approximations 
to the density evolution inferred from the simulation 
(perhaps surprisingly so, 
given the various simplifying assumptions of
our mass transfer model). 
This is illustrated in the right panel of Fig.~\ref{fig:core}, where 
we compare the density evolution expected from our model (fixing $a = 2.5)$
with the results from the simulation.

{Note that, other than the value of $a$, which 
is theoretically fairly tightly constrained by the arguments of 
Appendix~\ref{app:powerflux}  (cf. equation \ref{eq:calca}),
there are no free parameters 
in these fits. The   
energy distribution of halo particles
at $T=0$, and the energy transfer rates from gas to halo, are of course characteristic of a given system. They will thus vary from simulation to another, but they are entirely fixed for any one system (within numerical uncertainties in their inference from the simulation).}

\section{Conclusion}

Potential fluctuations from feedback driven gas can `heat' halo cusps, turning them into cores. 
This work aimed at quantifying this, by measuring the gas fluctuations and tracking the way they transfer energy to the central halo, forcing the outward migration of the dark matter. 
The interpretive framework we use is a model, first outlined in \citet{EZFC}, which predicts that these processes principally depend on the amplitude of the fluctuations, as measured by the normalisation of their  power spectrum, and the average gas density. The result being a standard diffusion process,  characterised by  a linear temporal increase in velocity variance and  energy of halo particles as a result of their interaction with the fluctuating gas field. As shown 
previously (\citealp{EZFCH}), in this picture the effect of the fluctuations reduces to standard two body relaxation in the case of a white noise power spectrum. It may thus, in this limit,  also describe halo heating {\it via}  dynamical friction from a system of compact, monolithic massive clumps moving among much lighter dark matter particles.  

To test this interpretative framework,
we  measure the density fluctuations in feedback-driven gas from a full hydrodynamic simulation of a model dwarf galaxy (Figs. \ref{fig:Gas_Map}-\ref{fig:densityprofiletime_all}), and obtain their density contrast power spectrum (Fig.~\ref{fig:powerspecavrg}). We find that the spectrum follows a power-law in wave number ($\propto k^{-n}$), with  an exponent $ 2 \lesssim n \lesssim  3 $, for the whole period of the simulation (spanning a Hubble time). 
Although the time-averaged density of the driven gas varies with radius, the variation is modest inside the initial NFW scale length, relative to a sharp drop outside. This suggests a characteristic density that may be used as input for the model.  
We also examine the velocity distribution of the gas and find that it is approximately fit by Maxwellians at larger radii (Appendix~\ref{Sec:Additional_Gauss}). The kinetic energy power spectrum is close to the Kolmogorov form over large range of scales (Fig.~\ref{fig:velocitypowerspecfit}). 
This reinforces the picture of a fully turbulent medium. 

With the input parameters directly measured from the simulation, we use our model  
to calculate the total energy transfer rate from fluctuating gas to the central halo. The result is compared with the actual energy input to the system of halo particles, as directly inferred from the simulation (Fig.~\ref{fig:Ein_main}). This is found to generally agree --- displaying the general linear increase, expected of a steady diffusion process ---  over almost a Hubble time.  
We also examine the radial 
distribution of the energy input rate (Fig. \ref{fig:Lovsglb}).  
We find that the energy 
transfer is indeed much better approximated as a global rather than local process; in the sense that it depends on the average gas energy density within the core region rather than the local density at each radius. 

The energy is initially transferred from the gas as kinetic energy to individual halo particles, but it is then redistributed through the self consistent gravity of the system, a process through which the core replaces the cusp. 
The process comes with a low energy cutoff in the halo phase space distribution function, as  particles migrate to higher energy levels. It is straightforward to link the level of this cutoff to the total energy input from the gas, and to the resulting change in  halo gravitational potential that comes with core formation (Appendix~\ref{sec:cut}).

The energy flow upwards is accompanied by a mass flow outwards. 
We empirically find an exponential decrease in halo mass with time, within a given radius, in the initial cusp. We then devise a simple approximate description of the mass flow, based on our model, from which the exponential form may be inferred. In this context, the exponential decay time scales with (and is of the order of) the local (energy) relaxation time, and the evolution of the corresponding theoretical density profile mimics that in the simulation (Fig.~\ref{fig:core}).

Strictly speaking, energy transfer {\it via} a standard diffusion process requires the gas force fluctuations to be normally distributed. We have verified this to be the case to a  good approximation (even if the larger {\it density} fluctuations can be lognormal rather than Gaussian at small radii, cf. Appendix~\ref{Sec:Additional_Gauss}). 
Another assumption of the model is the use of the `sweeping' approximations of turbulence theory, whereby the statistics of the  
spatial fluctuations are transferred    
to the time domain through large scale flows. The general consistency of the spatial power spectrum parameters with those that fit the mass dispersion in  the time domain (Fig.~\ref{fig:massdisper}), confirm that this may be an applicable approximation.

In general, the assumptions and predictions of the theoretical framework 
seem vindicated for the model galaxy studied here. This suggests a remarkably concise description of halo core formation from gaseous fluctuations, 
summarizing the effect of much complex `gastrophysics'  
in terms of the two principal parameters of gas density and fluctuation levels. 
Obvious extensions include considering different galaxy masses, as well as verifying that the model also predicts a lack of core formation in simulations that do not produce them. Success there 
would help delineate particular elements in the physical `subgrid' input that are most crucial in producing the required conditions for core formation, and examining how they compare with observations.  

It may also be possible to test the predictions regarding 
the level of the fluctuations required for core formation directly from observations. 
In particular, it has already been possible to derive surface density power spectra for larger,and relatively quiescent galaxies. Estimating these for actively star forming galaxies, and calibrating  them with three dimensional density contrast power spectra entering into our model calculations, may provide a direct test of our picture of core formation through gaseous fluctuations.

\section*{Acknowledgements}

We would like to thank the referee for a careful reading, thorough report and constructive criticism that helped improve our manuscript. 

\section*{Data Availability}


The simulation data underlying this article will be shared on reasonable request.



\bibliographystyle{mnras}
\bibliography{cuspcoreII} 

\begin{thebibliography}{}
\makeatletter
\relax
\def\mn@urlcharsother{\let\do\@makeother \do\$\do\&\do\#\do\^\do\_\do\%\do\~}
\def\mn@doi{\begingroup\mn@urlcharsother \@ifnextchar [ {\mn@doi@}
  {\mn@doi@[]}}
\def\mn@doi@[#1]#2{\def\@tempa{#1}\ifx\@tempa\@empty \href
  {http://dx.doi.org/#2} {doi:#2}\else \href {http://dx.doi.org/#2} {#1}\fi
  \endgroup}
\def\mn@eprint#1#2{\mn@eprint@#1:#2::\@nil}
\def\mn@eprint@arXiv#1{\href {http://arxiv.org/abs/#1} {{\tt arXiv:#1}}}
\def\mn@eprint@dblp#1{\href {http://dblp.uni-trier.de/rec/bibtex/#1.xml}
  {dblp:#1}}
\def\mn@eprint@#1:#2:#3:#4\@nil{\def\@tempa {#1}\def\@tempb {#2}\def\@tempc
  {#3}\ifx \@tempc \@empty \let \@tempc \@tempb \let \@tempb \@tempa \fi \ifx
  \@tempb \@empty \def\@tempb {arXiv}\fi \@ifundefined
  {mn@eprint@\@tempb}{\@tempb:\@tempc}{\expandafter \expandafter \csname
  mn@eprint@\@tempb\endcsname \expandafter{\@tempc}}}

\bibitem[\protect\citeauthoryear{{Agertz}, {Kravtsov}, {Leitner}  \&
  {Gnedin}}{{Agertz} et~al.}{2013}]{Agertz2013}
{Agertz} O.,  {Kravtsov} A.~V.,  {Leitner} S.~N.,   {Gnedin} N.~Y.,  2013,
  \mn@doi [\apj] {10.1088/0004-637X/770/1/25}, \href
  {https://ui.adsabs.harvard.edu/abs/2013ApJ...770...25A} {770, 25}

\bibitem[\protect\citeauthoryear{{Aguilar} \& {White}}{{Aguilar} \&
  {White}}{1985}]{AguWhite1985}
{Aguilar} L.~A.,  {White} S.~D.~M.,  1985, \mn@doi [\apj] {10.1086/163382},
  \href {https://ui.adsabs.harvard.edu/abs/1985ApJ...295..374A} {295, 374}

\bibitem[\protect\citeauthoryear{{Bar-Or}, {Kupi}  \& {Alexander}}{{Bar-Or}
  et~al.}{2013}]{BarOr2013}
{Bar-Or} B.,  {Kupi} G.,   {Alexander} T.,  2013, \mn@doi [\apj]
  {10.1088/0004-637X/764/1/52}, \href
  {https://ui.adsabs.harvard.edu/abs/2013ApJ...764...52B} {764, 52}

\bibitem[\protect\citeauthoryear{{Bar}, {Blum}  \& {Sun}}{{Bar}
  et~al.}{2021}]{BarBlumRotII21}
{Bar} N.,  {Blum} K.,   {Sun} C.,  2021, arXiv e-prints, \href
  {https://ui.adsabs.harvard.edu/abs/2021arXiv211103070B} {p. arXiv:2111.03070}

\bibitem[\protect\citeauthoryear{{Binney} \& {Tremaine}}{{Binney} \&
  {Tremaine}}{2008}]{BT}
{Binney} J.,  {Tremaine} S.,  2008, {Galactic dynamics}

\bibitem[\protect\citeauthoryear{{Bouch{\'e}} et~al.,}{{Bouch{\'e}}
  et~al.}{2022}]{Bouche2022}
{Bouch{\'e}} N.~F.,  et~al., 2022, \mn@doi [\aap]
  {10.1051/0004-6361/202141762}, \href
  {https://ui.adsabs.harvard.edu/abs/2022A&A...658A..76B} {658, A76}

\bibitem[\protect\citeauthoryear{{Boylan-Kolchin}, {Bullock}  \&
  {Kaplinghat}}{{Boylan-Kolchin} et~al.}{2011}]{BoylanTBF2011}
{Boylan-Kolchin} M.,  {Bullock} J.~S.,   {Kaplinghat} M.,  2011, \mn@doi
  [\mnras] {10.1111/j.1745-3933.2011.01074.x}, \href
  {https://ui.adsabs.harvard.edu/abs/2011MNRAS.415L..40B} {415, L40}

\bibitem[\protect\citeauthoryear{{Bullock} \& {Boylan-Kolchin}}{{Bullock} \&
  {Boylan-Kolchin}}{2017}]{Bullock_B2017}
{Bullock} J.~S.,  {Boylan-Kolchin} M.,  2017, \mn@doi [Annual Review of
  Astronomy and Astrophysics] {10.1146/annurev-astro-091916-055313}, \href
  {https://ui.adsabs.harvard.edu/\#abs/2017ARA&A..55..343B} {55, 343}

\bibitem[\protect\citeauthoryear{{Burkert}}{{Burkert}}{2000}]{Burkert2000}
{Burkert} A.,  2000, \mn@doi [ApJl] {10.1086/312674}, \href
  {http://adsabs.harvard.edu/abs/2000ApJ...534L.143B} {534, L143}

\bibitem[\protect\citeauthoryear{{Burkert}}{{Burkert}}{2020}]{Burkert20}
{Burkert} A.,  2020, arXiv e-prints, \href
  {https://ui.adsabs.harvard.edu/abs/2020arXiv200611111B} {p. arXiv:2006.11111}

\bibitem[\protect\citeauthoryear{{Coles} \& {Jones}}{{Coles} \&
  {Jones}}{1991}]{Coles_lognor91}
{Coles} P.,  {Jones} B.,  1991, \mn@doi [MNRAS] {10.1093/mnras/248.1.1}, \href
  {https://ui.adsabs.harvard.edu/abs/1991MNRAS.248....1C} {248, 1}

\bibitem[\protect\citeauthoryear{{Collins} \& {Read}}{{Collins} \&
  {Read}}{2022}]{Collins2022}
{Collins} M. L.~M.,  {Read} J.~I.,  2022, \mn@doi [Nature Astronomy]
  {10.1038/s41550-022-01657-4}, \href
  {https://ui.adsabs.harvard.edu/abs/2022NatAs...6..647C} {6, 647}

\bibitem[\protect\citeauthoryear{{Dekel} et~al.,}{{Dekel}
  et~al.}{2021}]{Dekel2021}
{Dekel} A.,  et~al., 2021, \mn@doi [\mnras] {10.1093/mnras/stab2416}, \href
  {https://ui.adsabs.harvard.edu/abs/2021MNRAS.508..999D} {508, 999}

\bibitem[\protect\citeauthoryear{{Del Popolo} \& {Le Delliou}}{{Del Popolo} \&
  {Le Delliou}}{2017}]{delPopolo2017}
{Del Popolo} A.,  {Le Delliou} M.,  2017, \mn@doi [Galaxies]
  {10.3390/galaxies5010017}, \href
  {https://ui.adsabs.harvard.edu/\#abs/2017Galax...5...17D} {5, 17}

\bibitem[\protect\citeauthoryear{{Deng}, {Hertzberg}, {Namjoo}  \&
  {Masoumi}}{{Deng} et~al.}{2018}]{DHHertz18}
{Deng} H.,  {Hertzberg} M.~P.,  {Namjoo} M.~H.,   {Masoumi} A.,  2018, \mn@doi
  [\prd] {10.1103/PhysRevD.98.023513}, \href
  {https://ui.adsabs.harvard.edu/abs/2018PhRvD..98b3513D} {98, 023513}

\bibitem[\protect\citeauthoryear{{Di Cintio}, {Brook}, {Macci{\`o}}, {Stinson},
  {Knebe}, {Dutton}  \& {Wadsley}}{{Di Cintio} et~al.}{2014}]{DiCintio2014}
{Di Cintio} A.,  {Brook} C.~B.,  {Macci{\`o}} A.~V.,  {Stinson} G.~S.,  {Knebe}
  A.,  {Dutton} A.~A.,   {Wadsley} J.,  2014, \mn@doi [\mnras]
  {10.1093/mnras/stt1891}, \href
  {https://ui.adsabs.harvard.edu/abs/2014MNRAS.437..415D} {437, 415}

\bibitem[\protect\citeauthoryear{{Dubinski} \& {Carlberg}}{{Dubinski} \&
  {Carlberg}}{1991}]{Dubinski1991}
{Dubinski} J.,  {Carlberg} R.~G.,  1991, \mn@doi [ApJ] {10.1086/170451}, \href
  {http://adsabs.harvard.edu/abs/1991ApJ...378..496D} {378, 496}

\bibitem[\protect\citeauthoryear{{El-Zant}}{{El-Zant}}{2008}]{El-Zant08}
{El-Zant} A.~A.,  2008, \mn@doi [ApJ] {10.1086/587022}, \href
  {https://ui.adsabs.harvard.edu/abs/2008ApJ...681.1058E} {681, 1058}

\bibitem[\protect\citeauthoryear{{El-Zant}, {Shlosman}  \& {Hoffman}}{{El-Zant}
  et~al.}{2001}]{Zant2001}
{El-Zant} A.,  {Shlosman} I.,   {Hoffman} Y.,  2001, \mn@doi [ApJ]
  {10.1086/322516}, \href {http://adsabs.harvard.edu/abs/2001ApJ...560..636E}
  {560, 636}

\bibitem[\protect\citeauthoryear{{El-Zant}, {Hoffman}, {Primack}, {Combes}  \&
  {Shlosman}}{{El-Zant} et~al.}{2004}]{Zant2004}
{El-Zant} A.~A.,  {Hoffman} Y.,  {Primack} J.,  {Combes} F.,   {Shlosman} I.,
  2004, \mn@doi [\apjl] {10.1086/421938}, \href
  {https://ui.adsabs.harvard.edu/abs/2004ApJ...607L..75E} {607, L75}

\bibitem[\protect\citeauthoryear{{El-Zant}, {Freundlich}  \&
  {Combes}}{{El-Zant} et~al.}{2016}]{EZFC}
{El-Zant} A.~A.,  {Freundlich} J.,   {Combes} F.,  2016, \mn@doi [MNRAS]
  {10.1093/mnras/stw1398}, \href
  {https://ui.adsabs.harvard.edu/\#abs/2016MNRAS.461.1745E} {461, 1745}

\bibitem[\protect\citeauthoryear{{El-Zant}, {Freundlich}, {Combes}  \&
  {Halle}}{{El-Zant} et~al.}{2020}]{EZFCH}
{El-Zant} A.~A.,  {Freundlich} J.,  {Combes} F.,   {Halle} A.,  2020, \mn@doi
  [MNRAS] {10.1093/mnras/stz3478}, \href
  {https://ui.adsabs.harvard.edu/abs/2020MNRAS.492..877E} {492, 877}

\bibitem[\protect\citeauthoryear{{Errani}, {Navarro}, {Pe{\~n}arrubia},
  {Famaey}  \& {Ibata}}{{Errani} et~al.}{2023}]{Errani_Penn2023}
{Errani} R.,  {Navarro} J.~F.,  {Pe{\~n}arrubia} J.,  {Famaey} B.,   {Ibata}
  R.,  2023, \mn@doi [\mnras] {10.1093/mnras/stac3499}, \href
  {https://ui.adsabs.harvard.edu/abs/2023MNRAS.519..384E} {519, 384}

\bibitem[\protect\citeauthoryear{{Evans}}{{Evans}}{1994}]{Evans1994}
{Evans} N.~W.,  1994, \mn@doi [\mnras] {10.1093/mnras/267.2.333}, \href
  {https://ui.adsabs.harvard.edu/abs/1994MNRAS.267..333E} {267, 333}

\bibitem[\protect\citeauthoryear{{Falkovich}}{{Falkovich}}{1994}]{FalkovichBottle1994}
{Falkovich} G.,  1994, \mn@doi [Physics of Fluids] {10.1063/1.868255}, \href
  {https://ui.adsabs.harvard.edu/abs/1994PhFl....6.1411F} {6, 1411}

\bibitem[\protect\citeauthoryear{{Famaey}, {Khoury}, {Penco}  \&
  {Sharma}}{{Famaey} et~al.}{2020}]{FamaueyBarint2020}
{Famaey} B.,  {Khoury} J.,  {Penco} R.,   {Sharma} A.,  2020, \mn@doi [\jcap]
  {10.1088/1475-7516/2020/06/025}, \href
  {https://ui.adsabs.harvard.edu/abs/2020JCAP...06..025F} {2020, 025}

\bibitem[\protect\citeauthoryear{{Flores} \& {Primack}}{{Flores} \&
  {Primack}}{1994}]{Flores1994}
{Flores} R.~A.,  {Primack} J.~R.,  1994, \mn@doi [ApJl] {10.1086/187350}, \href
  {http://adsabs.harvard.edu/abs/1994ApJ...427L...1F} {427, L1}

\bibitem[\protect\citeauthoryear{{Frenk} \& {White}}{{Frenk} \&
  {White}}{2012}]{Frenk2012}
{Frenk} C.~S.,  {White} S.~D.~M.,  2012, \mn@doi [Annalen der Physik]
  {10.1002/andp.201200212}, \href
  {http://adsabs.harvard.edu/abs/2012AnP...524..507F} {524, 507}

\bibitem[\protect\citeauthoryear{{Freundlich}, {Dekel}, {Jiang}, {Ishai},
  {Cornuault}, {Lapiner}, {Dutton}  \& {Macci{\`o}}}{{Freundlich}
  et~al.}{2020a}]{Freundlich2020}
{Freundlich} J.,  {Dekel} A.,  {Jiang} F.,  {Ishai} G.,  {Cornuault} N.,
  {Lapiner} S.,  {Dutton} A.~A.,   {Macci{\`o}} A.~V.,  2020a, \mn@doi [\mnras]
  {10.1093/mnras/stz3306}, \href
  {https://ui.adsabs.harvard.edu/abs/2020MNRAS.491.4523F} {491, 4523}

\bibitem[\protect\citeauthoryear{{Freundlich} et~al.,}{{Freundlich}
  et~al.}{2020b}]{Freundlich2020b}
{Freundlich} J.,  et~al., 2020b, \mn@doi [\mnras] {10.1093/mnras/staa2790},
  \href {https://ui.adsabs.harvard.edu/abs/2020MNRAS.499.2912F} {499, 2912}

\bibitem[\protect\citeauthoryear{{Gilman}, {Birrer}, {Nierenberg}, {Treu}, {Du}
   \& {Benson}}{{Gilman} et~al.}{2020}]{GilmanLensMisssat2020}
{Gilman} D.,  {Birrer} S.,  {Nierenberg} A.,  {Treu} T.,  {Du} X.,   {Benson}
  A.,  2020, \mn@doi [\mnras] {10.1093/mnras/stz3480}, \href
  {https://ui.adsabs.harvard.edu/abs/2020MNRAS.491.6077G} {491, 6077}

\bibitem[\protect\citeauthoryear{{Governato} et~al.,}{{Governato}
  et~al.}{2015}]{Governato2015}
{Governato} F.,  et~al., 2015, \mn@doi [\mnras] {10.1093/mnras/stu2720}, \href
  {https://ui.adsabs.harvard.edu/abs/2015MNRAS.448..792G} {448, 792}

\bibitem[\protect\citeauthoryear{{Grisdale}, {Agertz}, {Romeo}, {Renaud}  \&
  {Read}}{{Grisdale} et~al.}{2017}]{GrisRomReadTurb2017}
{Grisdale} K.,  {Agertz} O.,  {Romeo} A.~B.,  {Renaud} F.,   {Read} J.~I.,
  2017, \mn@doi [\mnras] {10.1093/mnras/stw3133}, \href
  {https://ui.adsabs.harvard.edu/abs/2017MNRAS.466.1093G} {466, 1093}

\bibitem[\protect\citeauthoryear{{Hernquist} \& {Ostriker}}{{Hernquist} \&
  {Ostriker}}{1992}]{Hernquist1992}
{Hernquist} L.,  {Ostriker} J.~P.,  1992, \mn@doi [ApJ] {10.1086/171025}, \href
  {http://adsabs.harvard.edu/abs/1992ApJ...386..375H} {386, 375}

\bibitem[\protect\citeauthoryear{{Inoue}, {Dekel}, {Mandelker}, {Ceverino},
  {Bournaud}  \& {Primack}}{{Inoue} et~al.}{2016}]{DekelQ2016}
{Inoue} S.,  {Dekel} A.,  {Mandelker} N.,  {Ceverino} D.,  {Bournaud} F.,
  {Primack} J.,  2016, \mn@doi [\mnras] {10.1093/mnras/stv2793}, \href
  {https://ui.adsabs.harvard.edu/abs/2016MNRAS.456.2052I} {456, 2052}

\bibitem[\protect\citeauthoryear{{Ir{\v{s}}i{\v{c}}}
  et~al.,}{{Ir{\v{s}}i{\v{c}}} et~al.}{2017a}]{ViDvielWDM2017}
{Ir{\v{s}}i{\v{c}}} V.,  et~al., 2017a, \mn@doi [\prd]
  {10.1103/PhysRevD.96.023522}, \href
  {https://ui.adsabs.harvard.edu/abs/2017PhRvD..96b3522I} {96, 023522}

\bibitem[\protect\citeauthoryear{{Ir{\v{s}}i{\v{c}}}, {Viel}, {Haehnelt},
  {Bolton}  \& {Becker}}{{Ir{\v{s}}i{\v{c}}} et~al.}{2017b}]{VidVielFDM}
{Ir{\v{s}}i{\v{c}}} V.,  {Viel} M.,  {Haehnelt} M.~G.,  {Bolton} J.~S.,
  {Becker} G.~D.,  2017b, \mn@doi [\prl] {10.1103/PhysRevLett.119.031302},
  \href {https://ui.adsabs.harvard.edu/abs/2017PhRvL.119c1302I} {119, 031302}

\bibitem[\protect\citeauthoryear{{Kim}, {Peter}  \& {Hargis}}{{Kim}
  et~al.}{2018}]{KimMisssat2018}
{Kim} S.~Y.,  {Peter} A. H.~G.,   {Hargis} J.~R.,  2018, \mn@doi [\prl]
  {10.1103/PhysRevLett.121.211302}, \href
  {https://ui.adsabs.harvard.edu/abs/2018PhRvL.121u1302K} {121, 211302}

\bibitem[\protect\citeauthoryear{{Kochanek} \& {White}}{{Kochanek} \&
  {White}}{2000}]{Kochanek2000}
{Kochanek} C.~S.,  {White} M.,  2000, \mn@doi [ApJ] {10.1086/317149}, \href
  {http://adsabs.harvard.edu/abs/2000ApJ...543..514K} {543, 514}

\bibitem[\protect\citeauthoryear{{Koposov} et~al.,}{{Koposov}
  et~al.}{2008}]{KopGilMisssat2008}
{Koposov} S.,  et~al., 2008, \mn@doi [\apj] {10.1086/589911}, \href
  {https://ui.adsabs.harvard.edu/abs/2008ApJ...686..279K} {686, 279}

\bibitem[\protect\citeauthoryear{{Kraichnan}}{{Kraichnan}}{1964}]{KraichSweep1964}
{Kraichnan} R.~H.,  1964, \mn@doi [Physics of Fluids] {10.1063/1.2746572},
  \href {https://ui.adsabs.harvard.edu/abs/1964PhFl....7.1723K} {7, 1723}

\bibitem[\protect\citeauthoryear{{Kritsuk}, {Norman}, {Padoan}  \&
  {Wagner}}{{Kritsuk} et~al.}{2007}]{KristukBottle2007}
{Kritsuk} A.~G.,  {Norman} M.~L.,  {Padoan} P.,   {Wagner} R.,  2007, \mn@doi
  [\apj] {10.1086/519443}, \href
  {https://ui.adsabs.harvard.edu/abs/2007ApJ...665..416K} {665, 416}

\bibitem[\protect\citeauthoryear{{Li}, {Dekel}, {Mandelker}, {Freundlich}  \&
  {Fran{\c{c}}ois}}{{Li} et~al.}{2022}]{Li2022}
{Li} Z.,  {Dekel} A.,  {Mandelker} N.,  {Freundlich} J.,   {Fran{\c{c}}ois} T.,
   2022, arXiv e-prints, \href
  {https://ui.adsabs.harvard.edu/abs/2022arXiv220607069L} {p. arXiv:2206.07069}

\bibitem[\protect\citeauthoryear{{Macci{\`o}}, {Paduroiu}, {Anderhalden},
  {Schneider}  \& {Moore}}{{Macci{\`o}} et~al.}{2012}]{Maccio2012b}
{Macci{\`o}} A.~V.,  {Paduroiu} S.,  {Anderhalden} D.,  {Schneider} A.,
  {Moore} B.,  2012, \mn@doi [MNRAS] {10.1111/j.1365-2966.2012.21284.x}, \href
  {http://adsabs.harvard.edu/abs/2012MNRAS.424.1105M} {424, 1105}

\bibitem[\protect\citeauthoryear{{Moore}}{{Moore}}{1994}]{Moore1994}
{Moore} B.,  1994, \mn@doi [New Astron.t] {10.1038/370629a0}, \href
  {http://adsabs.harvard.edu/abs/1994Natur.370..629M} {370, 629}

\bibitem[\protect\citeauthoryear{{Nadler} et~al.,}{{Nadler}
  et~al.}{2020}]{nadler2020}
{Nadler} E.~O.,  et~al., 2020, arXiv e-prints, \href
  {https://ui.adsabs.harvard.edu/abs/2020arXiv200800022N} {p. arXiv:2008.00022}

\bibitem[\protect\citeauthoryear{{Nadler}, {Birrer}, {Gilman}, {Wechsler},
  {Du}, {Benson}, {Nierenberg}  \& {Treu}}{{Nadler} et~al.}{2021}]{Nadler2021}
{Nadler} E.~O.,  {Birrer} S.,  {Gilman} D.,  {Wechsler} R.~H.,  {Du} X.,
  {Benson} A.,  {Nierenberg} A.~M.,   {Treu} T.,  2021, \mn@doi [ApJ]
  {10.3847/1538-4357/abf9a3}, \href
  {https://ui.adsabs.harvard.edu/abs/2021ApJ...917....7N} {917, 7}

\bibitem[\protect\citeauthoryear{{Navarro}, {Eke}  \& {Frenk}}{{Navarro}
  et~al.}{1996a}]{Navarro1996a}
{Navarro} J.~F.,  {Eke} V.~R.,   {Frenk} C.~S.,  1996a, MNRAS, \href
  {http://adsabs.harvard.edu/abs/1996MNRAS.283L..72N} {283, L72}

\bibitem[\protect\citeauthoryear{{Navarro}, {Frenk}  \& {White}}{{Navarro}
  et~al.}{1996b}]{nfw}
{Navarro} J.~F.,  {Frenk} C.~S.,   {White} S.~D.~M.,  1996b, \mn@doi [ApJ]
  {10.1086/177173}, \href {http://adsabs.harvard.edu/abs/1996ApJ...462..563N}
  {462, 563}

\bibitem[\protect\citeauthoryear{{Newton} et~al.,}{{Newton}
  et~al.}{2021}]{NewtonMisssat2021}
{Newton} O.,  et~al., 2021, \mn@doi [\jcap] {10.1088/1475-7516/2021/08/062},
  \href {https://ui.adsabs.harvard.edu/abs/2021JCAP...08..062N} {2021, 062}

\bibitem[\protect\citeauthoryear{{Nusser} \& {Silk}}{{Nusser} \&
  {Silk}}{2022}]{NusSilkTurb2022}
{Nusser} A.,  {Silk} J.,  2022, \mn@doi [\mnras] {10.1093/mnras/stab3121},
  \href {https://ui.adsabs.harvard.edu/abs/2022MNRAS.509.2979N} {509, 2979}

\bibitem[\protect\citeauthoryear{{Ogiya} \& {Burkert}}{{Ogiya} \&
  {Burkert}}{2015}]{OgiyaBurkTBF2015}
{Ogiya} G.,  {Burkert} A.,  2015, \mn@doi [\mnras] {10.1093/mnras/stu2283},
  \href {https://ui.adsabs.harvard.edu/abs/2015MNRAS.446.2363O} {446, 2363}

\bibitem[\protect\citeauthoryear{{Ogiya} \& {Nagai}}{{Ogiya} \&
  {Nagai}}{2022}]{OgiyaDF2022}
{Ogiya} G.,  {Nagai} D.,  2022, \mn@doi [\mnras] {10.1093/mnras/stac1311},
  \href {https://ui.adsabs.harvard.edu/abs/2022MNRAS.514..555O} {514, 555}

\bibitem[\protect\citeauthoryear{{Oman}, {Marasco}, {Navarro}, {Frenk},
  {Schaye}  \& {Ben{\'\i}tez-Llambay}}{{Oman} et~al.}{2019}]{Oman2019}
{Oman} K.~A.,  {Marasco} A.,  {Navarro} J.~F.,  {Frenk} C.~S.,  {Schaye} J.,
  {Ben{\'\i}tez-Llambay} A.,  2019, \mn@doi [\mnras] {10.1093/mnras/sty2687},
  \href {https://ui.adsabs.harvard.edu/abs/2019MNRAS.482..821O} {482, 821}

\bibitem[\protect\citeauthoryear{{Orkney} et~al.,}{{Orkney}
  et~al.}{2021}]{Orkney2021}
{Orkney} M. D.~A.,  et~al., 2021, \mn@doi [\mnras] {10.1093/mnras/stab1066},
  \href {https://ui.adsabs.harvard.edu/abs/2021MNRAS.504.3509O} {504, 3509}

\bibitem[\protect\citeauthoryear{{Pe{\~n}arrubia}}{{Pe{\~n}arrubia}}{2019a}]{PenarrubiaA}
{Pe{\~n}arrubia} J.,  2019a, \mn@doi [\mnras] {10.1093/mnras/stz338}, \href
  {https://ui.adsabs.harvard.edu/abs/2019MNRAS.484.5409P} {484, 5409}

\bibitem[\protect\citeauthoryear{{Pe{\~n}arrubia}}{{Pe{\~n}arrubia}}{2019b}]{PenarrubiaB2019}
{Pe{\~n}arrubia} J.,  2019b, \mn@doi [\mnras] {10.1093/mnras/stz2648}, \href
  {https://ui.adsabs.harvard.edu/abs/2019MNRAS.490.1044P} {490, 1044}

\bibitem[\protect\citeauthoryear{{Pe{\~n}arrubia}, {Pontzen}, {Walker}  \&
  {Koposov}}{{Pe{\~n}arrubia} et~al.}{2012}]{Penarrubia2012}
{Pe{\~n}arrubia} J.,  {Pontzen} A.,  {Walker} M.~G.,   {Koposov} S.~E.,  2012,
  \mn@doi [\apjl] {10.1088/2041-8205/759/2/L42}, \href
  {https://ui.adsabs.harvard.edu/abs/2012ApJ...759L..42P} {759, L42}

\bibitem[\protect\citeauthoryear{{Pontzen} \& {Governato}}{{Pontzen} \&
  {Governato}}{2014}]{Pontzen2014}
{Pontzen} A.,  {Governato} F.,  2014, \mn@doi [Nature] {10.1038/nature12953},
  \href {http://adsabs.harvard.edu/abs/2014Natur.506..171P} {506, 171}

\bibitem[\protect\citeauthoryear{{Read} \& {Erkal}}{{Read} \&
  {Erkal}}{2019}]{ReadMisssat2019}
{Read} J.~I.,  {Erkal} D.,  2019, \mn@doi [\mnras] {10.1093/mnras/stz1320},
  \href {https://ui.adsabs.harvard.edu/abs/2019MNRAS.487.5799R} {487, 5799}

\bibitem[\protect\citeauthoryear{{Read} \& {Gilmore}}{{Read} \&
  {Gilmore}}{2005}]{Read2005}
{Read} J.~I.,  {Gilmore} G.,  2005, \mn@doi [\mnras]
  {10.1111/j.1365-2966.2004.08424.x}, \href
  {https://ui.adsabs.harvard.edu/abs/2005MNRAS.356..107R} {356, 107}

\bibitem[\protect\citeauthoryear{{Read} \& {Trentham}}{{Read} \&
  {Trentham}}{2005}]{ReadBarFrac2005}
{Read} J.~I.,  {Trentham} N.,  2005, \mn@doi [Philosophical Transactions of the
  Royal Society of London Series A] {10.1098/rsta.2005.1648}, \href
  {https://ui.adsabs.harvard.edu/abs/2005RSPTA.363.2693R} {363, 2693}

\bibitem[\protect\citeauthoryear{{Read}, {Wilkinson}, {Evans}, {Gilmore}  \&
  {Kleyna}}{{Read} et~al.}{2006}]{ReadTBF2006}
{Read} J.~I.,  {Wilkinson} M.~I.,  {Evans} N.~W.,  {Gilmore} G.,   {Kleyna}
  J.~T.,  2006, \mn@doi [\mnras] {10.1111/j.1365-2966.2005.09959.x}, \href
  {https://ui.adsabs.harvard.edu/abs/2006MNRAS.367..387R} {367, 387}

\bibitem[\protect\citeauthoryear{{Read}, {Agertz}  \& {Collins}}{{Read}
  et~al.}{2016}]{Readsim2016}
{Read} J.~I.,  {Agertz} O.,   {Collins} M.~L.~M.,  2016, \mn@doi [MNRAS]
  {10.1093/mnras/stw713}, \href
  {https://ui.adsabs.harvard.edu/abs/2016MNRAS.459.2573R} {459, 2573}

\bibitem[\protect\citeauthoryear{{Read}, {Walker}  \& {Steger}}{{Read}
  et~al.}{2019}]{Read2019}
{Read} J.~I.,  {Walker} M.~G.,   {Steger} P.,  2019, \mn@doi [\mnras]
  {10.1093/mnras/sty3404}, \href
  {https://ui.adsabs.harvard.edu/abs/2019MNRAS.484.1401R} {484, 1401}

\bibitem[\protect\citeauthoryear{Salucci}{Salucci}{2019}]{Salucci:2018hqu}
Salucci P.,  2019, \mn@doi [Astron. Astrophys. Rev.]
  {10.1007/s00159-018-0113-1}, 27, 2

\bibitem[\protect\citeauthoryear{{Salucci}, {Turini}  \& {di Paolo}}{{Salucci}
  et~al.}{2020}]{SallucciBarint2020}
{Salucci} P.,  {Turini} N.,   {di Paolo} C.,  2020, \mn@doi [Universe]
  {10.3390/universe6080118}, \href
  {https://ui.adsabs.harvard.edu/abs/2020Univ....6..118S} {6, 118}

\bibitem[\protect\citeauthoryear{{Schmidt}, {Hillebrandt}  \&
  {Niemeyer}}{{Schmidt} et~al.}{2004}]{Scmidtbottle2004}
{Schmidt} W.,  {Hillebrandt} W.,   {Niemeyer} J.~C.,  2004, arXiv e-prints,
  \href {https://ui.adsabs.harvard.edu/abs/2004astro.ph..7616S} {pp
  astro--ph/0407616}

\bibitem[\protect\citeauthoryear{{Simon} \& {Geha}}{{Simon} \&
  {Geha}}{2007}]{SimonMisssat2007}
{Simon} J.~D.,  {Geha} M.,  2007, \mn@doi [\apj] {10.1086/521816}, \href
  {https://ui.adsabs.harvard.edu/abs/2007ApJ...670..313S} {670, 313}

\bibitem[\protect\citeauthoryear{{Spitzer}}{{Spitzer}}{1987}]{Spitzer1987}
{Spitzer} L.,  1987, {Dynamical evolution of globular clusters}

\bibitem[\protect\citeauthoryear{{Taylor}}{{Taylor}}{1938}]{TaylSwepp1938}
{Taylor} G.~I.,  1938, \mn@doi [Proceedings of the Royal Society of London
  Series A] {10.1098/rspa.1938.0032}, \href
  {https://ui.adsabs.harvard.edu/abs/1938RSPSA.164..476T} {164, 476}

\bibitem[\protect\citeauthoryear{{Taylor} \& {Navarro}}{{Taylor} \&
  {Navarro}}{2001}]{TayNav01}
{Taylor} J.~E.,  {Navarro} J.~F.,  2001, \mn@doi [ApJ] {10.1086/324031}, \href
  {https://ui.adsabs.harvard.edu/abs/2001ApJ...563..483T} {563, 483}

\bibitem[\protect\citeauthoryear{{Tennekes}}{{Tennekes}}{1975}]{TennekSweep1975}
{Tennekes} H.,  1975, \mn@doi [Journal of Fluid Mechanics]
  {10.1017/S0022112075000468}, \href
  {https://ui.adsabs.harvard.edu/abs/1975JFM....67..561T} {67, 561}

\bibitem[\protect\citeauthoryear{{Teyssier}}{{Teyssier}}{2002}]{Teyssier2002}
{Teyssier} R.,  2002, \mn@doi [\aap] {10.1051/0004-6361:20011817}, \href
  {https://ui.adsabs.harvard.edu/abs/2002A&A...385..337T} {385, 337}

\bibitem[\protect\citeauthoryear{{Teyssier}, {Pontzen}, {Dubois}  \&
  {Read}}{{Teyssier} et~al.}{2013}]{Teyssier2013}
{Teyssier} R.,  {Pontzen} A.,  {Dubois} Y.,   {Read} J.~I.,  2013, \mn@doi
  [MNRAS] {10.1093/mnras/sts563}, \href
  {http://adsabs.harvard.edu/abs/2013MNRAS.429.3068T} {429, 3068}

\bibitem[\protect\citeauthoryear{{Turner}, {Lovell}, {Zavala}  \&
  {Vogelsberger}}{{Turner} et~al.}{2021}]{CoreColSIDM21}
{Turner} H.~C.,  {Lovell} M.~R.,  {Zavala} J.,   {Vogelsberger} M.,  2021,
  \mn@doi [MNRAS] {10.1093/mnras/stab1725}, \href
  {https://ui.adsabs.harvard.edu/abs/2021MNRAS.505.5327T} {505, 5327}

\bibitem[\protect\citeauthoryear{{Warren}, {Quinn}, {Salmon}  \&
  {Zurek}}{{Warren} et~al.}{1992}]{Warren1992}
{Warren} M.~S.,  {Quinn} P.~J.,  {Salmon} J.~K.,   {Zurek} W.~H.,  1992,
  \mn@doi [ApJ] {10.1086/171937}, \href
  {http://adsabs.harvard.edu/abs/1992ApJ...399..405W} {399, 405}

\bibitem[\protect\citeauthoryear{{Widrow}}{{Widrow}}{2000}]{WidrowDF2000}
{Widrow} L.~M.,  2000, \mn@doi [\apjs] {10.1086/317367}, \href
  {https://ui.adsabs.harvard.edu/abs/2000ApJS..131...39W} {131, 39}

\bibitem[\protect\citeauthoryear{{Yang}, {Nadler}  \& {Yu}}{{Yang}
  et~al.}{2022}]{YangNad_SIDMDivers2022}
{Yang} D.,  {Nadler} E.~O.,   {Yu} H.-b.,  2022, arXiv e-prints, \href
  {https://ui.adsabs.harvard.edu/abs/2022arXiv221113768Y} {p. arXiv:2211.13768}

\bibitem[\protect\citeauthoryear{{Yu}, {Bian}, {Krumholz}, {Shi}, {Li}  \&
  {Chen}}{{Yu} et~al.}{2021}]{YUKrumTurb2021}
{Yu} X.,  {Bian} F.,  {Krumholz} M.~R.,  {Shi} Y.,  {Li} S.,   {Chen} J.,
  2021, \mn@doi [\mnras] {10.1093/mnras/stab1681}, \href
  {https://ui.adsabs.harvard.edu/abs/2021MNRAS.505.5075Y} {505, 5075}

\makeatother
\end{thebibliography}




\appendix


\section{Statistics of the gas fluctuations}
\label{Sec:Additional_Gauss}

For the description of the effect of the gas fluctuations 
on the halo particles
in terms of a diffusion 
process, leading to relaxation timescales of the form of Eq~(\ref{eq:relax}), to be complete, their statistics 
should be entirely described by averages and dispersions, as in a Gaussian random process. 
We wish to examine whether this is the case with the gas fluctuations considered in this work. 

\begin{figure}
    \centering
	\includegraphics[width=\linewidth]{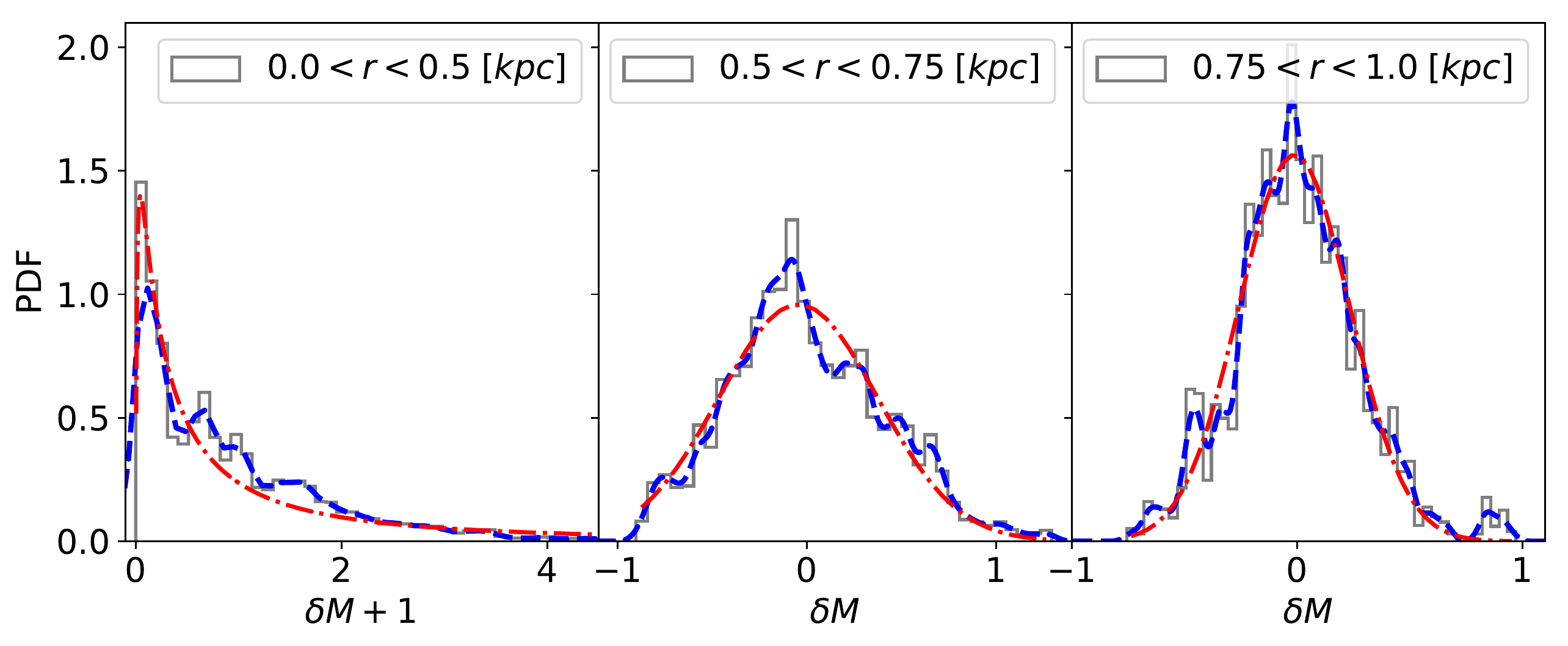}
    \caption{Probability distribution function (PDF) of the mass fluctuations within different radii $r$. The fluctuations are well-fit by Gaussians  for $r\ge 0.5 0.5 {\rm kpc}$ and by a lognormal distribution below. The fits are shown by the red dashed-dotted lines}
    \label{fig:massfluctpdf}
\end{figure}

\begin{figure}
    \centering
	\includegraphics[width=\linewidth]{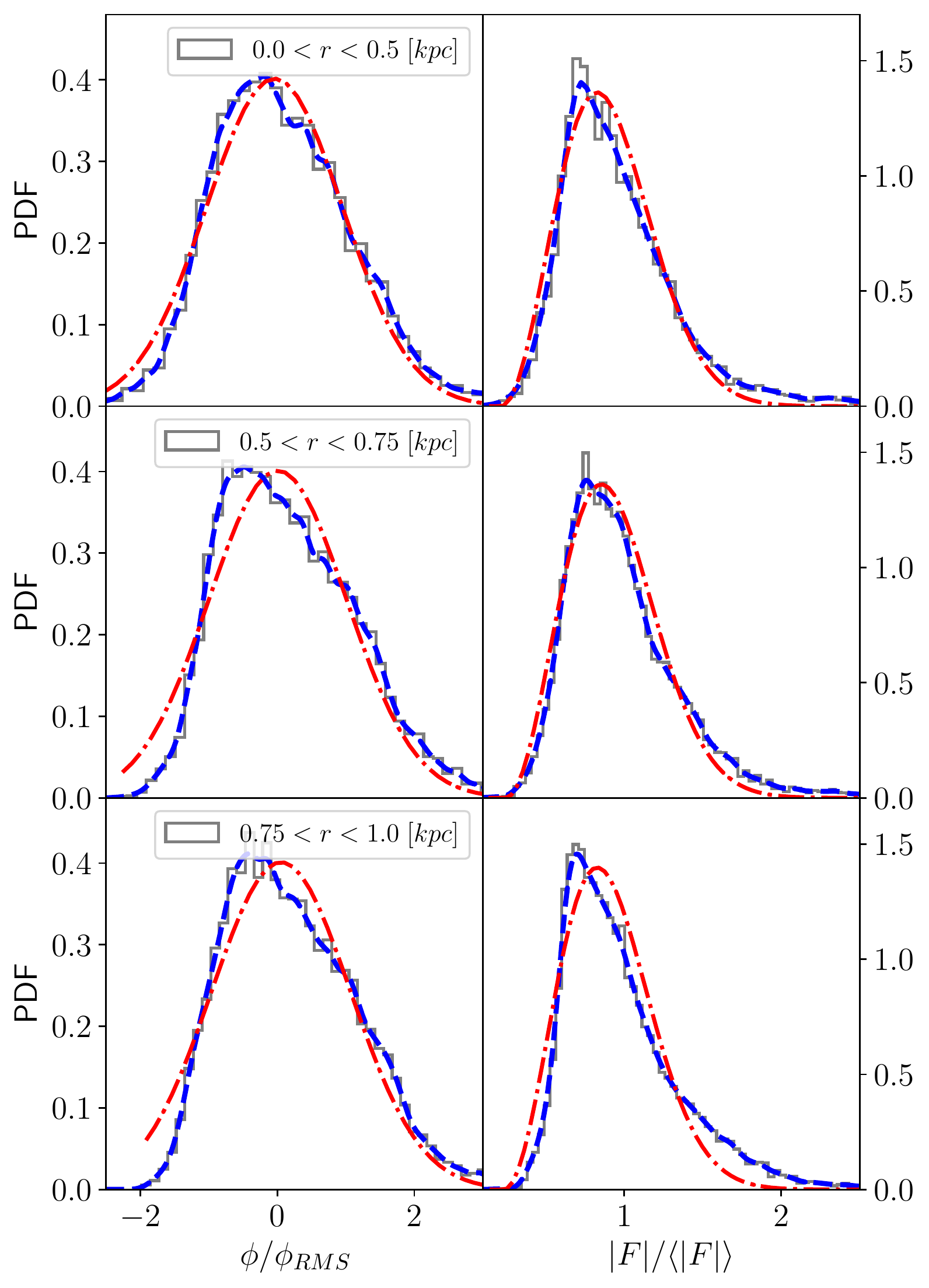}
    \caption{PDF of the fluctuations in the gas gravitational potential and the magnitude of the fluctuating force inside $r = 1.0\; {\rm kpc}$. The fits (dashed-dotted red lines) are Gaussian for the potential fluctuations and Maxwellian (i.e., assumed to emanate from a Gaussian distribution in each degree of freedom, as suggested by Fig.~\ref{fig:forcecomppdf}) for the absolute value of the force.}
    \label{fig:gravpotenmap}
\end{figure}

\begin{figure}
    \centering
	\includegraphics[width=\linewidth]{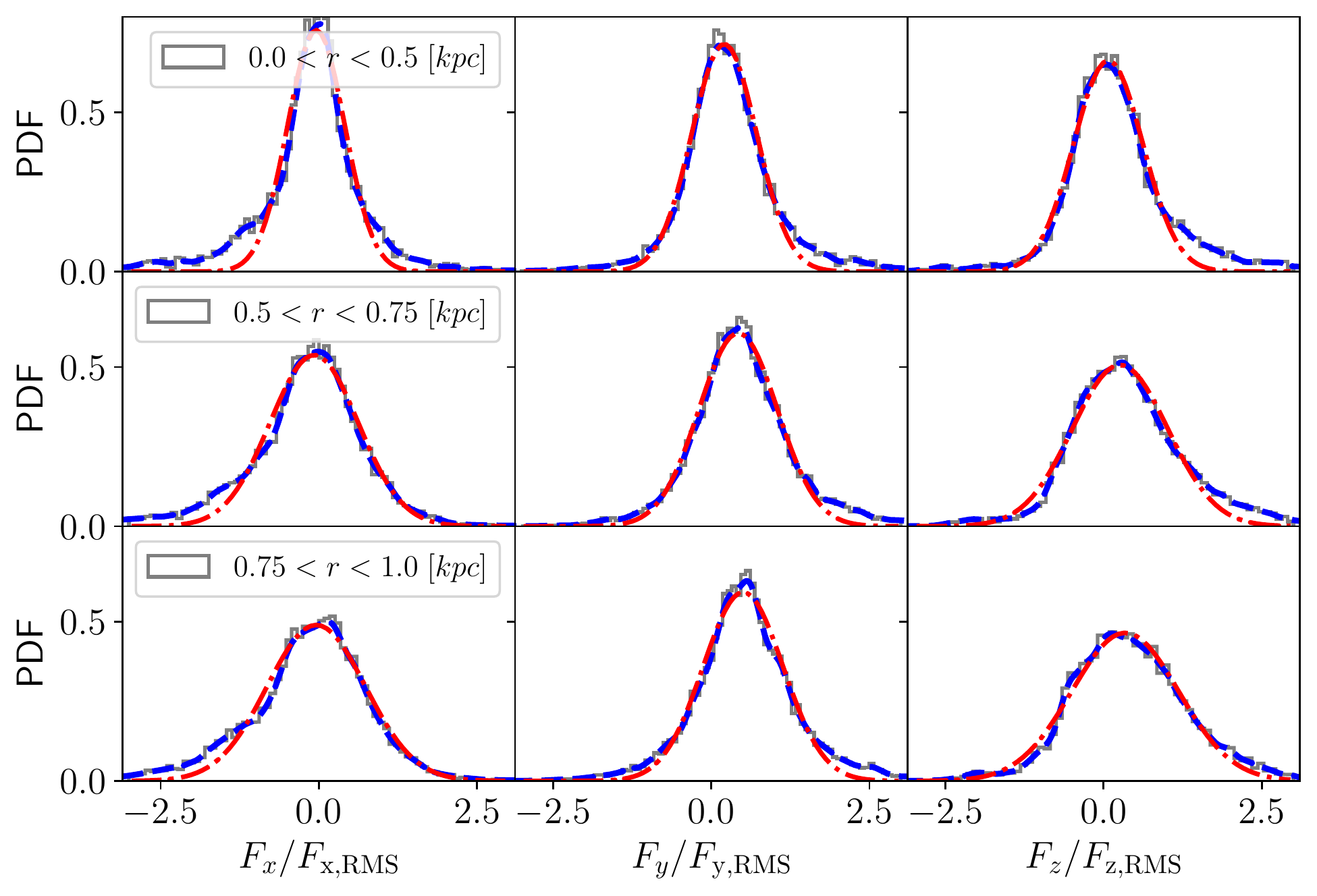}
    \caption{PDF of  X,Y,Z components of force fluctuations with Gaussian fits (red dashed-dotted lines).}
    \label{fig:forcecomppdf}
\end{figure}

\begin{figure}
	\includegraphics[width=\linewidth]{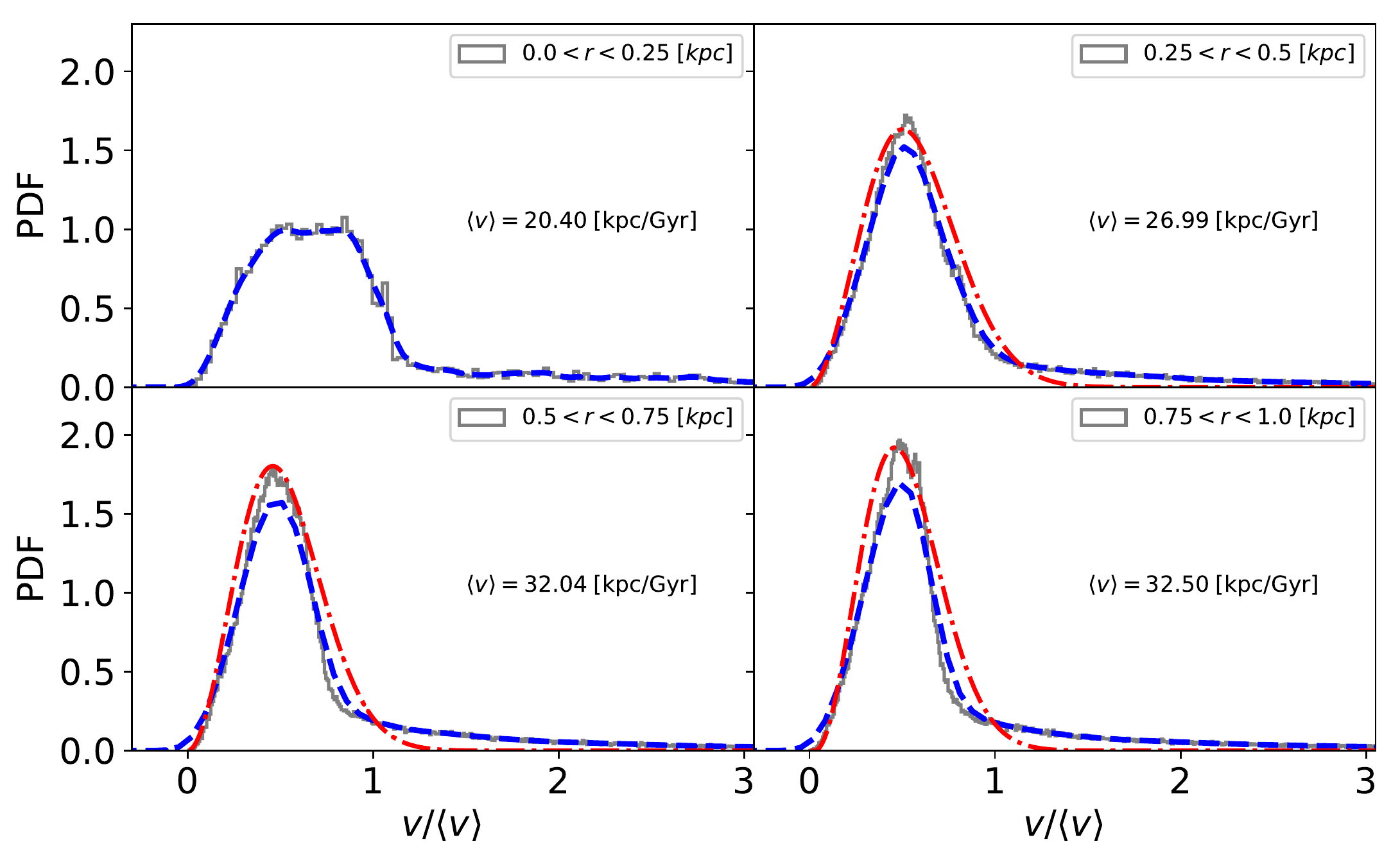}
    \caption{PDF of gas speeds, with Maxwellian fits at larger radial bins (red dashed-dotted lines).}
    \label{fig:velav}
\end{figure}

Fig.~\ref{fig:massfluctpdf} shows the statistics of the mass fluctuations in three different bins in the region where the core forms. For $r \gtrsim 0.5 r_s$ the fluctuations are well fit by Gaussians. 
Inside $0.5 r_s$ however they are better described by a lognormal distribution.  This is not surprising, given that the mass fluctuations at these scales are quite large (as may also be probed from Fig.~\ref{fig:massdisper}), and the density is bounded (by zero) from below but not necessarily from above.   A similar situation, with analogous description, also appears in the 
description of large scale structure at large density contrasts \citep{Coles_lognor91}. 

The density and mass fluctuations do not however affect the dynamics directly. Rather 
what affects CDM  trajectories are potential and force fluctuations. As the
potentials and forces are 
related (through the Poisson equation) to the density through integrals, their fluctuations 
are generally milder in gravitational systems. We now examine these.   

In order to obtain the gravitational field exerted by the gas perturbations on  DM particles, we first sample random points inside a spherical region centred at the DM halo's centre of potential. Then we calculate gas mass fluctuations defined as $\delta M  = M - \langle M \rangle$, where $\langle M \rangle$ is the average gas mass profile from the gas centre of mass. The gravitational potential of the gas perturbations is then obtained using:
\begin{equation}
\phi = \sum \frac{\delta M}{d}, 
\end{equation}
where $d$ is the distance between a DM particle and gas points.  

The gravitational force is directly obtained using the relations:
\begin{eqnarray}
F_x &=& \sum \frac{\delta M}{d^2} \frac{x-x_c}{d}, \\
F_y &=& \sum \frac{\delta M}{d^2} \frac{y-y_c}{d}, \\
F_z &=& \sum \frac{\delta M}{d^2} \frac{z-z_c}{d},
\end{eqnarray}
where $x_c, y_c, z_c$ are the coordinates of the gravitational field centre. The force amplitude is given as $F = \sqrt{F_x^2 + F_y^2 + F_z^2}$.

The results are shown in figures~\ref{fig:gravpotenmap} and~\ref{fig:forcecomppdf}. Apart from modest departures --- some long tail effects, as well as some skewness in the distribution of potential fluctuations --- the fluctuating potential and force field is indeed Gaussian. This is particularly clear in the distribution of the force components shown 
in Fig.~\ref{fig:forcecomppdf}.  

{We note nevertheless that the modest but non-negligible non-Gaussianities 
imply that higher order moments of velocity perturbations 
($\langle (\Delta v)^m \rangle$, $m > 2$) may contribute to the effect of gas fluctuations;  
the departure of halo particle trajectories from their 
collisionless counterparts may involve corrections beyond
the increase in velocity variance in the standard diffusion 
limit as described by (\ref{eq:vdisp}).   
In standard two body relaxation, 
the long velocity tails arise from 
strong, close encounters, and are alleviated when the force perturbations are caused by 
extended structures (\citealp{PenarrubiaB2019}). In our case, for power spectra $\mathcal{P} (k) \sim k^{-n}$  with $n>0$), the relaxation process is expected to be dominated by large scale modes (analogous to distant encounters the standard case); as equation~(\ref{eq:relax}) in fact emanates from an effective theory in which a higher scale $k_x$ is eliminated when $n > 0$ and $k_x \gg k_m$ (cf. \citealp{EZFC} Section 2). 
The long tails here are thus likely to be due to  halo
particle interactions with rare large scale fluctuations (with large perturbing forces; cf. equation~33 of~\citealp{EZFC}). 
Their effect may be limited however by the flattening of the power spectrum 
at $k < k_m$ (Fig.~\ref{fig:powerspecavrg}), and slow temporal variation of such modes (cf. Footnote~5).
Verifying this, and comparing with theoretical formulations
of anomalous diffusion 
(e.g.~\citealp{BarOr2013}), may be of interest for further study of the dynamical effects of 
stochastic fluctuations with power law spectra considered here.}

Finally we show in Fig.~\ref{fig:velav}
the distribution of speeds {of the gas fluid elements}. 
This is calculated from the speeds in all gas 
cells within the indicated radii and the result is averaged over all time snapshots (excluding the
first Gyr). 
At larger radial bins the speeds closely follow a Maxwellian distribution, (but with long tail at high speeds), with an average value saturating at  
$\simeq 30 ~{\rm km/s}$. The speeds are smaller as one moves towards the centre.


\section{Energy input and cutoff in the halo distribution function}
\label{sec:cut}

\begin{figure}
    \centering
	\includegraphics[width=\linewidth]{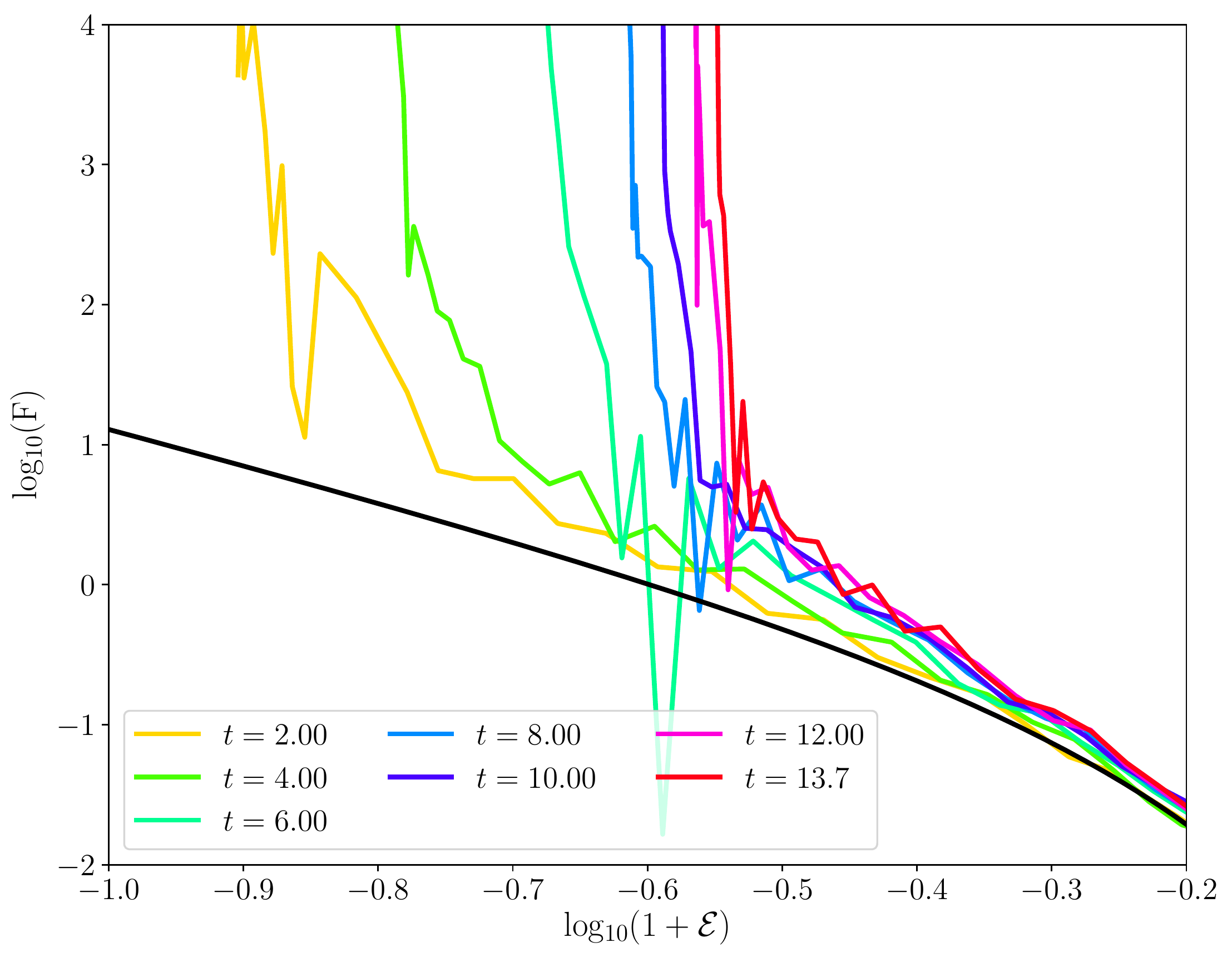}
    \caption{Dimensionless phase space distribution function $F$ in terms of dimensionless  energy  $\mathcal{E}$, measured from the minimal energy in the initial NFW potential. In the initial NFW cusp the phase space density of halo particles continuously increases  as the zero energy point of the NFW system (at $\mathcal{E} = -1$) is approached. As energy is transferred from the fluctuating gas, however, a low energy cutoff in the distribution emerges, as particles migrate towards progressively higher energies.}
    \label{fig:distfun}
\end{figure}

\begin{figure}
    \centering
	\includegraphics[width=\linewidth]{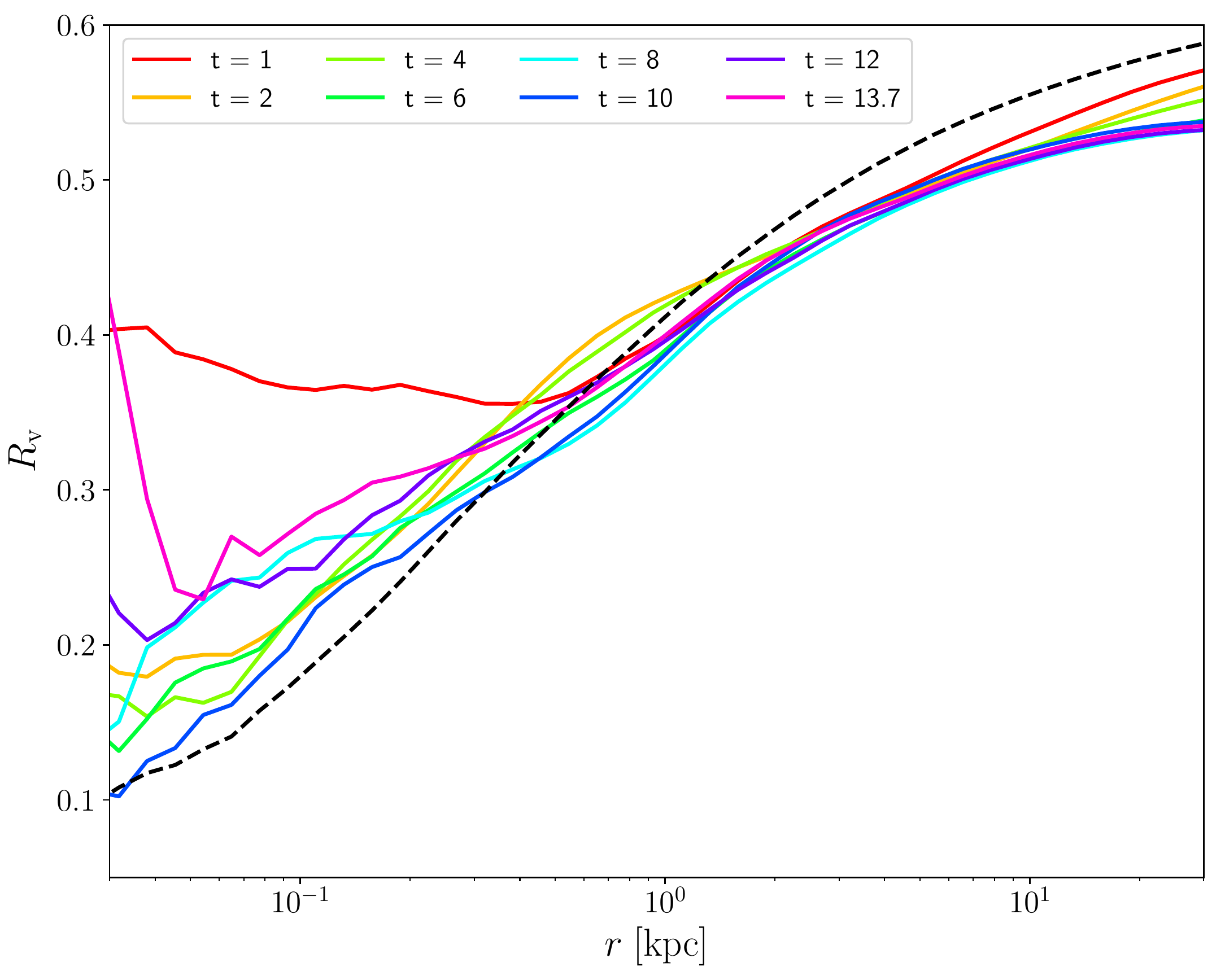}
    \caption{The ratio $R = T_r/\Phi_r$, where $T_r = M \langle T_i \rangle$, $\Phi_r = \frac{M}{2} \langle \Phi_i \rangle$, and $T_i$ and $\Phi_i$ are  halo particle specific kinetic energy and potential energies, 
and the averages are taken inside radii $r$.}
    \label{fig:vir}
\end{figure}

\begin{figure}
\centering
	\includegraphics[width=\linewidth]{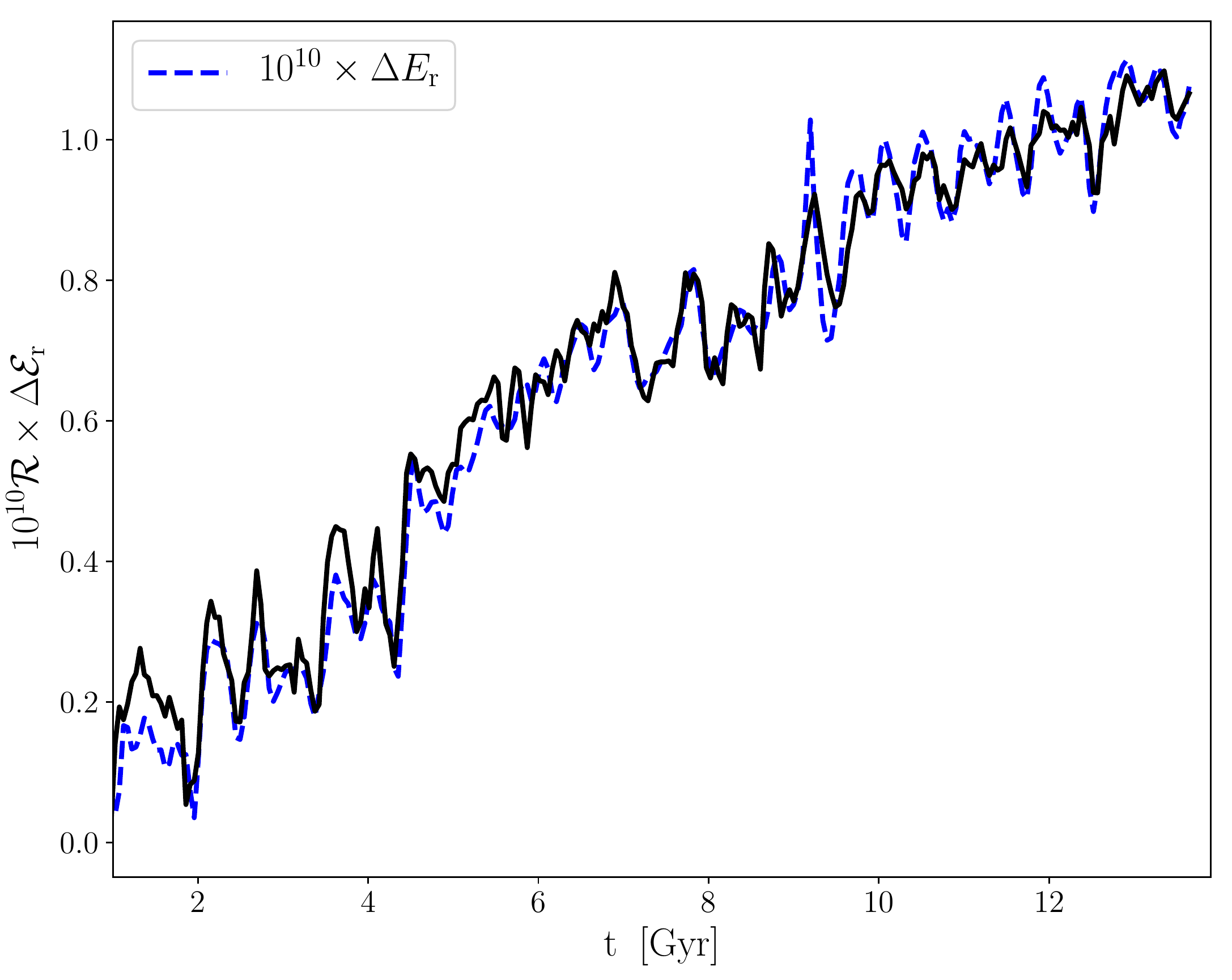}
    \caption{Comparison of total energy transferred from gas fluctuations  to the halo (as in Fig.~\ref{fig:Ein_main}) with the average energy 
    gained by dark matter particles evaluated {\it via} the cutoff in their distribution function $\mathcal{E}_c$. The latter is 
    inferred from Fig.~\ref{fig:distfun} and transformed through the factor $\mathcal{R}$ from   
    equation~\ref{eq:Rfac}.}
    \label{fig:part_v_tot}
\end{figure}

\begin{figure}
\centering
	\includegraphics[width=\linewidth]{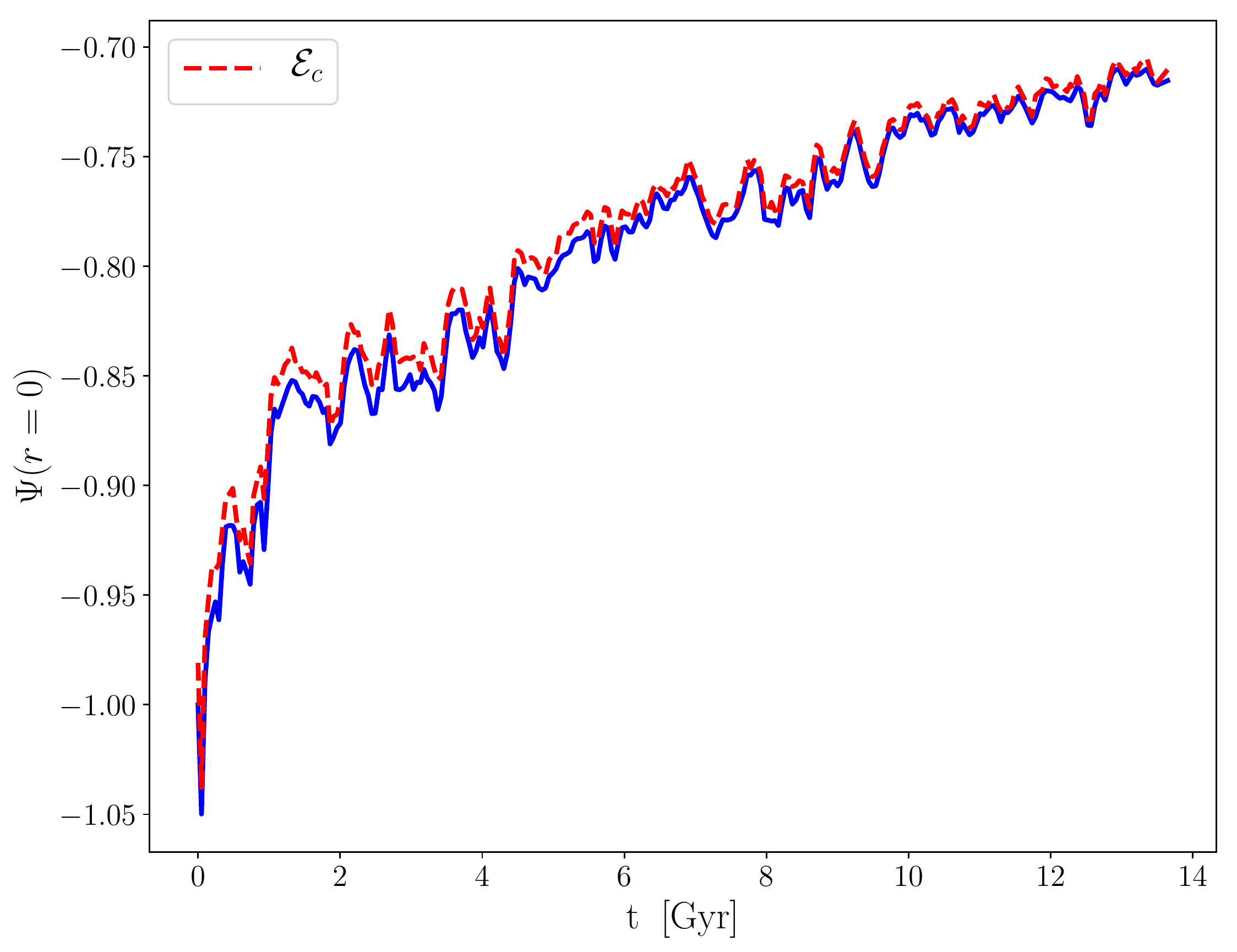}
    \caption{Comparison of the minimal dimensionless specific particle energy $\mathcal{E}_c$ 
    (corresponding to the energy cutoffs in Fig.~\ref{fig:distfun}) with that of the dimensionless potential $\Psi = \Phi/4 \pi G \rho_s r_s^2$ at $r=0$ (as in Fig.~\ref{fig:distfun} the normalisation corresponds to the initial NFW parameters listed in Table~1).}
    \label{fig:Psi_v_Ec}
\end{figure}

We obtain the phase space distribution $f$ 
of halo particles by using the Eddington formula for a spherical density distribution with isotropic velocities~(\citealp{BT}), by inserting numerical 
values for the dark matter potential and density 
inferred from the simulation. 
In Fig.~\ref{fig:distfun} we plot the dimensionless distribution function of halo particles, $F (\mathcal{E}) = f/(4 \pi G)^{3/2} r_s^3 \rho_s^{1/2}$, 
where $\mathcal{E}= E/4 \pi G \rho_s r_s^2$, 
against $1+\mathcal{E}$, which measures excess in energy above the minimum energy in the initial NFW system (i.e with zero point at $\Phi_{\rm NFW} (r = 0)$). 
Here one probes a cutoff in the distribution  
function  at progressively 
larger values of $1+ \mathcal{E}$; as the evolution progresses, 
halo particles migrate to higher energies 
as a result of their interaction with the fluctuating 
gas. Lower energy levels are emptied in the process. 

We wish to relate this effect to the total energy input
examined in Section~\ref{sec:relax}, which is initially pumped into the 
particle trajectories in the form 
of kinetic energy.  
For this purpose, we
may rewrite equation~(\ref{eq:Eind}) 
in terms of the average kinetic and potential energies of the 
 halo particles,
\begin{equation}
E_r = M  \left(\langle T_i \rangle + \frac{1}{2} \langle \Phi_i  \rangle \right),     
\end{equation}
where the averages are taken over all particles within radius $r$ and $M = M (<r)$
refers to the mass enclosed therein. Then we define $T_r = M \langle T_i \rangle$,
$\Phi_r = M \langle \Phi_i \rangle$ and $R_v (r) = - T_r/\Phi_r$ such that
\begin{equation}
 E_r = \frac{M}{2} \left(1-R_v\right) \langle \Phi_i \rangle.    
\end{equation}
Also, using $\langle E_i \rangle = \langle T_i \rangle + \langle \Phi_i \rangle$ and 
$\langle T_i \rangle/ \langle \Phi_i \rangle = -R_v/2$, 
\begin{equation}
\langle E_i \rangle =   \left(1- \frac{R_v}{2}\right)  \langle \Phi_i \rangle.
\label{eq:penergy}
\end{equation}
And thus, 
\begin{equation}
E_r = \frac{M}{2}~\left(\frac{1-R_v} {1-\frac{R_v}{2}}\right)~\langle E_i \rangle.    
\label{eq:relateein}
\end{equation}
Fig.~\ref{fig:vir} shows the `virial ratio' $R_v$ of the collection of halo particles within
radius $r$. Within $r_s$, it is seen to vary from $\sim$$0.1$, in the inner region, to $\sim$$0.4$ at $r \simeq r_s$. The latter is the relevant value if we  are interested in evaluating the average energy  
input within the initial NFW cusp. 

To compare with 
the energy input inferred from Fig.~\ref{fig:Ein_main},  we first estimate the cutoff $\mathcal{E}_c$ through the intersection of the distribution function with 
a horizontal line drawn around the 
maximum values in the distribution function in Fig.~\ref{fig:distfun} 
(we don't explicitly draw this; it can be considered to correspond 
to the upper horizontal limiting line of the figure panel at $F= 10^4$).

As with Fig.~\ref{fig:Ein_main}, we subtract the average over
$0.5 \le t \le 1.5 ~{\rm Gyr}$ from the numbers thus inferred and start our comparison at $t = t_{\rm diff} = 1~{\rm Gyr}$ (which is again our $T=0$ point). 
We denote the result 
by $\Delta \mathcal{E}_{r}$, and associate  it 
with the average energy input per  particle 
$\langle \Delta E_i \rangle = (4 \pi G \rho_s r_s^2)~\Delta \mathcal{E}_r$. The full transformation is then $\Delta E_r = \mathcal{R}~\Delta \mathcal{E}_r$, with 
\begin{equation}
\mathcal{R} =   2 \pi G \rho_s r_s^2~M~\left(\frac{1-R_v}{1-\frac{R_v}{2}}\right).
\label{eq:Rfac}
\end{equation}
Fig.~\ref{fig:part_v_tot} shows the result of matching the total energy input at the 
saturation radial level  (converging lines in Fig.~\ref{fig:Ein_main}), with  
$\mathcal{R}~\Delta \mathcal{E}_r$, 
obtained as just described, assuming $R_v =  0.4$ and $M = M (< 1 {\rm kpc})$
in equation~(\ref{eq:Rfac}).  

Finally we show that the dimensionless 
minimal cutoff energy scale $\mathcal{E}_c$ in fact corresponds to the corresponding dimensionless potential ($\Psi$) at $r=0$  (Fig.~\ref{fig:Psi_v_Ec}). This illustrates the process of energy redistribution; even if the initial  energy is transferred from the 
gas to the halo particles in the form of kinetic energy, particles may still end up with zero kinetic energy, but still with increased energy as a result of a shallower self consistent potential well, corresponding to the formed core.

\section{Fokker Planck flux for power law and exponential differential energy distributions}

\label{app:powerflux}

{\ In this appendix we examine the three major  simplifications made 
to arrive at  
equation~(\ref{eq: approxFP}). 
Namely, that the differential energy 
distribution may be approximated by its average value (equation~\ref{eq:a_p}); 
that the energy space flux of dark matter particles, migrating into higher energy levels under the  influence of stochastic gas fluctuations, 
may be described in terms of the first order diffusion coefficient 
(as in equation \ref{eq:Mflux}); and that the flux through energy surface $E$ due to the self consistent evolution of the system may likewise be absorbed in the parameter $a$. 

In particular, we show how the assumption 
of absorbing the complexity of the situation in a single parameter $a$ 
may be reasonable when the 
differential energy distribution may be approximated by a power law 
(which we will find to be the 
case in the region where the halo cusp is transformed into a core). 
We also show that this parameter is positive, allowing for outward mass flow, for systems with density decreasing with radius. 

We finally discuss the qualitatively different behaviour at intermediate radii, where the differential energy distribution may be approximated instead by 
an exponential. In this latter case, the competing effects of the 
fluxes associated with the first and second order diffusion coefficients 
may cancel out, and the Fokker Planck flux may in fact vanish. 
}

\subsection{Approximating the differential energy distribution by its average}

{
Consider a spherical density distribution 
$\rho_p \propto r^{\gamma}$, with $-2 < \gamma < 0$. 
If the distribution is infinite in radial 
extent, the  potential $\Phi (r)$, defined in terms of the work done to 
take a particle to infinity, diverges. But 
$\phi (r)  \propto r^{-\beta}$, associated with the
work done in moving a particle from the centre to radius $r$, is well defined, and  
from the Poisson equation $\gamma = - (\beta + 2)$. 
One can also define the specific energy $E_p = \phi (r) + \frac{1}{2} v^2$.
In the notation  used in this work, $E_p$ will correspond 
to $E- E_0$ when $\Phi (r)$
is well defined --- as in the central region of a density distribution that, 
though well approximated by $\rho_p \sim r^{\gamma}$ in the central region, 
eventually falls off sufficiently steeply with increasing radius. In this  
case $\phi(r) = \Phi(r) - \Phi (0)$.

For  such power law systems, the mass weighed distribution function of a  
configuration with isotropic velocities can be written as 
$g = g_0 E_p^{2/\beta - 1/2}$ and the density of states 
$p = p_0 E_p^{1/2 - 3/\beta}$, where $g_0$ and $p_0$ are positive 
constants (\citealp{Evans1994, El-Zant08}). 
In general, the density of states corresponds to the 
area  that encloses a phase space volume
\begin{equation}
q (< E) = \frac{16 \pi^2}{3} \int_0^{r_{\rm max}} 
 [2 (E- \Phi)]^{3/2} r^2 d E,  
 \label{eq:q}
\end{equation}
where $r_{\rm max} (E)$is the maximum radius 
that a particle with specific energy $E$ can reach. 
Thus $p(E) = \partial q (< E)/\partial E$. 

The differential energy distribution is 
thus $\partial M (< E_p)/\partial E_p = p g = p_0 g_0 E_p^{-1/\beta}$, which for 
a power law distribution embedded in a larger halo tends to  
\begin{equation}
    \frac{\partial M (< E)}{\partial E} = p g = p_0 g_0 (E - E_0)^{-1/\beta},
 \label{eq:diffenpower}   
\end{equation}
from where it also follows that 
\begin{equation}
\frac{\partial M (< E)}{\partial E} = (1- 1/\beta) \frac{M (< E)}{E-E_0}. 
\label{eq:ap}
\end{equation}

Therefore, for such systems, equation~(\ref{eq:a_p}) becomes exact with $a_p = 1-1/\beta = (3 + \gamma)/(2+\gamma)$, as mentioned in Section~\ref{sec:massmod}. 
}

\subsection{Use of first order diffusion coefficient}

{

As the process of core formation proceeds on a significantly larger 
timescale than the dynamical time, one may use the orbit averaged 
Fokker Planck equation to describe the cumulative effect of small 
 perturbations emanating from stochastic fluctuations (e.g.~\citealp{Spitzer1987, BT}). 
 In flux conservation form, 
 the mass change inside energy level $E$ can then be written as~(\ref{eq: approxFP})
for an isotropic system (i.e., with $g = g (E)$) may then be written as (\citealp{El-Zant08})
\begin{equation}
\frac{\partial M (<E)}{\partial E} = - F_E + g \frac{\partial q}{\partial E}, 
\label{eq:fullmassflux}
\end{equation}
which says that a decrease in  total
mass of (in our case dark matter) particles with energy less than $E$
is due to the Fokker Planck flux out of the volume enclosed by $E$ 
in addition to changes in that volume resulting from the evolution in the structure 
of the system. 

A major simplification leading to 
equation~(\ref{eq: approxFP}) involves 
the use of the first order energy diffusion coefficient $D[\Delta E] = \langle \Delta E \rangle/T$, 
the value of which is inferred 
directly from the simulation, instead of $F_E$.
Theoretically however, terms associated with the second order coefficient 
$D[(\Delta E)^2] = \langle (\Delta E)^2 \rangle/T$, 
constitute a comparable contribution, as both coefficients involve the velocity 
variance  $\langle (\Delta v)^2 \rangle$. 
Nevertheless, as we now show, if $D [\Delta E]$ can be considered an energy independent 
constant, 
as assumed in equation~(\ref{eq:Dcoef}), the total flux is simply proportional to
$D [\Delta E]$~\footnote{In the terminology of Sections~\ref{sec:relaxpar} and~\ref{sec:enin},  equation~(\ref{eq:Dcoef}) effectively 
assumes a global energy input, with effective gas density $\rho_0$
corresponding to the average density within the saturation radius 
$r_{\rm sat}$, and parameters corresponding to those in equation~\ref{eq:Ein}}. 
Furthermore, for power law systems, the inclusion of $D[(\Delta E)^2]$ 
in the evaluation of the total flux leads 
to a simple rescaling that can be absorbed in a parameter $a$.

In general, for a spherical system with isotropic velocities, the orbit averaged Fokker Planck flux can be written as (e.g.~\citealp{El-Zant08})~\footnote{Note that  $\langle \Delta E \rangle$ and $\langle (\Delta E)^2 \rangle$ are defined there directly as the rates of changes of energy and its variance, and therefore are themselves the diffusion coefficients referred to here by $D [\Delta E]$ and $D [(\Delta E)^2]$.} 
\begin{equation}
   F_E =  \left(D_E  -D_{EE}  \frac{\partial}{\partial E}\right) g,
\end{equation}
where
\begin{equation}
    D_E = p  D [\Delta E] - \frac{1}{2} \frac{\partial}{\partial E} 
    \left(p D[(\Delta E)^2]\right),  
\end{equation}
and 
\begin{equation}
    D_{EE} = \frac{1}{2} p D[(\Delta E)^2]. 
\end{equation}

As dark matter particles are too light to exchange energy through dynamical 
friction with the fluctuating gas field,   
 $\langle \Delta E \rangle = \frac{1}{2} \langle (\Delta v)^2 \rangle$. Also, to leading order in perturbation $\Delta {\bf v}/{\bf v}$,  $\langle (\Delta E)^2 \rangle = \langle ({\bf v} . \Delta {\bf v})^2 \rangle = \frac{1}{3} v^2 \langle (\Delta v)^2 \rangle$ (e.g.~\citealp{PenarrubiaA}).
If we assume that $D[\Delta E$] is independent of energy
so will $\langle (\Delta v)^2 \rangle$. 
In this case, the first order energy diffusion coefficient is already orbit averaged, and averaging the second order one simply involves 
evaluating the mean of $v^2$ over the energy shell of surface area $p$, 
which we denote by $\langle v^2 \rangle_E$. 
The Fokker Planck 
flux across that shell is then given by 
\begin{equation}
F_E = D [\Delta E]  \left[p g  - \frac{1}{3} 
\frac{\partial}{\partial E} \left(\langle v^2 \rangle_E pg\right)\right].
\label{eq:fullflux}
\end{equation}
The total flux is thus simply proportional 
to the first order coefficient 
$D [\Delta E]$.  

Furthermore, for a power law region embedded in a larger halo, the density 
of states may be approximated by  $p =  p_0 (E- E_0)^{1/2 - 3/\beta}$, 
and the corresponding $\langle v^2 \rangle_E = 6 \beta (3 \beta - 6)^{-1} (E-E_0)$, and thus
\begin{equation}
F_E = p_0 g_0 D [\Delta E] \left( \frac{6 + \beta}{6 - 3 \beta} \right) (E - E_0)^{-1/\beta}, 
\end{equation} 
 From equations (\ref{eq:diffenpower}), and~(\ref{eq:ap}) this can be rewritten as  
\begin{equation}
F_E = 
\left(\frac{(6 + \beta)(\beta -1)}{\beta (6 - 3 \beta)} \right) 
 \frac{M (< E)}{E-E_0} D [\Delta E]. 
\label{eq:powerflux}
\end{equation}

Thus, for power law systems, the total flux is proportional to 
to the flux associated with $D [\Delta E]$
through a constant scaling factor (in brackets), that depends
on the steepness of the power law. 
It is also always positive for systems 
with density decreasing with radius ($\beta > -2$), allowing for mass 
outflow from stochastic fluctuations. 

}

\subsection{Self consistent evolution and evaluation of parameter $a$}
\begin{figure}
    \centering
	\includegraphics[width=\linewidth]{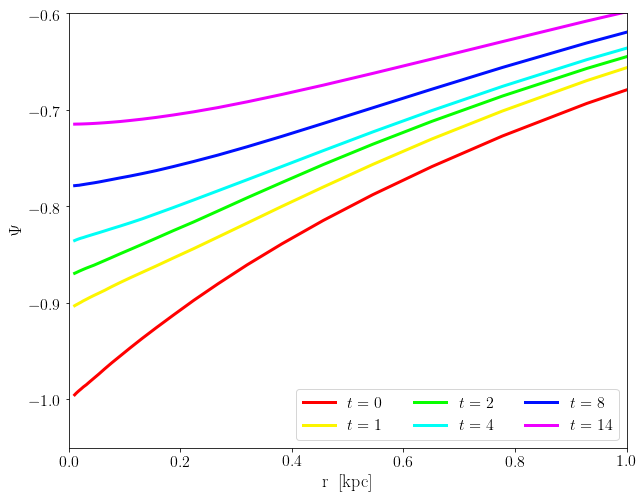}
    \caption{Dimensionless potential (defined as in Fig.~\ref{fig:Psi_v_Ec}) as a function of radius and time, as the central halo region evolves 
    under the influence of the gas flcutuations.}
    \label{fig:PsiEvol}
\end{figure}

\begin{figure}
    \centering
	\includegraphics[width=\linewidth]{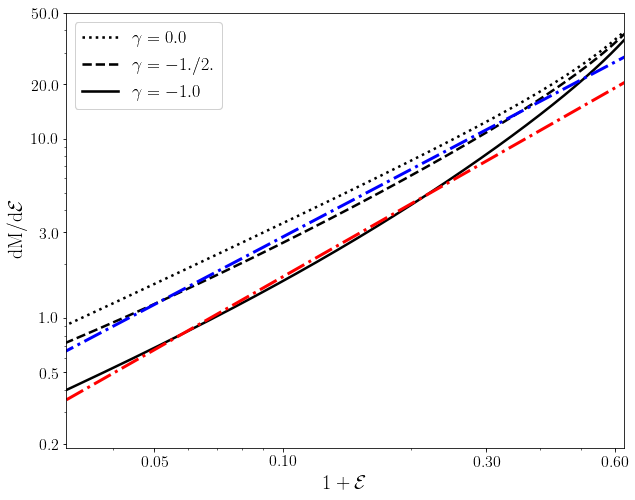}
    \caption{Differential energy distributions in the inner regions of three density profiles with  
    $\rho = \rho_s  r^{\gamma} (1 + r/r_s)^{-3-\gamma}$, where $\gamma =-1$ for the NFW profile. The dimensionless energy 
    $\mathcal{E} = E/ 4\pi G \rho_s r_s^2$
    is scaled such that the three profiles converge to the same densities as $r \gg r_s$  (thus $C = 1$ 
    for the NFW,  $C = 3/2$ for $\gamma = -1/2$ and $C = 2$ for $\gamma = 0$; cf. \citealp{WidrowDF2000} for more details regarding these models).  
    The dashed dotted lines are power law fits.  The one  
    fitting the NFW has slope $1.35$. It approximates the differential energy  distribution over energy scales 
    $ 0.03 \lesssim \mathcal {E} +1  \lesssim 0.33$, which enclose 
    zero velocity spheres with radii of about $50 {\rm pc}$ and $1 {\rm kpc}$ in our simulation (with $r_s = 0.88 {\rm kpc}$ as in Table~1), which correspond, respectively, to about the smallest radius inside which the halo density may be reliably resolved and the region inside which the gas fluctuations have significant effect on the halo profile. The dashed dotted line fitting the shallower profiles ($\gamma = -1,2, 0$) has only slightly flatter slope of $1.25$.}
    \label{fig:Diffen_power}
\end{figure}

{

To estimate the second term on the right hand side of equation~(\ref{eq:fullmassflux}) 
we note that, from equation~(\ref{eq:q}),  
\begin{equation}
    \frac{\partial q}{\partial t} = - 16 \pi^2 \int_0^{r_{\rm max}} 
   \frac{\partial \Phi}{\partial t} [2 (E- \Phi)]^{1/2} r^2 d E.
\end{equation}
Therefore the orbit averaged change of potential due to the self consistent evolution at energy $E$ is then 
\begin{equation}
\langle \dot{\Phi}  (r) \rangle_E = - \frac{1}{p} \frac{\partial q}{\partial t}. 
\end{equation}

Recall that in Fig.~\ref{fig:Psi_v_Ec} we found that 
the change of potential at the centre of the halo ($r=0$),  resulting from the gas fluctuations, corresponds
to the average energy gained per particle, taking into account the self consistent 
evolution of the system. Fig.~\ref{fig:part_v_tot}, on the other 
hand, showed that that this 
corresponds to the average energy per unit mass 
pumped into the central halo multiplied by a factor 
$\frac{2- R_v} {1-R_v}$, with $R_v = 0.4$; i.e., by a factor factor 
of about $2.7$. Thus 
$\dot{\Phi}_{E = E_0} = 2.7 \times D [\Delta E]$.  
But, as can be seen  Fig.~\ref{fig:PsiEvol}, the change in the 
potential with time is significantly smaller at larger radii.  
Thus $\dot{\Phi}_{E > E_0} < 2.7 \times D [\Delta E]$; 
at $r = 1 {\rm kpc}$ (i.e., about the radius inside which the core forms)
the total change is a bit more than a quarter of the value at $r=0$. 
As the radial variation is not very steep we will approximate 
the change in energy due to evolution of the self consistent potential 
with an effective value, such that $\dot{\Phi}_{\rm eff} = \eta D [\Delta E]$, 
where from the above considerations $0.7 < \eta < 2.7$.
As we will see below, the mass particles within energy shell $E$ increases steeply with energy away from the zero point in the central power law regions, with $M (<E) \sim (E-E_0)^{2.3}$. An effective value, replacing  
 $\langle \dot{\Phi} (r) \rangle_E$, 
would thus correspond to more weight over larger energies, 
and therefore larger radii. 
We may thus expect $\eta$ to be closer to its lower, rather than upper, limit. 

Valid values of $\eta$ are actually tightly constrained 
by the mass and density evolution in our model galaxy
(Fig.~\ref{fig:core}), and the differential energy distribution in the inner halo. 
To see this we replace $\partial q/\partial t$ by $- p \eta D [\Delta E]$,
and use equation~(\ref{eq:powerflux}), 
to rewrite equation~(\ref{eq:fullmassflux}) as
\begin{equation}
\frac{\partial M (< E)}{\partial t} = -
\left(\frac{(6 + \beta)(\beta -1)}{\beta (6 - 3 \beta)} + \eta \right) 
 \frac{M (< E)}{E-E_0} D [\Delta E]. 
\end{equation}
This is equivalent to equation~(\ref{eq: approxFP}) if $a$ is equal to the term 
in brackets. That is, 
\begin{equation}
a = \frac{(6 + \beta)(\beta -1)}{\beta (6 - 3 \beta)} + \eta.
\label{eq:calca}
\end{equation}

To determine $\beta$ we need to show that the differential energy distribution of the inner halo may be fit by a power law and determine its index, in which case $\beta$ is fixed by equation~(\ref{eq:diffenpower}). 
Fig.~\ref{fig:Diffen_power} shows  
that in the region where the core 
replaces the cusp the differential energy 
distribution of an NFW profile 
may  indeed be fit by a power law, with index of about $1.35$, which corresponds to $\beta = - 0.74$. 
{Furthermore, as can also be seen from Fig.~\ref{fig:Diffen_power} the differential energy slopes of models 
with the same asymptotic physical density distribution for $r \gg r_s$, but have
milder inner cusps ($\gamma = - 1/2$), or even cores ($\gamma = 0$), can be approximated by similar power laws (with $\beta \simeq-0.8$),  suggesting that the power law approximation may hold as the cusp transforms into core.}

Furthermore, in  
Section~\ref{sec:massmod} we found that a value of 
$2.5 \lesssim a \lesssim 2.8$ can be used to obtain solutions to equation~(\ref{eq: approxFP}) 
that describe the mass and density evolution fairly well (Fig.~\ref{fig:core}, right hand panel). 
For the aforementioned values of $\beta$,  
this corresponds to 
$1 \lesssim \eta \lesssim 1.4$, which is within the expected range inferred above. 

}

\subsection{The vanishing of the flux for exponential differential energies (and constant dispersion)}
\label{app:exp}

\begin{figure}
    \centering
	\includegraphics[width=\linewidth]{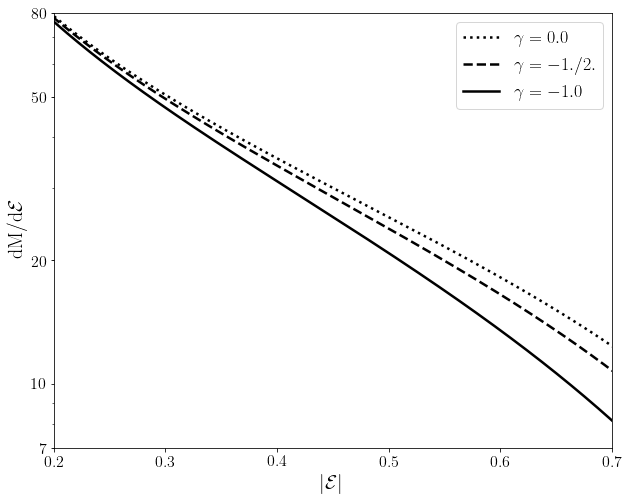}
    \caption{The differential energy distributions of the models 
    of Fig.~\ref{fig:Diffen_power} in the intermediate energy range,
    where the quasi-linear variation  on the log-linear plot suggests an exponential scaling, 
    particularly for 
    $0.35 \lesssim |\mathcal{E}| \lesssim 0.65$, corresponding to zero velocity spheres
    with radii of about 1.4 and 4.7 kpc in our model galaxy.}
    \label{fig:diffmassexp}
\end{figure}

{

At larger energies, corresponding to intermediate radii beyond $r_s$, 
the differential energy distribution is better approximated by an exponential, 
rather than a power law, as Fig.~\ref{fig:diffmassexp} illustrates.  
At such radii, 
the halo is nearly isothermal, with density $\rho_p \sim 1/r^2$, and $\langle v^2 \rangle_E$   approximately constant.
Assuming this, 
 and inserting $\partial M(< E)/\partial E \propto e^{\beta E}$ into~(\ref{eq:fullflux}), 
it is easily seen that the flux is nullified if $\beta = 1/\sigma^2$,
where $\sigma^2 =  \frac{1}{3} \langle v^2 \rangle$ is the one dimensional velocity variance.
This may explain the formation of cores up to the region where the 
differential energy distribution may be approximated by a power law, where the flux is positive, 
but that the response to fluctuations is more robust when 
the exponential is a better approximation. 

}


\bsp	
\label{lastpage}
\end{document}